\documentclass[useAMS,usenatbib]{mn2e}

\usepackage{epsfig}
\usepackage{amsmath}
\usepackage{amssymb}
\usepackage{ifthen}
\usepackage{txfonts}
\usepackage{rotating}
\usepackage{url}
\usepackage{varioref}
\usepackage{verbatim}
\usepackage{latexsym}
\usepackage{graphicx}

\DeclareSymbolFont{cmletters}{OML}{cmm}{m}{it}
\DeclareMathSymbol{v}{\mathalpha}{cmletters}{"76}

\usepackage{multirow}

\newcommand*{\bfrac}[2]{\genfrac{}{}{0pt}{}{#1}{#2}}


\DeclareSymbolFont{cmletters}{OML}{cmm}{m}{it}

\DeclareMathSymbol{v}{\mathord}{cmletters}{"76}

\def\be{\begin{equation}}
\def\ee{\end{equation}}

\newcommand{\detg}{{\sqrt{-g}}}
\newcommand{\del}{{\partial}}

\newcommand{\bR}{{\bf{R}}}
\newcommand{\dF}{{^{^*}\!\!F}}

\newcommand{\rholab}{{\rho}}

\newcommand{\uvec}{{\underline{u}}}
\newcommand{\bvec}{{\underline{b}}}
\newcommand{\Bvec}{{\underline{B}}}

\newcommand{\Fvec}{{F}}

\newcommand{\xvec}{{\underline{x}}}
\newcommand{\Avpotvec}{{\underline{A}}}

\newcommand{\rvec}{{\underline{r}}}
\newcommand{\vvec}{{\underline{v}}}
\newcommand{\etavec}{{\underline{\eta}}}
\newcommand{\reluvec}{{\underline{\tilde{u}}}}

\newcommand{\alf}{Alfv\'en}

\newcommand{\erg}{{\rm\,erg}}

\newcommand{\gdet}{\sqrt{-g}}

\newcommand{\cut}[1]{\hbox{}}

\DeclareSymbolFont{cmletters}{OML}{cmm}{m}{it}
\DeclareMathSymbol{v}{\mathalpha}{cmletters}{"76}

\usepackage{subfigure}
\usepackage{ifthen}
\usepackage[usenames,dvipsnames]{color}
\usepackage{graphicx}

\usepackage{hyperref}

\newcommand\aj{\rmfamily{AJ}}%
\newcommand\araa{\rmfamily{ARA\&A}}%
\newcommand\apj{\rmfamily{ApJ}}%
\newcommand\apjl{\rmfamily{ApJ}}%
\newcommand\apjs{\rmfamily{ApJS}}%
\newcommand\apss{\rmfamily{Ap\&SS}}%
\newcommand\aap{\rmfamily{A\&A}}%
\newcommand\mnras{\rmfamily{MNRAS}}%
\newcommand\prd{\rmfamily{Phys.~Rev.~D}}%
\newcommand\pasj{\rmfamily{PASJ}}%
\newcommand\nat{\rmfamily{Nature}}%
\newcommand\jqsrt{\rmfamily{J.~Quant.~Spec.~Radiat.~Transf.}}%
\defcitealias{bz77}{BZ77}

\newcommand{\emath}{{\rm e}}
\newcommand{\eff}{{\eta}}

\newcommand{\Qone}{Q_{\theta,\rm MRI}}
\newcommand{\Qx}{Q_{x,\rm MRI}}
\newcommand{\Qoneweak}{Q_{\theta,\rm weak,MRI}}
\newcommand{\Qtwo}{{S_{d,\rm MRI}}}

\newcommand{\Qtwoweak}{{S_{d,\rm weak,MRI}}}
\newcommand{\Qthree}{{Q_{\phi,\rm MRI}}}
\newcommand{\Qthreeweak}{{Q_{\phi,\rm weak,MRI}}}

\renewcommand{\vec}[1]{\bmath{#1}}

\newcommand{\rhorest}{{\rho_{0}}} % rest-mass density
\newcommand{\ug}{{e_{\rm gas}}} % gas internal energy density
\newcommand{\pg}{{p_{\rm gas}}} % gas pressure

\newcommand{\sB}{{\mathcal{B}}}

\newcommand{\tu}{{\tilde{u}}}
\newcommand{\tv}{{\tilde{v}}}
\newcommand{\tQ}{{\tilde{Q}}}

\newcommand{\SM}{{\rm SM}}

\newcommand{\SR}{{\rm SR}}

\title[3D GRMHD Simulations of Super-Eddington Accretion]
{Three-Dimensional General Relativistic Radiation Magnetohydrodynamical Simulation of Super-Eddington Accretion, using a new code HARMRAD with M1 Closure}

\author[J.~C.~McKinney,
A.~Tchekhovskoy,
A.~Sadowski,
R.~Narayan]
{Jonathan C. McKinney$^1$\thanks{\hbox{E-mail: jcm@umd.edu~(JCM)}},
Alexander Tchekhovskoy$^2$,
Aleksander Sadowski$^3$,
Ramesh Narayan$^3$
\\
 $^1$University of Maryland at College Park, Dept. of Physics, Joint Space-Science Institute, 1117 John S. Toll Building \#082, College Park, MD 20742, USA \\
 $^2$Astronomy Department and Theoretical Astrophysics Center, University of California, Berkeley, 601 Campbell Hall, Berkeley, CA 94720, USA; Einstein Fellow \\
 $^3$Harvard-Smithsonian Center for Astrophysics, 60 Garden St., Cambridge, MA 02134, USA \\
  }{
}

\begin{document}
\date{Accepted 2014.  Received 2013; in original form 2013.}
\pagerange{\pageref{firstpage}--\pageref{lastpage}} \pubyear{2013}
\maketitle

\label{firstpage}

\begin{abstract}

  Black hole (BH) accretion flows and jets are dynamic hot
  relativistic magnetized plasma flows whose radiative opacity can
  significantly affect flow structure and behavior.  We describe a
  numerical scheme, tests, and an astrophysically relevant application
  using the M1 radiation closure within a new three-dimensional (3D)
  general relativistic (GR) radiation (R) magnetohydrodynamics (MHD)
  massively parallel code called HARMRAD.  Our 3D GRRMHD simulation of
  super-Eddington accretion (about $20$ times Eddington) onto a
  rapidly rotating BH (dimensionless spin $j=0.9375$) shows sustained
  non-axisymmemtric disk turbulence, a persistent electromagnetic jet
  driven by the Blandford-Znajek effect, and a total radiative output
  consistently near the Eddington rate.  The total accretion
  efficiency is of order $20\%$, the large-scale electromagnetic jet
  efficiency is of order $10\%$, and the total radiative efficiency
  that reaches large distances remains low at only order $1\%$.
  However, the radiation jet and the electromagnetic jet both emerge
  from a geometrically beamed polar region, with super-Eddington
  isotropic equivalent luminosities.  Such simulations with HARMRAD
  can enlighten the role of BH spin vs.\ disks in launching jets, help
  determine the origin of spectral and temporal states in x-ray
  binaries, help understand how tidal disruption events (TDEs) work,
  provide an accurate horizon-scale flow structure for M87
  and other active galactic nuclei (AGN), and isolate whether AGN
  feedback is driven by radiation or by an electromagnetic, thermal,
  or kinetic wind/jet.  For example, the low radiative efficiency and
  weak BH spin-down rate from our simulation suggest that BH growth
  over cosmological times to billions of solar masses by redshifts of
  $z\sim 6$--$8$ is achievable even with rapidly rotating BHs and ten
  solar mass BH seeds.

\end{abstract}


\section{Introduction}
\label{sec_intro}
\newcommand{\MBH}{{M}}
\newcommand{\MBHO}{{M_i}}
\newcommand{\Mdot}{{\dot{M}_0}}
\newcommand{\Mdotedd}{{\dot{M}_{\rm Edd}}}

Modern black hole (BH) accretion theory has identified radiative
cooling and transport as having a significant effect on accretion disk
states and temporal behaviors.  Without radiation, equations like the
test field limit of the general relativistic (GR) magnetohydrodynamic
(MHD) equations can be solved to obtain a single solution that applies
to arbitrary black hole mass $\MBH$ and mass accretion rate
$\Mdot$ because only two physical constants, the speed of light $c$
and gravitational constant (times mass) $G\MBH$, appear
independently. Introduction of a radiative scale via the electron
scattering Thomson cross section $\sigma_T$ (giving scattering opacity
$\kappa_{\rm es}=\sigma_{\rm T}/m_p$, with Planck's constant per
electron mass, $\hbar/m_e$ and proton mass $m_p$) and radiation
constant ($a_{\rm rad}$, with $\hbar$ appearing alone) forces
$\Mdot$ and $\MBH$ to be specific physical values.  Then, a useful
scale that measures the importance of radiative effects in accretion
disks is the Eddington luminosity, as due to a radial balance between
a radiative force $F_{\rm rad} = L\sigma_T/(4\pi c r^2)$ and gravity
$F_{\rm grav} \approx G\MBH m_p/r^2$ for radiative luminosity $L$, radius
$r$.  The Eddington luminosity is given by
\begin{equation}\label{Ledddef}
L_{\rm Edd}=\frac{4\pi G \MBH c}{\kappa_{\rm es}} \approx 1.3\times 10^{46} \frac{\MBH}{10^8M_{\odot}}{\erg~{\rm s}^{-1}},
\end{equation}
which can be used to normalize quantities like the mass accretion rate
($\Mdot c^2$) and $L$.  As done for this paper's abstract, one can
also choose to normalize $\Mdot$ by $\Mdotedd =
(1/\eta_{\rm NT})L_{\rm Edd}/c^2$, where $\eta_{\rm NT}$ is the
nominal accretion efficiency for the Novikov-Thorne thin disk solution
\citep{Novikov:1973:IBH} (commonly, a fixed $\eta_{\rm NT}=0.1$ is
used, but we include the spin dependence).

At very low accretion rates, $L/L_{\rm Edd}\ll 10^{-2}$, e.g., Sgr
A$^*$ the super-massive BH (SMBH) at our Galactic Center
\citep*{nym95}, the plasma becomes a radiatively inefficient accretion
flow (RIAF) with some dissipated energy advecting into the BH and the
rest ejected into a wind.  These flows could be advection-dominated
accretion flows (ADAFs)
\citep{Ichimaru:1977:BBA,nar94,nar95b,acgl96,Abramowicz:1995:TEA,pg98},
convection-dominated accretion flows (CDAFs) \citep{nia00,qg00}, and
advection-dominated inflow-outflow solutions (ADIOSs)
\citep{Blandford:1999:FGA,beg11}.  Such flows are optically thin and
geometrically thick (disk height ($H$) to cylindrical radius ($R$)
ratio of $|H/R|\sim 0.5$--$0.9$). Analytical and semi-analytical
models agree well with the primary spectral features of, e.g., SgrA*
\citep{yuan03}.  Still, modern GRMHD simulations with physical cooling
suggest that accretion rate of Sgr A* is actually near the limit of
the regime where radiative cooling may be important
\citep{dibietal12}, while systems like M87 that are normally
associated with low luminosity systems may have important radiative
cooling \citep{2011ApJ...735....9M,dibietal12}.

RIAFs have been studied with various GRMHD codes (with no radiative
transfer)
\citep[e.g.,][]{dev03,2003ApJ...589..444G,anninosetal05,2007A&A...473...11D},
and these simulations seek to find thermodynamically and dynamically
self-consistent solutions and to determine what free parameters (such
as net magnetic flux) set the results \citep{narayan12,2012MNRAS.423.3083M,tm12a}.
Some effects of radiation have been included by performing radiative
transfer during post-processing to produce observables
\citep[e.g.][]{schnittmanetal06,shcherbakovetal10}, which is valid
when the radiation has no dynamical importance.  In a few cases,
physically-motivated local cooling has been included
\citep[e.g.,][]{fm09,dibietal12}, which is permissable if the gas is
quite optically thin.  Simulations, however, have not been performed
in a way that self-consistently determines the thermodynamic structure
of the disk, which would be controlled by significant cooling or
regulated by turbulent dissipation and winds.  Instead, simulations
with various initial conditions lead to evolved disk states of various
disk thicknesses, such as $|H/R|\sim 0.1$--$0.15$
\citep{hk01,dev03,beckwith08b,bhk08,bhk09trans} and $|H/R|\sim 0.2$
\citep{hb02,mhm00,mm03,2004ApJ...611..977M,fragile07,mb09} and $|H/R|\sim 0.3$--$0.4$
\citep{ina03,pmntsm10}, as well as radiatively efficient,
geometrically thick flows with $|H/R|\sim 0.6$--$1.0$
\citep{sto01,2001ApJ...554L..49H,igu02,pmw03,ppmgl10,2012MNRAS.423.3083M}.

For higher accretion rates, $10^{-2}\lesssim L/L_{\rm Edd} \lesssim
0.3$, the disk cools efficiently, and the inner accretion disk can
collapse into an optically thick geometrically thin accretion disk
\citep{ss73,Novikov:1973:IBH,thorne74,esin97,esin98,mm03,mccl06,done07}.
In this regime, it remains uncertain whether such a
radiation-dominated disk is stable
\citep{1974ApJ...187L...1L,1978ApJ...221..652P,2009ApJ...691...16H,2013ApJ...778...65J}.
Example systems include transient black hole binaries (BHBs) that
approach near-Eddington accretion rates near the peak of their
outbursts \citep*{mccl06,remi06,done07}, during which they produce a
thermal black-body-like spectrum consistent with standard
$\alpha$-disk theory \citep{ss73,novikovthorne73,frank02} that allows
one to measure black hole spins \citep{mccl11,straub11}.  However,
typical $\alpha$-disk models assume an averaged vertical structure
with limited treatment of the radiation.  Simulations have only so-far
included an ad hoc cooling function that leads to $|H/R|\sim
0.05$--$0.1$
\citep{shafee08,rf08,rm09,noble09,pennaetal10,nkh10,sorathia10,nobleetal11,bas11}.

Near and beyond the Eddington luminosity limit, $L\gtrsim 0.3 L_{\rm
  Edd}$, the accretion flow become geometrically thick and optically
thick, and in this regime the photons can advect or remain trapped
within the flow.  The ``slim disk'' model treats this regime
\citep{abra88,sad09} (but for issues see
\citealt{2002ApJ...574..315O}) and additional physics such as thermal
conduction has been added to such models \citep{2013Ap&SS.346..341G},
but it does not include multi-dimensional effects or magnetic fields.
Super-Eddington accretion may help explain ultra-luminous X-ray
sources as highly super-Eddington stellar-mass BHs
\citep{wata01,2005PASJ...57..513W}, without requiring intermediate
mass BHs (as required if accretion were limited to Eddington rates)
\citep{miller04}.  Also, a few black hole x-ray binaries spend
significant periods of time with $L\gtrsim L_{\rm Edd}$ (e.g., SS433,
\citealt{margon79,margon84,2010PASJ...62L..43T}; GRS1915+105,
\citealt*{fb04}).  In addition, tidal disruption events (TDEs) require
such high accretion rates with geometrically-thick accretion disks
\citep{cb13} and strong magnetic fields
\citep{2011Sci...333..203B,Burrows+11,Levan+11,Zauderer+11,2013arXiv1301.1982T}.

This important regime may determine SMBH mass growth in the Universe
from $z\sim 10$--$20$ to $z\sim 6$--$8$ leading to black holes with
masses of $10^{9}M_\odot$
\citep{collin02,barth03,willot05,fan06,willot10,mort11,2011MNRAS.417.2562K,fab12}
and control the evolution of black hole mass and spin
\citep{2004ApJ...602..312G,2005ApJ...620...69V,2008ApJ...684..822B} as
well as set the degree of active galactic nuclei (AGN) feedback
\citep{2005Natur.433..604D,2005MNRAS.361..776S}.

On the one hand, often it is assumed that mass accretion is limited by
Eddington, since otherwise radiation blows off a massive wind and
prevents accretion (see discussion in \citealt{2009PASJ...61..783T}).
Then one expects the mass accretion rate to be limited to Eddington
($\Mdot\lesssim \Mdotedd$), the luminosity to be limited to Eddington
($L\le L_{\rm Edd}$), and the radiative efficiency ($\eta_{\rm rad}$)
to be rougly given by Novikov-Thorne values, which vary from
$\eta_{\rm rad}\sim 0.057,0.065,0.082,0.26$ for $a/M=0,0.2,0.5,0.99$,
respectively.  Then, the disk luminosity is given by $L = \eta_{\rm
  rad} \Mdot c^2$.  The black hole mass grows from $\MBHO$ due to the
residual mass not converted to energy, so giving $\dot{\MBH} =
(1-\eta_{\rm rad})\Mdot$.  Thus, $\dot{\MBH} \le (\MBH/t_{\rm
  Edd})(1-\eta_{\rm rad})/\eta_{\rm rad}$ where $t_{\rm Edd}\equiv
\kappa c/(4\pi G)\approx 4.5\times 10^8{\rm yr}$.  Hence,
\begin{equation}
\frac{\MBH}{\MBHO} < \exp\left[\frac{t}{t_{\rm Edd}}(1/\eta_{\rm rad}-1)\right] ,
\end{equation}
which is exponentially sensitive to efficiency (and hence
exponentially sensitive to black hole spin).  Over $t\sim 0.8{\rm
  Gyr}$ (roughly $z\sim 20$ to $z\sim 6$), Eddington-limited accretion
with $a/M\approx 0.99$ leads to $\MBH/\MBHO\lesssim 142$, a severe
restriction to growth.  One requires $a/M\lesssim 0.5$ and seeds of
$\MBHO\approx 10M_{\odot}$ to reach $\MBH\gtrsim 4\times
10^9M_{\odot}$, as needed to produce quasars at $z\sim 6$--$8$.

On the other hand, photon-trapping can allow for super-Eddington mass
accretion rates \citep{2007ApJ...670.1283O}, such that only $L$ might
be limited to near Eddington values.  So one should split the
efficiency due to radiation and accretion, given by $\eta_{\rm acc}$,
such that $\dot{\MBH} = (1-\eta_{\rm acc})\Mdot$.  Then
\begin{equation}\label{BHgrow}
\frac{\MBH}{\MBHO} < \exp\left[\frac{t}{t_{\rm Edd}}(1-\eta_{\rm acc})/\eta_{\rm rad}\right] .
\end{equation}
In this case, if $a/M=1$ such that $\eta_{\rm acc}\approx 0.43$, then
for (e.g.) $\Mdot c^2\sim 9L_{\rm Edd}$ one expects $\eta_{\rm
  rad}\sim 0.43/9\approx 0.05$.  In this case, one can easily have
$a/M\approx 1$ and seeds of $\MBHO\approx 10M_{\odot}$ and reach
$\MBH\sim 10^{10}M_{\odot}$, as could be required by the most massive
quasars at redshifts of $z\sim 6$--$8$.  So even mild modifications to
accretion rates and luminosities can exponentially affect the growth
of black holes.  This highlights the importance of including
general relativity (to properly account for spin that controls both
efficiencies), radiation (that at least determines the radiative
efficiency), as well as magnetic fields (that control the spin down ;
\citealt{2004ApJ...602..312G}).  So to understand the accretion
physics in such systems, radiation GRMHD models (that
self-consistently couple gas, radiation and magnetic fields) are
crucial.

In order to study such a complex and sensitive interaction between GR,
radiation, and magnetic fields, modern black hole (BH) accretion disk
theory has relied upon several approximations, which often involve
approximate closure schemes (e.g. magnetohydrodynamics (MHD),
flux-limited radiative diffusion) to capture unresolved spatial,
unresolved temporal scales and unresolved physical processes.  Such
closure schemes can be more efficient than evolving distribution
functions or individual particles/rays, and such methods have been
successful in the study of radiative disks.

The flux-limited diffusion approximation is a closure that only allows
isotropic emission relative to the fluid frame, which over-constrains
disk emission in the highly radiative regions near the BH for the disk
surface where the optical depth is order unity.  This formalism has
been included as an explicit scheme in some codes
\citep{farrisetal08,zanottietal11,fragileetal12}, but
implicit-explicit Runge-Kutta \citep{pr05}
or fully implicit type numerical schemes
are required to maintain stability for all optical depths
(e.g. \citealt{roedigetal12}).  Using a flux-limited diffusion approximation,
small patches of radiatively efficient thin accretion disks have been
simulated using the local shearing box approximation
\citep[][]{turner2003,kro07,blaes2007,blaes2011,hirose09a,hirose09b}.
Also, using a non-relativistic code and flux-limited diffusion,
super-Eddington accretion flows have been simulated and their spectra
computed \citep{ohsuga03,Ohsuga+05,2006ApJ...640..923O,ohsuga09,om11,kawashima12,2012EPJWC..3906005M,2013PASJ...65...88T,2013arXiv1306.1871Y}.

To model a radiative disk around a rotating black hole
self-consistently, the closure method should be able to handle both
the optically thick (disk interior) and optically thin (corona-jet)
limits in full GR including rotating black holes.  One might treat the
radiation more accurately than flux-limited diffusion using the
``instant light'' approximation
\citep{hayesnorman03,gonzalesetal07,jiangetal12,davis12,2013ApJ...778...65J},
but currently such schemes do not allow relativistic radiative fluxes
that are natural near the black hole
\citep{jiangetal12,davis12,2013ApJ...778...65J}.

In the work described here, we have implemented the so-called M1
closure scheme
\citep{levermore84,dubrocafeugeas99,gonzalesetal07,2013MNRAS.429.3533S,2013ApJ...772..127T},
which is similar to other schemes that use a truncated moment
formalism \citep{2012PThPh.127..535S}.  M1 closure allows a limited
treatment of anisotropic radiation and works well at all optical
depths.  Using the Koral code, the M1 method has been shown to work
well to handle axisymmetric GR hydrodynamic (HD) flows around black
holes \citep{2013MNRAS.429.3533S} as well as axisymmetric GRMHD flows
around rotating black holes \citep{2013arXiv1311.5900S}.  We have
implemented M1 into the GRMHD code HARM \citep{2003ApJ...589..444G},
leading to the code we call HARMRAD, that adds to the Koral feature
set the additional abilities to handle full 3D in spherical polar
coordinates, higher-order reconstruction for both gas and radiation,
and efficient parallel computation using distributed nodes on
supercomputers.

The structure of the paper is as follows: The equations solved are
presented in \S\ref{sec:goveqns} for the gas fluid and in
\S\ref{sec:goveqnsrt} for the radiation, numerical methods are
presented in \S\ref{sec:nummethods}, and radiative tests of the method
are presented in \S\ref{s.tests}.  Results for our fiducial 3D GRMHD
model of a super-Eddington accretion flow around a rotating black hole
are presented in \S\ref{sec:fiducial}.  We summarize our method and
results in \S\ref{sec:summary}.

%%% Local Variables: 
%%% mode: latex
%%% TeX-master: "ms"
%%% End: 

\section{Governing MHD Equations}
\label{sec:goveqns}

We solve the GRMHD equations for a radiative magnetized fluid in the
test-field limit for the fluid in an arbitrary stationary space-time.
Internal coordinates $\xvec^\alpha\equiv
(t,\xvec^{(1)},\xvec^{(2)},\xvec^{(3)})$ on a uniform grid map to
arbitrary set of coordinates (Cartesian, spherical polar, etc.), e.g.,
for spherical polar coordinates: $\rvec^\alpha\equiv
(t,r,\theta,\phi)$. We write tensors in an orthonormal basis gas-fluid
frame using ``widehats'' as $\widehat R$ that just means the
components of $R$ have been transformed to the gas fluid frame in an
orthonormal basis (i.e. for vectors, $\widehat u \equiv
u^{\hat{\mu}}$).  We write radiation fluid frame quantities as
$\bar{u}$.  Quasi-orthonormal vectors are denoted as $u_\mu\approx
\sqrt{|g_{\mu\mu}|} u^\mu$. When necessary to distinguish from
orthonormal versions, contravariant (covariant) vectors are denoted as
$\uvec^\mu$ ($\uvec_\mu$), while higher-ranked coordinate basis
tensors have no underbar.  We work with Heaviside-Lorentz units, often
set $c=GM=1$ when no explicit units are given, and let the horizon
radius be $r_{\rm H}$.

Mass conservation is given by
\begin{equation}
\nabla_\mu (\rho_0 \uvec^\mu) = 0 ,
\end{equation}
where $\rho_0$ is the rest-mass density, $\uvec^\mu$ is the
contravariant 4-velocity, and $\rho=\rho_0 \uvec^t$ is the lab-frame
mass density.

Energy-momentum conservation is given by
\begin{equation}\label{emomeq}
\nabla_\mu  T^\mu_\nu = G_\nu ,
\end{equation}
where $G_\nu$ is an external 4-force
and the stress energy tensor $T^\mu_\nu$ includes both matter (MA)
and electromagnetic (EM) terms:
\begin{eqnarray}\label{MAEM}
{T^{\rm MA}}^\mu_\nu &=& (\rho_0 + \ug + \pg ) \uvec^\mu \uvec_\nu + \pg \delta^\mu_\nu \nonumber ,\\
{T^{\rm EM}}^\mu_\nu &=& b^2 \uvec^\mu \uvec_\nu + p_b\delta^\mu_\nu - \bvec^\mu \bvec_\nu \nonumber ,\\
T^\mu_\nu &=& {T^{\rm MA}}^\mu_\nu + {T^{\rm EM}}^\mu_\nu .
\end{eqnarray}
The MA term can be decomposed into a particle (PA) term: ${T^{\rm
    PA}}^\mu_\nu = \rho_0 \uvec_\nu \uvec^\mu$ and an enthalpy (EN)
term.  The MA term can be reduced to a free thermo-kinetic energy
(MAKE) term, which is composed of free particle (PAKE) and enthalpy
(EN) terms:
\begin{eqnarray}\label{MAKEs}
{T^{\rm MAKE}}^\mu_\nu &=& {T^{\rm MA}}^\mu_\nu - \rho_0 \uvec^\mu \etavec_\nu/\alpha ,\\
{T^{\rm PAKE}}^\mu_\nu &=& (\uvec_\nu - \etavec_\nu/\alpha)\rho_0 \uvec^\mu  \nonumber ,\\
{T^{\rm EN}}^\mu_\nu &=&  (\ug + \pg)\uvec^\mu \uvec_\nu + \pg \delta^\mu_\nu \nonumber ,
\end{eqnarray}
such that ${T^{\rm MAKE}}^\mu_\nu = {T^{\rm PAKE}}^\mu_\nu+{T^{\rm
    EN}}^\mu_\nu$.  Here, $\ug$ is the internal energy density and
$\pg=(\Gamma-1)\ug$ is the ideal gas pressure with adiabatic index
$\Gamma$. The contravariant fluid-frame magnetic 4-field
is given by $\bvec^\mu$, which is related to the lab-frame 3-field via
$\bvec^\mu = \Bvec^\nu h^\mu_\nu/\uvec^t$ where $h^\mu_\nu = \uvec^\mu
\uvec_\nu + \delta^\mu_\nu$ is a projection tensor, and
$\delta^\mu_\nu$ is the Kronecker delta function.  The magnetic energy
density ($u_b$) and pressure ($p_b$) are $u_b=p_b=\bvec^\mu
\bvec_\mu/2 = b^2/2$.  The total pressure is $p_{\rm tot} = \pg +
p_b$, and plasma $\beta\equiv \pg/p_b$.  The 4-velocity of a zero
angular momentum observer (ZAMO) is $\etavec_\mu=\{-\alpha,0,0,0\}$
where $\alpha=1/\sqrt{-g^{tt}}$ is the lapse.  The 4-velocity relative
to this ZAMO is $\reluvec^\mu = \uvec^\mu - \gamma \etavec^\mu$ where
$\gamma=-\uvec^\alpha \etavec_\alpha$.

A corresponding entropy conservation equation
(that can be used instead of the energy equation) is given by
the evolution of the specific gas entropy $s_{\rm gas,s}$
as determined by the comoving time-rate of change
$Ds_{\rm gas,s}/Dt\equiv \partial_{\tau} s_{\rm gas,s}$.
Using baryon conservation one obtains
\begin{equation}\label{eq:S}
\nabla_\mu(\rho_0 u^\mu s_{\rm gas,s}) = \rho_0 \frac{D s_{\rm gas,s}}{Dt} = \rho_0 \frac{\partial s_{\rm gas,s}}{\partial\tau}  = G_S
\end{equation}
where the entropy density is given by $S\equiv \rho_0 s_{\rm gas,s}$,
and where the right-hand-side ($G_S$) corresponds to a source or sink
of entropy.  To obtain machine accurate entropy conservation, the
specific entropy must be per unit volume.  For example, for an ideal
gas the specific entropy constant $K=P/\rho_0^\gamma$ is constant at
constant specific entropy, but using such an entropy tracer only leads
to entropy conservation truncation error.  Using instead $s_{\rm
  gas,s} = \log(P^n/\rho_0^{n+1})$ with $n=1/(\gamma-1)$, such that
the entropy density is $s_g=\rho_0 s_{\rm gas,s}$, leads to entropy
conservation at machine round-off error.

Magnetic flux conservation is given by the induction equation
\begin{equation}
\partial_t(\detg \Bvec^i) = -\partial_j[\detg(\Bvec^i \vvec^j - \Bvec^j \vvec^i)] ,
\end{equation}
where $g={\rm Det}(g_{\mu\nu})$ is the metric's determinant, and the
lab-frame 3-velocity is $\vvec^i = \uvec^i/\uvec^t$.  No explicit
viscosity or resistivity are included, but we use the energy
conserving HARM scheme so all dissipation is captured
\citep{2003ApJ...589..444G,mck06ffcode}.

Apart from any physical source term giving a non-zero $G_\mu$, the
energy-momentum conservation equations are only otherwise modified due
to so-called numerical density floors that keep the numerical code
stable as described in detail in Appendix~A of \citet{2012MNRAS.423.3083M}.  The
injected densities are tracked and removed from all calculations.

\newcommand{\bea}{\begin{eqnarray}}
\newcommand{\eea}{\end{eqnarray}}
\newcommand{\pdder}[2]{\frac{\partial^2 #1}{\partial#2^2}}
\newcommand{\pder}[2]{\frac{\partial#1}{\partial#2}}
\newcommand{\der}[2]{\frac{{\rm d}#1}{{\rm d}#2}}
\newcommand{\derln}[2]{\ensuremath{\frac{{\rm d\,ln}\, #1}{{\rm d\,ln}\, #2}}}
\newcommand{\pderln}[2]{\ensuremath{\frac{\partial\,\rm ln\,#1}{\partial\,\rm ln\,#2}}}

\newcommand{\AS}[1]{\textbf{\color{Magenta}#1}}
\newcommand{\RN}[1]{\textbf{\color{Green}#1}}
\newcommand{\AT}[1]{\textbf{\color{Blue}#1}}
\newcommand{\YZ}[1]{\textbf{\color{Orange}#1}}
\newcommand{\koral}{\texttt{koral}}
\newcommand{\harmrad}{\texttt{harmrad}}

\def\bE{\bar{E}}
\def\bR{\bar{R}}
\def\bu{\bar{u}}

\section{Governing Radiative Transfer Equations}
\label{sec:goveqnsrt}

For a radiation stress-energy tensor $R^\mu_\nu$,
total energy-momentum conservation ($\nabla \cdot (T+R)=0$) for the MHD fluid and radiation
can be written using the 4-force density $G^\nu$ as
\bea\label{eq.cons2}
{T^\mu_\nu}_{;\mu}&=&G_\nu,\\\nonumber
{R^\mu_\nu}_{;\mu}&=&-G_\nu,
\eea

The radiation stress-energy tensor can be obtained from its
simple form in an orthonormal frame where it is comprised of
various moments of the specific intensity $I_\nu$,
as discussed in \citet{2013MNRAS.429.3533S}.
E.g., in an orthonormal fluid frame it takes the following form,
\be
\widehat R= \left[ \begin{array}{cc} \widehat E & \widehat F^i \\ \widehat F^j &
    \widehat P^{ij}
\end{array} \right],
\ee
where, for frequency $\nu$ and solid angle $\Omega$, the orthonormal fluid-frame quantities
\bea
\hspace{1in}\widehat E&=& \int \widehat I_\nu {\,\rm d \nu \,d\Omega},\\
\hspace{1in}\widehat F^i&=& \int \widehat I_\nu {\,\rm d \nu \,d\Omega \,N^i},\\
\hspace{1in}\widehat P^{ij}&=& \int \widehat I_\nu {\,\rm d \nu \,d\Omega \,N^i\,N^j}
\eea
are the radiation energy density, the radiation flux and the radiation
pressure tensor, respectively, and $N^i$ is a unit vector in direction
$x^i$.

The radiation stress-energy tensor allows one to obtain the radiation 4-force,
$G^\mu$ as given by \citep{mihalasbook},
\be
G^\mu=\int(\kappa_{\nu,\rm tot} I_\nu - \eta_\nu){\,\rm d \nu \,d\Omega \,N^i},
\ee
which in the orthonormal fluid frame becomes
\be\label{eq.Gff}
\widehat G=
\left[ \begin{array}{c}
 \kappa_{\rm abs} \widehat E - \lambda \\
 \kappa_{\rm tot} \widehat F^i
\end{array} \right],
\ee where the gas-fluid frame energy density emission rate of the gas
is given by $\lambda$, and for a given absorption opacity, Kirchhoff's
law gives that $\lambda = \kappa_{\rm abs} 4\pi \widehat{B}$ for
$\widehat{B}_{\rm gas} = a_{\rm rad} T_{\rm gas}^4/(4\pi)$.  Here,
$\widehat{B}_{\rm gas}=\sigma_{\rm rad} T_{\rm gas}^4/\pi$ is the integrated Planck
function corresponding to the gas temperature $T_{\rm gas}$, $\sigma_{\rm rad}$ is the
Stefan-Boltzmann constant, $\kappa_{\nu,\rm tot}$ and $\eta_\nu$
denote the frequency-dependent opacity and emissivity coefficients,
respectively, while $\kappa_{\rm abs}$ and $\kappa_{\rm sca}$ are the
frequency integrated absorption and scattering opacity coefficients,
respectively, and the total opacity is given by $\kappa_{\rm tot} =
\kappa_{\rm abs} + \kappa_{\rm sca}$.

Using covariance (or boosting from the lab-frame to fluid orthonormal frame),
the covariant 4-force is then
\be\label{eq:Gcov}
G^{\mu}= -(\kappa_{\rm abs} R^\mu_\alpha u^\alpha + \lambda u^\mu) - \kappa_{\rm es}(R^\mu_\alpha u^\alpha + R^\alpha_\beta u_\alpha u^\beta u^\mu) .
\ee
The corresponding entropy source term is obtained from
\be
T_{\rm gas} \frac{ds_{\rm gas,s}}{d\tau} = \frac{dq_{\rm gas}}{d\tau} 
\ee
such that in covariant form one has
\be\label{eq:GS}
(\rho_0 s_{\rm gas,s} u^\mu)_{;\mu} = -\frac{1}{T_{\rm gas}} G^\mu u_\mu \equiv G_S
\ee
that specifies Eq.~(\ref{eq:S}) in the presence of a radiation 4-force.

\subsection{Closure scheme}
\label{s.closure}

To close the above set of equations we need a prescription to compute
the second moments of the angular radiation intensity distribution.
Specifically, we need $R^{\mu\nu}$ only knowing the radiative energy density
and fluxes in some frame (e.g. $R^{tt}$ and $R^{ti}$ in the lab-frame).

The simplest approach is the Eddington approximation,
which assumes a nearly isotropic radiation field in the gas fluid frame,
which in the gas fluid frame is given by
\be\label{eq.eddapr}
\widehat P^{ij}=\frac13\widehat E \delta^{ij}.
\ee
However, the radiation is only isotropic in the optically
thick limit, so this closure does not handle optically thin flows.

To handle general optical depths, we use the M1 closure \citep{levermore84},
which assumes the radiation satisfies the Eddington closure in an independent radiation frame
within which radiation fluxes vanish.
Thus, in the radiation frame, $\bR^{tt}=\bE$,
$\bR^{ii}=\bE/3$, and all other components of $\bR$ are zero.
In the radiation rest frame, the radiation stress tensor can be written
as
\begin{equation}
\bR^{\mu\nu} = \frac{4}{3}\bE\, \bu^\mu_{\rm rad} \bu^\nu_{\rm rad} + \frac{1}{3}\bE\, g^{\mu\nu},
\label{eq:R}
\end{equation}
where $\bu^\mu_{\rm rad}$ is the radiation frame's 4-velocity.
Using general covariance (or boosting into the lab-frame),
the covariant expression is
\begin{equation}
R^{\mu\nu} = \frac{4}{3}\bE\, u^\mu_{\rm rad} u^\nu_{\rm rad} + \frac{1}{3}\bE\, g^{\mu\nu},
\label{eq:Rlab}
\end{equation}
The quantity $\bE = u^\mu_{\rm rad} u^\nu_{\rm rad} R_{\mu\nu}$ is the radiation energy density
as measured in the radiation rest frame.

For an orthonormal Cartesian basis, the above formulation reduces to the
standard formulae \citep{levermore84,dubrocafeugeas99,gonzalesetal07}.
For instance, the radiation pressure tensor $\widehat P^{ij}$
in the fluid frame has the form,
\be
\widehat P^{ij}=\left(\frac{1-\xi}{2} \delta^{ij}+\frac{3\xi-1}{2} \frac{f^if^j}{|f|^2}\right)\widehat E,
\ee
where $f^i=\widehat F^i/\widehat E$ is the reduced radiative flux and $\xi$ is the
Eddington factor given by
\citep{levermore84},
\be
\xi=\frac{3+4f^if_i}{5+2\sqrt{4-3f^if_i}}.
\ee

In the extreme ``optically thick limit'', $\widehat F^i \approx 0$,
and then $f^i =
0$, $f^if_i = 0$ and $\xi = 1/3$, which gives the expected
Eddington approximation,
\be
\widehat P^{ij}_{\tau \gg 1}=
\left[ \begin{array}{ccc}
1/3&0&0\\
0&1/3&0\\
0&0&1/3\\
\end{array} \right]\widehat E.
\ee
In the opposite extreme ``optically thin limit'', $\widehat F^1=\widehat E$,
i.e., a uni-directional radiation field directed along the x-axis, we
have $f^i = \delta^i_1$, $f^if_i = 1$ and $\xi = 1/3$, which gives
\be
\widehat P^{ij}_{\tau \ll 1}=
\left[ \begin{array}{ccc}
1&0&0\\
0&0&0\\
0&0&0\\
\end{array} \right]\widehat E,
\ee
which gives the expected intensity distribution of a
Dirac $\delta$-function parallel to the flux vector.

The M1 closure scheme thus handles both
optical depth extremes well, and smoothly and stably interpolates
between these extreme optical depths.
However, because M1 treats the radiation as isotropic in a single
frame, it cannot handle general anisotropic intensity distributions.
So, at locations where multiple radiation fluids interact,
M1 isotropizes the radiation in an averaged
radiation frame.  M1 is expected to be an ok approximation for expanding radiation fields
like from accretion disks, but a convergent radiation field will
lead to ``photon collisions'' even in optically thin regions.
In any case, M1 closure will provide a
superior treatment of radiation in the optically thin regions near and
above the disk photosphere, as compared to the Eddington approximation or
flux-limited diffusion.

\section{Numerical Methods: HARMRAD}
\label{sec:nummethods}

The core of HARMRAD is built upon HARM.  The GRMHD code HARM is based
upon a conservative shock-capturing Godunov scheme with 3rd order (2nd
or 4th order choosable) Runge-Kutta time-stepping, Courant factor
$0.5$, LAXF (HLL choosable, but less stable for highly magnetized
flows) fluxes, simplified wave speeds, PPM-type interpolation for
primitive quantities ($P=\{\rhorest,\ug,\reluvec^i,\Bvec^i\}$), a
staggered magnetic field representation, and any regular grid warping
\citep{2003ApJ...589..444G,nob06,2007MNRAS.379..469T}.  Treatment of the numerical density
floors, 3D polar axis, conserved to primitive inversion attempts and
reductions to simpler equations, and connection coefficients
($\Gamma^\lambda_{\nu\kappa}$) is provided in the appendix in
\citet{2012MNRAS.423.3083M}.  As with HARM, HARMRAD is based upon a
hybrid OpenMP - MPI framework use ROMIO for parallel I/O operating at
up to order 32,000 cores on Kraken at $>70\%$ efficiency.  Double
precision floats are used for all tests, although long double
precision (including for all needed transcendental functions) can be
used to test precision issues.

The internal code units set scales for length of $\tilde{L} = GM/c^2$,
time $\bar{T}=\bar{L}/c$, velocity $\bar{V}=\bar{L}/\bar{T}$,
$\bar{\rho}=1$ for mass density in cgs units (i.e. grams) for a
fiducial arbitrary choice of $\MBH=10M_{\odot}$, $\bar{M}=\bar{\rho}
\bar{L}^3$, $\bar{E}=\bar{M} \bar{V}^2$, $\bar{U}=\bar{\rho}
\bar{V}^2$, and gas temperature $\bar{T}_{\rm gas}= m_b c^2/k_b$ for
baryon mass $m_b$ and Boltzmann's constant $k_b$.  This gives an
opacity scaling of $\bar{\kappa} = \bar{L}^2/\bar{M}$, and a radiation
constant scaling of $\bar{a}_{\rm rad} = \bar{U}/\bar{T}_{\rm gas}^4$.
We sometimes report mass fluxes, energy fluxes, time, density,
magnetic field, etc. as per unit Eddington to simplify these scalings
for the reader, and these are obtained by simply rescaling $\Mdotedd$
by only $c$, $G$, and $\MBH$.

The MHD and radiation conservation laws are evolved using a method of
lines using a Runge Kutta approach.  We consider a set of $q=13$
quantities:
\begin{enumerate}
\item primitive: $P_q=\{\rho_0, \ug, \reluvec^\mu_{\rm gas}, B^i, \bar{E},
\reluvec^\mu_{\rm rad}, \ug\}$.
\item conserved: $U_q = \detg\{\rhorest u^t , \rhorest u^t_{\rm gas} + {T^t}_t, {T^t}_i, B^i, {R^t}_\nu, S u^t\}$,
\item fluxes: $F_q = \detg\{\rhorest u^j , {T^j}_\nu, B^i v^j - B^j
v^i, {R^j}_\nu, S u^j\}$,
\item geometry: $\SM_q = \detg\{0,{T^\kappa}_\lambda {\Gamma^\lambda}_{\nu\kappa},
0,0,0,{R^\kappa}_\lambda {\Gamma^\lambda}_{\nu\kappa},0\}$,
\item GAS-RAD 4-force: $\SR_q = \detg\{0,G_\nu,0,0,0,-G_\nu, G_S\}$,
\end{enumerate}
Each $U_q$, $F_q$, $\SM_q$, $\SR_q$ can be obtained in closed form
as functions of $P_q$.
Magnetic field primitives/conserved/source quantities sit at cell
faces, while magnetic field fluxes (EMFs) sit at the corners of each
2D plane that passes through the cell center.  All non-magnetic
primitives/conserved/source quantities sit at cell centers, while
non-magnetic fluxes sit at cell faces.

%%%%%%%%%%%%%%%%%%%%%%%%%%%%%%%%%%%%%%%%%%%%%%%
\subsection{Implicit-Explicit Runge-Kutta}
\label{sec:imex}

For large 4-forces relative to the conserved quantities (e.g. for
large optical depths), the radiative 4-force $G_\nu$ become stiff,
making explicit integration practically impractical
\citep[e.g.,][]{zanottietal11}.  To generally handle the 4-force in
all regimes, we treat the 4-force term using an implicit-explicit
(IMEX) Runge-Kutta (RK) scheme to evolve the equations forward in
time on each full timestep of size $dt$. For intermediate or full
steps, let the explicit-only (X) terms and IMEX (M) terms be written
as
\begin{eqnarray}
X^i_q &\equiv& \Delta_j {\mathcal{F}^{i}}_q^j/dx^j + \SM^{i}_q ,\\
M^i_q &\equiv& \SR^{i}_q ,
\end{eqnarray}
where the first equation comes from a Riemann solver with $j$ summed over all spatial dimensions.
For any $q$, let $U^{i}$ be the value of $U$, such that $X^{i} \equiv X(U^{i})$ and $M^{i} \equiv M(U^{i})$.
Then, IMEX RK schemes take the form \citep{pr05}
\begin{eqnarray}\label{Uaeq2}
U^{i}   &=& U^n + dt \sum_{j=1}^{i-1} {\tilde{a}}_{ij} X^j     + dt \sum_{j=1}^{\nu} a_{ij} M^j \\
U^{n+1} &=& U^n + dt \sum_{i=1}^{\nu} {\tilde{\omega}}_{i} X^i + dt \sum_{i=1}^{\nu} \omega_{i} M^i ,
\end{eqnarray}
where $U^{i}$ are the auxiliary intermediate values of the IMEX
Runge-Kutta scheme and $U^{n+1}$ is the final full step solution.  The
matrices $\tilde{A}= (\tilde{a}_{ij})$ and $A= (a_{ij})$ are $\nu
\times \nu$ matrices (for a $\nu$-stage IMEX scheme), such that the
resulting scheme is explicit in $F$ (i.e. $\tilde{a}_{ij} = 0$ for $j
\geq i$) and implicit in $M$. An IMEX Runge-Kutta scheme is
characterized by these two matrices and the coefficient vectors
$\tilde{\omega}_i$ and $\omega_i$.  Since simplicity and efficiency in
solving the implicit part at each step is important, we consider
diagonally implicit Runge-Kutta (DIRK) schemes (i.e. $a_{ij}=0$ for $j
> i$) for stiff terms.

We use a set of IMEX coefficients that generates so-called strongly
stable preserving (SSP) (formally known as total variational
diminishing (TVD)) schemes
\citep{Pareschi00implicit-explicitrunge-kutta,pr05}, such that the
total variation in $U_q$ always diminishes.  We also restrict
ourselves to schemes that are asymptotically preserving (AP), such
that in the infinitely stiff limit the RK method preserves
higher-order accuracy and becomes an explicit RK method for the
non-radiative GRMHD equations -- as desired.  For efficiency, we also
restrict ourselves to diagonally implicit Runge-Kutta (DIRK) schemes
with (i.e. $a_{ij}=0$ for $j > i$) for the stiff terms that reduces
the number of calculations required.  Several applications of the IMEX
method demonstrate its accuracy and robustness
\citep{2009MNRAS.394.1727P,K11_138,roedigetal12,2013ApJ...764..122T,2013ApJ...772..127T}.
Improvements in IMEX schemes beyond our implementation are discussed
elsewhere
\citep{DBLP:journals/siamsc/BoscarinoR09,2011AIPC.1389.1315B,2013arXiv1306.4926B,journals/appml/Trenchea14}.

The IMEX scheme can be written many ways, and here we expand the form
for implementation into HARMRAD.  The 2nd order IMEX scheme has 2
explicit stages (given in square brackets) and 2 implicit stages and
one final result for $U^{n+1}$, given by (IMEX2)
\begin{eqnarray}
U^{0} &=& [U^n] + \gamma dt M^{0} \\
U^{1} &=& [\frac{3\gamma-1}{\gamma} U^n + \frac{1-2\gamma}{\gamma} U^0 + dt F^{0}] + \gamma dt M^{1} \\
U^{n+1} &=& [\frac{1}{2} U^{n} + \frac{1}{2} U^{1} + \frac{dt}{2} F^{1} + dt \gamma M^{0} + dt \frac{1-\gamma}{2} M^{1}] ,
\end{eqnarray}
where $\gamma=1-1/\sqrt{2}$.
The steps are written so explicit steps only need use the immediately prior $U^{i}$ or $U^n$ to avoid needing
to store intermediate $F^{i}$, while one should progressively store each $M^i$ as computed so available for any substep.

Similarly, another 2nd order IMEX scheme has 3 explicit stages and 3 implicit stages, given by (IMEX2B)
\begin{eqnarray}
U^{0} &=& [U^n] + \frac{dt}{4} M^{0} \\
U^{1} &=& [U^n + \frac{dt}{2} F^{0}] + \frac{dt}{4} M^{1} \\
U^{2} &=& [U^{1} + \frac{dt}{2} F^{1}  + \frac{dt}{3} M^{0} + \frac{dt}{12} M^{1}] + \frac{dt}{3} M^{2} \\
U^{n+1} &=& [\frac{1}{3} U^{n} + \frac{2}{3} U^{2} +\frac{dt}{3} F^{2} + \frac{dt}{9} (M^{0} + M^{1} + M^{2})] ,
\end{eqnarray}
which again is written to avoid storing intermediate $F^{i}$.

Likewise, the 3rd order IMEX scheme has 3 explicit stages, 4 implicit stages, given by  (IMEX3)
\begin{eqnarray}
U^{0} &=& [U^n] + \alpha dt M^{0} \\
U^{1} &=& [2 U^n  - U^{0}] +  \alpha dt M^{1} \\
U^{2} &=& [U^n + dt F^{1}  + (1-\alpha) dt M^{1}] + \alpha dt M^{2} \\
U^{3} &=& [\frac{3}{4}U^n + \frac{1}{4}U^{2} + \frac{dt}{4} F^{2}  + \beta dt M^{0} \\
      &+& \frac{(-1+\alpha+4\eta)dt}{4} M^{1} + \frac{(2-5\alpha-4(\beta+\eta))dt}{4} M^{2}] \nonumber\\
      &+& \alpha dt M^{3} \nonumber\\
U^{n+1} &=& [\frac{1}{3} U^{n} + \frac{2}{3} U^{3} +\frac{2dt}{3} F^{3} + \frac{-2\beta dt}{3} M^{0} \\
      &+& \frac{(1-4\eta)dt}{6} M^{1} + \frac{(-1+4\alpha+4(\beta+\eta))dt}{6} M^{2} \nonumber\\
      &+& \frac{4(1-\alpha)dt}{6} M^{3}] \nonumber,
\end{eqnarray}
which again is written to avoid storing intermediate $F^{i}$.
Here $\alpha \approx 0.24169426078821$, $\beta\approx 0.06042356519705$, and $\eta\approx 0.12915286960590$ \citep{pr05}.

For equal implicit and explicit stages as well as equal coefficients
for either implicit or explicit terms, one obtains the standard
mid-point non-TVD 2nd order RK (RK2M), TVD 2nd order RK (RK2), TVD 3rd
order RK (RK3), and 4th order non-TVD RK (RK4) methods \citep{shu88},
where then the implicit and explicit substeps use the exact same
timestep coefficients for substeps and final solution.  This leads to
the implicit terms being treated to 2nd order in time and 1st order in
space for optically thick fast waves.

In this paper, for simplicity and for historial reasons, we only
consider such simplified IMEX Runge-Kutta methods.  In particular, we
use the RK3 method with the same stages and coefficients for both
explicit and implicit terms.  The full higher-order IMEX2, IMEX2B, and
IMEX3 will be considered for other applications beyond the scope of
this paper.

\subsection{HARMRAD Algorithm}

\label{s.algorithm}
During each sub-step of the Runge-Kutta time integration, the code
carries out the following steps in the given order:
\begin{enumerate}
\item $P_q$ on the evolved domain are mapped into ghost cells however boundary conditions required.
\item $P_q$ at cell centers is used to compute $U_q$ at cell centers.
\item $\SM_q$ geometry sources are computed at cell centers.
\item $P_q$ is interpolated from cell centers to faces in each direction in each dimension giving $P_L,P_R$ at each face in each dimension.
\item For each L and R, $F_q$ is computed at faces from $P_q$ at faces of cube in three-dimensions
\item For each L and R, $v_w$ wavespeeds ($v_+$ for positive direction, $v_-$ for negative direction) are computed at faces for each $P_q$.
\item Using $F_q,U_q,v_+, v_-$ at each face, Godunov flux is computed using the 2-state HLL solution:
\be\label{eq.hll}
\mathcal{F}_q=\frac{v_{\rm min} F_{q,R} + v_{\rm max} F_{q,L} - v_{\rm min} v_{\rm max} (U_{q,R}-U_{q,L})}{v_{\rm min} + v_{\rm max}},
\ee
where $v_{\rm min} = -{\rm min}(0,v_{-,R},v_{-,L})$, $v_{\rm max} = {\rm max}(0,v_{+,R},v_{+,L})$, or LAXF solution:
\be\label{eq.lax}
\mathcal{F}_q=\frac12( F_{q,R} + F_{q,L} - v_{\rm tot} (U_{q,R}-U_{q,L}) ),
\ee
where $v_{\rm tot} = {\rm max}(v_{\rm min},v_{\rm max})$.
\item $B^i,v^j$ at cell faces are interpolated to cell centers and to the corner of each 2D plane passing through the cell center.
\item Godunov EMF flux is computed using 4-state Riemann LAXF/HLL solution using $B^i,v^j$ at corners \citep{2007A&A...473...11D}.
\item $U_{q,i}$ is set for both cell centered and staggered $U_q$'s.
\item Magnetic field primitives $P_q^{i+1}$ are obtained, so implicit solver uses the final magnetic field as guess.
\item IMEX solution is found based upon explicit $U^n_q$, $X^i$ and IMEX radiation source term $M^i$ that converges to give $U_{q,(i+1)}$ and $P_{q,(i+1)}$.  For each step, MHD and RAD inversions from conserved to primitive quantities are performed.
\item Backup methods are employed if implicit solver fails.
\end{enumerate}

\subsection{Primitive Spatial Reconstruction}

As in HARM, PPM interpolation (with no contact steepener, but with
shock flattener) \citep{col84} is used to reconstruct primitive
quantities at different spatial locations.  For spherical polar
coordinates, for the radial and $\phi$ directions, we interpolate
$\detg \reluvec^i_{\rm gas}$, $\detg \reluvec^i_{\rm rad}$, and $\detg
B^i$, and otherwise we interpolate $P_q$.  The PPM reconstruction's
monotonized slopes use the monotonized central (MC) limiter. The PPM
flattener (with shock parameter SP0=$0.75$ as in the Flash code;
\citealt{fry00}) is applied separately on GAS and RAD quantities, but
each GAS and RAD flattener is formed as a linear interpolation based
upon the optical depth $\tau$, so that by $\tau=1/2$, each GAS and RAD
use a single flattener value set as the maximum of the GAS and RAD
flattener values.  The PPM flattener for the GAS quantities uses the
specific mass flux $\detg u^i_{\rm gas}$ and gas pressure, and the RAD
quantities use $\detg u^i_{\rm rad}$ and radiation pressure. The
pressures used in the flattener are linearly interpolated for each GAS
and RAD to become a total GAS+RAD pressure in the limit that
$\tau=1/2$.  On Runge-Kutta sub-steps, the maximum flattener value
over any prior sub-steps is used for the current sub-step.

\subsection{Characteristic wavespeeds}
\label{s.wavespeeds}

Godunov schemes, like the LAXF scheme, require knowledge of the
maximal characteristic wave speeds of the system ($v_w$ in
Eq.~\ref{eq.lax}).  The wave speed calculation
can be quite approximate, because the value only enters
as a numerical grid dissipation that improves the stability of the method in handling discontinuities.
Each GAS and RAD fluxes are computed separately, which preserves stability while
avoids excessive artificial numerical viscosity
when the characteristic wavespeeds are not separated.

The GAS's fast magnetosonic characteristics are computed as in HARM
(see the approximate dispersion in section 3.2 of
\citealt{2003ApJ...589..444G}), which solves for the lab-frame characteristic
3-velocity ($v_{w,\rm gas}$) in terms of the covariant quantities such
as the metric, gas-fluid 4-velocity, and gas-fluid frame magnetosonic
speed $c_{\rm ms}$.  For GAS quantities, the wavespeeds are computed
at faces for the left and right states before forming the 2-state
Riemann solution, while for the magnetic field fluxes (EMFs) the wave
speeds are interpolated from those face values to the EMF location
before forming the 4-state Riemann solution.

For the M1-closure scheme, the radiation characteristic moves at a
uniform value of $c_{\rm rad}$ in the radiation frame, so we simply
use the HARM method to obtain the lab-frame velocity $v_{w,\rm rad}$
from $c_{\rm rad}$ by replacing the gas-fluid 4-velocity with the
radiation-fluid 4-velocity.  We do not solve for the Jacobian's
eigenvalues as done in Koral (see section 3.2 in
\citealt{2013MNRAS.429.3533S}).

To determine $c_{\rm rad}$, we consider the optically thin and thick
limits to determine how this wavespeed is used to compute the flux
using the Godunov scheme.  In the optically thin limit, $c_{\rm
  rad}=\pm 1/\sqrt{3}$.  In the limit of large optical depths, we
follow the Koral code by using the effective wave speed rather than
the actual wave speed.  The radiative energy density, when decoupled from gas
(e.g., for $\kappa_{\rm abs} \ll 1$ but $\kappa_{\rm tot} \gg 1$), has
a diffusion coefficient $D$ given by (see Section~\ref{RADPULSEPLANAR})
\be\label{eq.Ddiff} D=\frac1{3\kappa_{\rm tot}}.  \ee In this limit
the distribution of radiative energy density should remain stationary
($\partial/\partial t \rightarrow 0$).  On the other hand, the
optically thin value of $c_{\rm rad}$ is near the speed of light
\citep{gonzalesetal07}. If such large wave speeds are incorporated
into a numerical scheme they will result in large, unphysical,
numerical diffusion. To limit this effect, we modify the radiative
wave speeds in the fluid frame according to
\bea\label{eq.wavespeedlimit}
\hspace{.9in }a^i_R&\rightarrow &{\rm min}\left(a^i_R,\frac4{3\tau^i}\right),\\\nonumber
\hspace{.9in }a^i_L&\rightarrow &{\rm max}\left(a^i_L,-\frac4{3\tau^i}\right),
\eea
where $a^i_R$ and $a^i_L$ are the maximal right- and left-going
radiative wave speeds in the fluid frame in the direction $i$, and
$\tau^i=\kappa_{\rm tot} {\rm dx^i}$
is the total optical depth of a given cell in that direction,
where ${\rm dx^i}$ is the orthonormal cell size in each direction.

The smaller the characteristic wave speed in Eq.~(\ref{eq.lax}), the
weaker the numerical diffusion. This choice
of the wave speed limiter (Eq.~\ref{eq.wavespeedlimit}) is motivated
by the fact that, for a diffusion equation of the form
$y_{,t}=Dy_{,xx}$, the maximum allowed time step for an explicit
numerical solver is
\be
\Delta  t =\frac{(\Delta x)^2}{4D}.
\ee
This expression, combined with Eq.~(\ref{eq.Ddiff}), gives
\be
\frac{\Delta x}{\Delta t}=\frac{4}{3\kappa_{\rm tot} \Delta x}=\frac{4}{3\tau},
\ee
which is the limiter introduced in Eq.~(\ref{eq.wavespeedlimit}).
Essentially, we set $v_{w,\rm rad}$ to the velocity of diffusion in the optically
thick limit.

For each Runge-Kutta timestep, in each direction, the timestep is to
$dt_i = dx_i/v^i_w$ for $v^i_w$ as the larger of $v^i_{\rm gas}$ and $v^i_{\rm
  rad}$, where $v_{w,\rm gas}$ and $v_{w,\rm rad}$ were obtained as
effectively located at cell centers by taking the maximum of
characteristic speed at the left or right of each cell face.  These
characteristic speeds were constructed as lab-frame internal HARM
coordinate basis 3-velocities using the above-mentioned HARM
dispersion relation to convert $c_{\rm gas},c_{\rm rad}$ to $v_{w,\rm
  gas},v_{w,\rm rad}$.  The $dx_i$ are the uniform (constant) grid
internal cell sizes in each dimension.  The overall timestep is set as
$dt = 1/(1/dt_1 + 1/dt_2 + 1/dt_3)$.

%%%%%%%%%%%%%%%%%%%%%%%%%%%%%%%%%%%%%%%%%%%%%%%%%%%%%%%%%%%%%%%%%%%%
\subsection{MHD Inversion of Conserved to Primitive Quantities}
%
%
%%%%%%%%%%%%%%%%%%%%%%%%%%%%%%%%%%%%%%%%%%%%%%%%%%%%%%%%%%%%%%%%%%%%

The gas MHD inversion involves taking known conserved quantities $U_q$
and obtaining primitive quantities $P_q$.  Our method uses a single
non-linear equation as a function of one MHD variable plus the MHD
conserved quantities in order to obtain a solution to gas primitives
$P^{n+1}_q(U^{n+1}_q)$ for the explicit case
\citep{nob06,mm07,2013MNRAS.429.3533S} or steps within the implicit
solver.

\newcommand{\OneDW}{{$\mathrm{1D}_W$ }}
\newcommand{\OneDWc}{{$\mathrm{1D}_W$}}
\newcommand{\OneDWb}{{$\mathrm{\mathbf{1D}}_\mathbf{W}$ }}
\newcommand{\OneDWi}{{$\mathrm{\mathit{1D}}_\mathit{W}$ }}
\newcommand{\OneDVsq}{{$\mathrm{1D}^\star_{v^2}$ }}
\newcommand{\OneDVsqc}{{$\mathrm{1D}^\star_{v^2}$}}
\newcommand{\OneDVsqb}{{$\mathrm{\mathbf{1D}}^\star_\mathbf{v^2}$ }}
\newcommand{\OneDVsqi}{{$\mathrm{\mathit{1D}}^\star_\mathit{v^2}$ }}
\newcommand{\OneDVsqOrig}{{$\mathrm{1D}_{v^2}$ }}
\newcommand{\OneDVsqOrigc}{{$\mathrm{1D}_{v^2}$}}
\newcommand{\OneDVsqOrigb}{{$\mathrm{\mathbf{1D}}_\mathbf{v^2}$ }}
\newcommand{\OneDVsqOrigi}{{$\mathrm{\mathit{1D}}_\mathit{v^2}$ }}

First, we calculate $p(\rhorest,\ug) = p(\rhorest,\chi_{\rm gas})$ where
$\chi_{\rm gas}=\ug+\pg$ (e.g. ideal gas of $\pg = \left(\Gamma - 1\right)
\ug$), $w = \rhorest + \ug + \pg$, ZAMO relative Lorentz factor
$\gamma_{\rm gas} = \sqrt{1 + g_{ij} \tu^i_{\rm gas} \tu^j_{\rm gas}}$, ZAMO relative 4-velocity
$\tu^\mu_{\rm gas} = u^\mu_{\rm gas} - \gamma_{\rm gas} \eta^\mu$, $\tv^\mu_{\rm gas} = \tu^\mu_{\rm gas}/\gamma_{\rm gas}$ such
that $\gamma_{\rm gas}^2=1/(1-\tv_{\rm gas}^2)$, ZAMO field $\sB^\mu = \eta_\mu
\dF^{\mu\nu} = \alpha B^\mu$ for HARM magnetic field variable $B^\mu$,
fluid-frame field $b^\mu = {h^\mu}_\nu \sB^\nu/\gamma$, with
projection tensor ${h^\mu}_\nu = \delta^\mu_\nu + {u^\mu}_{\rm gas} {u_\nu}_{\rm gas}$, and
new variable $W\equiv w\gamma_{\rm gas}^2$ or $W' = W-D = D(\gamma_{\rm gas}-1) + \chi_{\rm gas}
\gamma_{\rm gas}^2$.  The $\sB^i$s are both primitive and conserved variables.

The ZAMO mass density is given by
\begin{equation}\label{PRIM1}
D = \gamma_{\rm gas} \rhorest ,
\end{equation}
and
\begin{equation}\label{bsqexp}
b^2 = {\frac{1}{\gamma_{\rm gas}^2}} \left[\sB^2 + \left(\sB^\mu {u_\mu}_{\rm gas}\right)^2\right]
\end{equation}
and
\begin{equation}\label{ndbeq}
n_\mu b^\mu = - {u_\mu}_{\rm gas} \sB^\mu .
\end{equation}
Using $\eta_\mu \sB^\mu = 0$, then
\begin{eqnarray}\label{PRIM2ORIG}
Q_\mu &\equiv& -\eta_\nu {T^\nu}_\mu = \gamma_{\rm gas} \left(w + b^2\right) {u_\mu}_{\rm gas} - \left(\pg + b^2/2\right) n_\mu + (\eta_\nu b^\nu)b_\mu \nonumber\\
      &=& (W+\sB^2) \tv_{\mu \rm gas} - (\tv_{\nu \rm gas} \sB^\nu) \sB_\mu \\
      &-& (-W + \pg - (1/2)\sB^2 (1+\tv^2_{\rm gas}) + (1/2)(\tv_{\nu \rm gas} \sB^\nu)^2)\eta_\mu \nonumber .
\end{eqnarray}
Using $\eta_\mu \sB^\mu = 0$ and $S\equiv Q_\mu \sB^\mu = {v_\mu}_{\rm gas} \sB^\mu W$, $\eta_\mu \tv^\mu_{\rm gas} = 0$, and $\sB^t=\tv^t_{\rm gas}=0$, then
the ZAMO energy density is given by
\begin{eqnarray}\label{EMHD}
-E &=& \eta^\mu Q_\mu = -W + \pg - \sB^2 \frac{1+\tv^2_{\rm gas}}{2} + \frac{S^2}{2W^2} \nonumber\\
   &=& -W + \pg - \sB^2 \frac{1+\tv^2_{\rm gas}}{2} + \frac{(\tv^k_{\rm gas} \sB_k)^2}{2} .
\end{eqnarray}
such that
\be\label{PRIM2}
Q_\mu = (W+\sB^2) \tv_{\mu \rm gas} + E\eta_\mu - (\tv_{\nu \rm gas} \sB^\nu) \sB_\mu .
\ee

As discussed in \cite{mm07}, one can avoid catastrophic cancellation issues in Eq.~(\ref{EMHD})
by using
\begin{equation}\label{eq:Wprime}
W'= \frac{D\tu_{\rm gas}^2}{1 + \gamma_{\rm gas}} + \chi_{\rm gas}\gamma_{\rm gas}^2
\end{equation}
instead of using $W = W'+D$, where $\chi_{\rm gas} \equiv \rhorest\epsilon_{\rm gas} + \pg$
and $\epsilon_{\rm gas}=(\rhorest + \ug)/\rhorest$.
Catastrophic cancellations for non-relativistic velocities
can be avoided by replacing $\gamma_{\rm gas}-1$ in any expression with $\tu^2_{\rm gas}/(\gamma_{\rm gas} + 1)$.

Next, using ${j^\mu}_\nu = \delta^\mu_\nu + \eta^\mu \eta_\nu$, compute the ZAMO momentum
\be\label{Qtilde}
\tilde{Q}^\nu \equiv {j^\nu}_\mu Q^\mu = Q^\nu + \eta^\nu (-E) = (W+\sB^2)\tv^\nu_{\rm gas} - ( \tv_{\mu \rm gas} \sB^\mu)\sB^\nu ,
\ee
and, since $u^\mu_{\rm gas}\sB_\mu/\gamma_{\rm gas} = \tu_\mu \sB^\mu/\gamma_{\rm gas} = \tv_{\mu \rm gas} \sB^\mu = (Q_\mu \sB^\mu)/W = (\tilde{Q}_\mu \sB^\mu)/W$, one can solve Eq.~(\ref{Qtilde})
for
\begin{equation}\label{tveq}
\tv^i_{\rm gas} = \frac{1}{W+\sB^2} \left[ \tilde{Q}^i
         +   \frac{ \left(Q_\mu \sB^\mu\right)  \sB^i }{W} \right]
\end{equation}
Also, one can obtain
\begin{equation}\label{tQeq}
\tilde{Q}^2 =
\tv^2_{\rm gas} \left(\sB^2 + W\right)^2
- {\left(Q_\mu \sB^\mu\right)^2 \left(\sB^2 + 2 W\right)\frac{1}{W^2}}.
\end{equation}
that can be solved for
\begin{equation}\label{vsqeq}
\tv^2_{\rm gas} = \frac{ \tQ^2 W^2 + \left(Q_\mu \sB^\mu\right)^2
\left( \sB^2 + 2 W \right) }{ \left(\sB^2 + W \right)^2 W^2 }  .
\end{equation}

A one dimensional inversion scheme is
derived by regarding Eq. (\ref{EMHD}) as a single nonlinear equation
in the only unknown $W$ (or $W'$).  Then one uses Eq. (\ref{tveq}) to
obtain $\tv^i_{\rm gas}$, then compute $\gamma_{\rm gas}$, then obtain $\rhorest$ from the
definition of $D$, then compute $\pg$ from the definition of $W$.
More details are provided in \citet{mm07}.

%%%%%%%%%%%%%%%%%%%%%%%%%%%%%%%%%%%%%%%%%%%%%%%%%%%%%%%%%%%%%%%%%%%%
\subsection{RAD Inversion of Conserved to Primitive Quantities}
%
%
%%%%%%%%%%%%%%%%%%%%%%%%%%%%%%%%%%%%%%%%%%%%%%%%%%%%%%%%%%%%%%%%%%%%

The radiative inversion is based upon the ZAMO frame, as compared to
the lab-frame in Koral.  For the radiative inversion using the M1
closure, one can solve for the radiation $P^{n+1}_q(U^{n+1}_q)$
analytically when given the radiative conserved quantities
$U_q^{n+1}$.

First, we let $p_{\rm rad}(e_{\rm rad}) = e_{\rm rad}/3$ (as for an ideal gas with $\Gamma=4/3$),
$\gamma_{\rm rad} = \sqrt{1 + g_{ij} \tu^i_{\rm rad} \tu^j_{\rm rad}}$,
$\tu_{\rm rad}^\mu = u_{\rm rad}^\mu - \gamma_{\rm rad} \eta^\mu$, $\tv_{\rm rad}^\mu  = \tu_{\rm rad}^\mu/\gamma_{\rm rad}$,
and $v_{\rm rad}^\mu = u_{\rm rad}^\mu/\gamma_{\rm rad}$, so that $\tv_{\rm rad}^2 = v_{\rm rad}^2 + 2\gamma_{\rm rad} -1$, and $W_{\rm rad} \equiv 4 p_{\rm rad} \gamma_{\rm rad}^2$.
For M1's ${R^\mu}_\nu = p_{\rm rad} (4u_{\rm rad}^\mu {u_{\nu}}_{\rm rad} + \delta^\mu_\nu)$, then
\begin{eqnarray}\label{RADPRIM1OLD}
U_\mu &=& -{R^\nu}_\mu \eta_\nu = \alpha {R^t}_\mu \\\nonumber
      &=& p_{\rm rad} (4\gamma_{\rm rad} u_{\mu \rm rad} - \eta_\mu) = W_{\rm rad} \tv_{\mu \rm rad} - (-W_{\rm rad} + p_{\rm rad})\eta_\mu ,
\end{eqnarray}
and
\begin{eqnarray}\label{RADPRIM2}
-E_{\rm rad} &=& -{R^\nu}_\mu \eta_\nu \eta^\mu = U_\mu \eta^\mu = \alpha {R^t}_\mu \eta^\mu \\\nonumber
            &=& -p_{\rm rad} (4\gamma_{\rm rad}^2 - 1) = -W_{\rm rad} + p_{\rm rad} ,
\end{eqnarray}
so that $p_{\rm rad} = -E_{\rm rad} + W_{\rm rad}$ and
\begin{equation}\label{RADPRIM1}
U_\mu = W_{\rm rad} \tv_{\mu \rm rad} + E_{\rm rad}\eta_\mu ,
\end{equation}
and
\begin{eqnarray}\label{RADPRIM3}
\tilde{U}^\mu &=& {j^\mu}_\nu U^\nu = (\delta^\mu_\nu + \eta^\mu \eta_\nu) U^\nu = U^\mu + \eta^\mu (-E_{\rm rad}) \\\nonumber
              &=& W_{\rm rad} (v_{\rm rad}^\mu - \eta^\mu) = W_{\rm rad} \tv_{\rm rad}^\mu ,
\end{eqnarray}
such that
\begin{equation}\label{RADPRIM4}
U^2 = W_{\rm rad}^2 v_{\rm rad}^2 - p_{\rm rad}^2 + 2p_{\rm rad} W_{\rm rad}
\end{equation}
and
\begin{equation}\label{RADPRIM5}
\tilde{U}^2 = W_{\rm rad}^2\tv_{\rm rad}^2 .
\end{equation}

One could solve Eq.~(\ref{RADPRIM2}) for $W_{\rm rad}$ using an iterative approach.
After obtaining a sufficiently
accurate $W_{\rm rad}$, one obtains the primitives from $\tv_{\rm rad}^\mu$ from
Eq.~(\ref{RADPRIM3}), then compute $\gamma_{\rm rad}$, then obtain $p_{\rm rad}$ from the
definition of $W_{\rm rad}$.

The M1 radiation inversion can be treated analytically
\citep{2013MNRAS.429.3533S}, where we solve these equations
differently than in Koral.  First, one has
\be
E_{\rm rad} = p_{\rm rad} (4\gamma_{\rm rad}^2-1)
\ee
and
\be
\tilde{U}^\mu = 4p_{\rm rad} \gamma_{\rm rad}^2 \tv_{\rm rad}^\mu .
\ee
Solving these equations for $p_{\rm rad}$ and $\tv_{\rm rad}^\mu$ (via $\tv_{\rm rad}^2$ or $\gamma_{\rm rad}^2$) gives
\be\label{prsol}
p_{\rm rad} = \frac{E_{\rm rad}}{4\gamma_{\rm rad}^2-1}
\ee
and
\be\label{tvrsol}
\tv_{\rm rad}^\mu = \frac{\tilde{U}^\mu(4\gamma_{\rm rad}^2-1)}{4E_{\rm rad}\gamma_{\rm rad}^2} = \frac{\tilde{U}^\mu}{4p_{\rm rad}\gamma_{\rm rad}^2}
\ee
where the correct root to choose is
\be\label{gammarvsy}
\gamma_{\rm rad}^2 = \frac{2-y+\sqrt{4-3y}}{4(1-y)} ,
\ee
where
\be\label{ysol}
y \equiv \frac{\tilde{U}^2}{E_{\rm rad}^2} ,
\ee
which is only allowed to range from $y=0$ to $y=1$ for
$\gamma_{\rm rad}=1$ to $\infty$, respectively.

Note that $\tilde{Q}^\mu+\tilde{U}^\mu$ and $E + E_{\rm rad}$ are
constant, so any instance of $E_{\rm rad}$ or $\tilde{U}^\mu$ can be
replaced by the total value minus the MHD value.  So, any occurrence
of $\gamma_{\rm rad}$ can be written in terms of quantities only
dependent upon MHD variables, and so $p_{\rm rad}$ (and so $W_{\rm
  rad}$) and $\tv_{\rm rad}^\mu$ can be similarly written.  One needs
to have some means to ensure that $0\le y \le 1$ and $p_{\rm rad}\ge
0$.

We have three approaches to limiting the radiation when $E_{\rm rad}\le 0$ or
$0<y\gtrsim 1$.  The ``BASIC'' limiter forces $\bar{E}=10^{-300}$ and
$\reluvec^i=0$ when $E_{\rm rad}\le 0$, $\reluvec^i_{\rm rad}=0$ when $0>y>-\epsilon_m$
for machine precision $\epsilon_m$, and if $y\ge y_{\rm max}$ then
relative 4-velocities are rescaled to ensure $y=y_{\rm max}$
corresponding to a $\gamma_{\rm max}$ via Eq.~(\ref{gammarvsy}).  The
``TYPE1'' and ``TYPE2'' limiters are similar, except if $y>y_{\rm
  max}$, then $U_{\rm abs} = (1/2) (\sqrt{|\tilde{U}^2|} + |E_{\rm rad}| +
10^{-150})$ and $\reluvec^i = \gamma \tilde{U}^i/U_{\rm abs}$ is first
set, and then this is rescaled to give the desired $\gamma_{\rm rad, max}$.
If ``TYPE2'' or if ``TYPE1'' with an original $y>1-100\epsilon_m$,
then we solve for $E_{\rm rad} = 10^{-150} + \sqrt{\tilde{U}^2/y_{\rm max}}$
then obtain $p_{\rm rad}$ and $\bar{E}$ as usual.  If using ``TYPE1'', then we check
whether this $\bar{E}$ is larger than the original estimate, and if so
we use the smaller value.  In summary, the ``BASIC'' limiter is the
most conservative for an actual solution to use during the evolution
since it helps avoids run-away energy gains.  However, for a smooth
Newton stepping, ``TYPE2'' is best since it avoids drop-outs to small
$\bar{E}$ that would cause the Newton stepping difficulties and
failure to recover the Newton stepping.

\subsection{Implicit radiative source terms}
\label{s.radsource}

To find an implicit solution (i.e. when Eq.~(\ref{Uaeq2}) has $U^i$ on
the left-hand side and $M^i$ on the right hand side for the same $i$
in the IMEX scheme), we use a 4D Newton scheme on a subset of the
equations of motion. This is possible because of the constraint
provided by total energy-energy momentum conservation between the gas
and fluid in a given cell.  We avoid a fully implicit scheme
\citep[e.g.,][]{krumholzetal07,jiangetal12}, which would require us to
use a more expensive 8D Newton scheme (to include geometry source
terms and expensive multi-cell Riemann flux differences and spatial
interpolations within the Newton solver).

We numerically solve one of the following two equations
\bea
&&\hspace{1cm}T^t_{\nu,(i+1)}-T^t_{\nu,(i)}=\Delta t ~G_{\nu,(i+1)},\label{eq.source3}\\
&&\hspace{1cm}R^t_{\nu,(i+1)}-R^t_{\nu,(i)}= - \Delta t~ G_{\nu,(i+1)}, \label{eq.source4}
\eea
where for the IMEX scheme, $i$ corresponds to all explicit contributions (i.e. initial+flux+geometry+explicit radiation),
while $i+1$ contains all implicit contributions (i.e. implicit radiation).
When solving the first equation, we find updates to $R^{\mu\nu}$
via the constraint that $T^t_{\mu,(i+1)}=T^t_{\mu,(i)}-\Delta R^t_{\mu,(i)}$
with $\Delta R^t_{\mu,(i)} = (R^t_{\mu,(i+1)}-R^t_{\mu,(i)})$.
When solving the second equation, we find updates to $T^{\mu\nu}$
via the constraint that $R^t_{\mu,(i+1)}=R^t_{\mu,(i)}-\Delta T^t_{\mu,(i)}$
with $\Delta T^t_{\mu,(i)} = (T^t_{\mu,(i+1)}-T^t_{\mu,(i)})$.

In cases when the energy-momentum conservation gives no solution
or $\ug<0$, then we replace the energy equation
\be
\hspace{1cm}T^t_{t,(i+1)}-T^t_{t,(i)}=\Delta t ~G_{t,(i+1)},\label{eq.source5}\\
\ee
with the entropy equation
\be
\hspace{1cm}[S u^t]_{(i+1)} - [S u^t]_{(i)}=\Delta t ~G_{S,(i+1)},\label{eq.source6}\\
\ee
while if also the entropy has no solution, we ignore energy and entropy
conservation (cold MHD limit).

Note that $\rhorest u^t_{\rm gas}$ is constant for the implicit equations, so its
addition to $T^t_t$ does not modify changes in $R^t_t$ when computing
$\Delta R^t_t$ or modify the value of $G_\mu$.

\subsection{Implicit Solver Methods}
\label{s.radsourcesmethods}

We use a Newton method to solve these energy/entropy/cold MHD
equations and estimate the Jacobian matrix numerically.

When the gas dominates the radiation or visa versa, then the finite
machine precision in the dominant fluid component can lead to
arbitrarily large numerical changes in the sub-dominant fluid
component.  E.g., when $|R^t_t|\gg|\rhorest u^t_{\rm gas} + T^t_t|$ or $\ug\gg \widehat E$ and
under other conditions, one should iterate the sub-dominant gas-fluid
quantities.  Further, one can choose to iterate conserved quantities
or primitive quantities in the implicit solver.  HARMRAD obtains the best solution out of iterating
one of four sets of quantities: $R^t_\mu$, $T^t_\mu$,
$\ug,\reluvec^i_{\rm gas}$, or $\bar{E},\reluvec^i_{\rm rad}$.

For each of the four sets of iterated quantities one requires
different steps to be taken to determine the error function that
enters into the Newton method.

For the method based upon iterating $R^t_\mu$ (called URAD), the steps are:
\begin{itemize}
\item Set $\Delta T^t_\mu = - \Delta R^t_\mu$.
\item Compute $G_s$ using prior primitives ($P_n$) to compute $G_\mu$ and $T_{\rm gas}$.
\item Perform an inversion from conserved quantities (latest solution for $T^t_{\mu,(i+1)},R^t_{\mu,(i+1)}$) to gas+radiation primitives (latest solution for $P_{i+1}$)
\item Recompute all conserved quantities from $P_{i+1}$ for consistency (in case of inversion failure/modification of solution).
\end{itemize}
This method requires an expensive Newton-inside-Newton calculation as
well as a numerical Jacobian that itself requires 4 MHD inversions per
overall Newton step.  That numerical Jacobian is computed with a
one-sided difference with a fixed difference size, which is prone to
arbitrarily large errors (see numerical recipes).  A centered
difference would be more accurate, but then 8 MHD inversions would be
required per Newton step.
Also, this method is unable to obtain an accurate entropy source term (although it will eventually converge),
and iterating $R^t_\mu$ can directly lead to out-of-bounds values
leading to no solution for the inversion to radiation primitives.

For a method based upon iterating $T^t_\mu$ (called UMHD), the steps are:
\begin{itemize}
\item Set $\Delta R^t_\mu = - \Delta T^t_\mu$.
\item Compute $G_s$ using prior primitives ($P_n$) to compute $G_\mu$, $u^\mu$, and $T_{\rm gas}$.
\item Perform an inversion from conserved quantities (latest solution for $T^t_{\mu,(i+1)},R^t_{\mu,(i+1)}$) to gas+radiation primitives (latest solution for $P_{i+1}$)
\item Recompute all conserved quantities for consistency.
\end{itemize}
This method is unable to obtain an accurate entropy source term
(although it will eventually converge),
and iterating $T^t_\mu$ can directly lead to out-of-bounds values
leading to no solution for the inversion to gas primitives.

For a method based upon iterating $S u^t, T^t_i$ (called ENTROPYUMHD), the steps are:
\begin{itemize}
\item Invert $S u^t,T^t_i$ to gas primitives (latest gas variables in $P_{i+1}$)
\item Recompute full $T^t_\mu$.
\item Set $\Delta R^t_\mu = - \Delta T^t_\mu$.
\item Invert $R^t_\mu$ to radiation primitives (latest radiation variables in $P_{i+1}$).
\item Recompute $R^t_\mu$ for consistency.
\end{itemize}
This method obtains an accurate entropy source for each implicit step,
but like the UMHD method, it iterating $T^t_i$ can directly lead to out-of-bounds values
leading to no solution for the inversion to gas primitives.

For a method based upon iterating the radiation primitives ($\bar{E}, \reluvec^\mu_{\rm rad}$) (called PRAD), the steps are:
\begin{itemize}
\item Compute $R^t_\mu$ from radiation $P_{i+1}$.
\item Estimate $G_S$ using prior primitives ($P_n$) to compute $G_\mu$, $u^\mu$, and $T_{\rm gas}$.
\item Set $\Delta T^t_\mu = - \Delta R^t_\mu$.
\item Invert $T^t_\mu$ to gas primitives (latest gas variables in $P_{i+1}$).
\item Recompute $T^t_\mu$ for consistency.
\end{itemize}
This method estimates $G_S$ and is slow as the UMHD method, but it
at least does not iterate out of bounds like the URAD method.  However, it cannot be used when the
gas is very sub-dominant due to machine precision issues.

For a method based upon iterating the gas primitives ($\ug, \reluvec^\mu_{\rm gas}$) (called PMHD), the steps are:
\begin{itemize}
\item Obtain $\rho_0 = U_1/u^t$ from the newly updated $P_{i+1}$.
\item Compute $S u^t,T^t_\mu$ from gas $P_{i+1}$.
\item Set $\Delta R^t_\mu = - \Delta T^t_\mu$.
\item Invert $R^t_\mu$ to radiation primitives (latest radiation variables in $P_{i+1}$).
\item Recompute $R^t_\mu$ for consistency.
\end{itemize}
This method has none of the flaws mentioned in the previous methods
and is fast because the radiation inversion is analytic and simple.
Its only flaw is that it cannot be used when the radiation is very sub-dominant due to
machine precision issues.

Now, after one (or all) of these methods are used, we have consistent
values for $U_q$ and $P_q$ from which we can compute an error function
required by the Newton method.  So, the next steps are:
\begin{itemize}
\item Compute 4-force $G_{\mu,(i+1)}(P_{q,(i+1)})$.
\item Compute the error function for each type of variable as $E_q = (U_{q,(i+1)} - U_{q,(i)})F_q + dt S_q F_q$.
\item Compute the normalized error $e_q$, by dividing by sum of absolute value of any signed terms that appear in $E_q$ and all its sub-expressions.  For each gas and radiation terms, spatial norms are merged in an orthonormal basis to form a single spatial norm for all dimensions.
\end{itemize}
For the $q$ conserved/primitive quantities, a similar set of extra
factors $F_q=\{1, 1, 1, 1, 1, 1, T_{\rm gas}\}$ multiply the overall error
function.  The temperature is factored into the entropy equation using
$F_{13}$ in order to generate a more regular/linear functional
behavior near roots to make it easier to find the solution, as found
experimentally using HARMRAD and test code in Mathematica using both
an analytical and numerical Jacobian.  Also, for this entropy term
with the $T_{\rm gas}$ factor, we include in the norm the energy norm,
which helps normalize the error in $T_{\rm gas} dS$.

The ``total normalized error'' is computed as
\be\label{totalerror}
e_T = (1/4)\sum_q e_q ,
\ee
which is computed over all quantities to produce an error independent
of the method or iterated quantities used as well as to account for
the error in the un-iterated quantities that could be large despite
small iterated quantities in different regimes. The ``iterated
normalized error'' is computed from iterated quantities as
\be\label{itererror}
e_I = (1/4)\sum_{q_k} e_{q_k} ,
\ee
over only those $q$'s that were iterated ($q_k$'s).  E.g., $k=2,3,4,5$
in $U_k$ for the UMHD method, $k=9,10,11,12$ in $U_k$ for the URAD
method, and the same $k$'s in $P_k$ for each of the PMHD and PRAD
methods, respectively.  For the energy methods, entropy is ignored in
the error.  For entropy methods, energy is ignored.  For cold MHD
methods, both energy and entropy errors are ignored.  This generates
the appropriate total and iterated error for each method.

Once the solution for $U_{q,(i+1)}$ is obtained, the implicit
radiation source term is given by
\begin{equation}
\SR_{q,(i+1)} = \frac{1}{dt_i}(U_{q,(i+1)} - U_{q,(i)})
\end{equation}
for each implicit Runge-Kutta sub-step of size $dt_i$.  This then
provides all terms required for the radiation GRMHD method to complete
a single Runge-Kutta sub-step.

%%%%%%%%%%%%%%%%%%%%%%%%%%%%%%%%%%%%%%%%%%%%%%%%%%%%%%%%%%%%%%%
\subsection{4D Newton-Raphson Scheme}
%
%
%
%
%%%%%%%%%%%%%%%%%%%%%%%%%%%%%%%%%%%%%%%%%%%%%%%%%%%%%%%%%%%%%%%

The implicit solver involves computing the $(k+1)$-th approximation to
a set of iterated dependent variables $\vec{x}$ using a Newton-Raphson
method.  For the various methods, the iterated quantities are one of
the $q=1$--$13$ quantities in the list of quantities given by
$\tilde{U}_q=U_q/\detg = \{\rhorest u^t , \rhorest u^t_{\rm gas} +
{T^t}_t, {T^t}_i, B^i, {R^t}_\nu, S u^t\}$.  For the energy-based or
entropy-based URAD,UMHD,PRAD,PMHD methods, we iterate
$\{\tilde{U}_{9,10,11,12},\tilde{U}_{2,3,4,5},P_{9,10,11,12},P_{2,3,4,5}\}$,
respectively.  The cold MHD based method simply uses a reduced 3D
method with iterated quantities from one of
$\{\tilde{U}_{10,11,12},\tilde{U}_{3,4,5},P_{10,11,12},P_{3,4,5}\}$.
The error function is independent from the iterated quantities, and
for each energy-based URAD,UMHD,PRAD,PMHD methods, the error function
quantities are one of
$\{\tilde{U}_{9,10,11,12},\tilde{U}_{2,3,4,5},\tilde{U}_{9,10,11,12},\tilde{U}_{2,3,4,5}\}$,
respectively, while for each entropy-based URAD,UMHD,PRAD,PMHD
methods, the error function quantities are one of
$\{\tilde{U}_{13,10,11,12},\tilde{U}_{13,3,4,5},\tilde{U}_{13,10,11,12},\tilde{U}_{13,3,4,5}\}$,
respectively, while for the cold MHD based URAD,UMHD,PRAD,PMHD
methods, the error function quantities are one of
$\{\tilde{U}_{10,11,12},\tilde{U}_{3,4,5},\tilde{U}_{10,11,12},\tilde{U}_{3,4,5}\}$,
respectively.  Then, the Newton update is obtained via
\begin{equation}\label{eq:newton}
  \vec{x}^{(k+1)} = \vec{x}^{(k)} - D\vec{E}(\vec{x}) \cdot \left(\frac{\partial \vec{E}(\vec{x})}{\partial \vec{x}}\right)^{-1}|_{\vec{x}=\vec{x}^{(k)}} .
\end{equation}
for damping factor $D$.

The Newton step requires computation of the Jacobian
$\partial\vec{E}/\partial\vec{x}$, which is simply obtained as a
finite difference of the error function away from a reference value of
$x$ from the latest estimate of $P_{i+1}$ or $U_{i+1}$.  A one-sided
finite difference (using reference and an offset) is used for
URAD,PRAD due the expense of the error function requiring the 1D MHD
inversion, while the UMHD,ENTROPYUMHD,PMHD methods use a two-sided
finite difference (using two offsets away from the reference value) if
the ``stages'' approach (see below) is used.  A fixed normalized
error offset of $10^{-8}$ is used, until $e_I$ from Eq.~(\ref{itererror})
drops below $10^{-9}$ in which case a normalized error offset of $10^{-10}$
is used.  If the Jacobian calculation hits a conserved to primitive
inversion failure, larger or smaller offsets are attempted until no
inversion problem occurs up until a normalized error offset of $0.3$.
Any MHD inversions for the Jacobian are computed with a tolerance of $10^{-2}$
times the Jacobian difference used, since any more accuracy is wasted
in the finite difference.  Any RAD inversions for the Jacobian use the
TYPE2 limiter on the inversion, which avoids both offsets giving the
same $f$ (and so singular inversion) when otherwise the BASIC or TYPE1
limiters would lead to $\bar{E}\to 0$ as the solution.

\subsection{Quickly Choosing Optimal Implicit Solution}
\label{s.radsourceswitch}

In total there are 6 Newton methods for each energy and entropy based
methods, while there is a single cold MHD method.  Arbitrary use of
all these approaches would be costly, so one needs to estimate which
is optimal to use to get the smallest error in the shortest time.

First, we consider quantities: (1) energy gas vs. radiation values of
$U_{q,i}$, (2) energy gas vs. radiation values of $\Delta_r U_q
= (U_{q,i} - U_q^n)/(|U_{q,i}| + |U_q^{n}|)$, and (3)
$\ug$ vs. $\widehat{E}$.  If any of the radiation versions of these
quantities have absolute magnitudes smaller than $10\epsilon_m$, then
we assume the radiation is in an ``extreme radiation sub-dominant
regime'' that the numerical machine precision errors in gas conserved
quantity calculations will ruin the ability to find an implicit
solution.  Hence, in this situation, we first try the URAD/PRAD
methods.  Otherwise, we use no other pre-conditions to control the
order of the method attempted, and in particular we otherwise first
use the fastest PMHD method as described next.

To optimize performance, a ``normal'' Newton method for a maximum of
$20$ steps is attempted first using the fast energy-based PMHD method
as the first attempted iterate/error choice, then the URAD method is
attempted if the PMHD solution is unacceptable, then the PRAD method
is attempted if the URAD solution is unacceptable.  Newton steps are
undamped ($D=1$), although damping has been found to help seek errors
closer to machine precision and to avoid cycle behaviors, at greater
computational cost.  For the PRAD,URAD methods, the total number of
MHD inversion steps is also monitored and not allowed to go beyond
$20\times 6$ total steps to avoid excessive steps.

Then, the PMHD, URAD, PRAD methods are attempted using a ``stage''
approach that takes (stage 1) one momentum-only Newton step, then
(stage 1) converges using energy-only Newton stepping, then (stage 3)
seeks the full 4D solution for a maximum of $40$ steps.  For this
``staged'' approach, the Newton steps are damped by a factor of $D=1/2$
for the first momentum step, $D=1/2$ for the first energy step, and
$D=1/4$ for the first full 4D step.  This avoids large changes when
transitioning to using a different set of equations to step.  The
``staged'' approach is avoided if the radiation is not in the
``extreme radiation sub-dominant regime,'' described already.  The
total number of MHD inversion steps is limited to $40\times 6$ total
steps to avoid excessive steps.

The guess for the Newton method, for any attempted method, is set as
the solution from the previous attempt with the smallest error ($e_T$)
as long as that previous attempt gave a solution with $e_T<10^{-4}$.
If no prior guess exist or have $e_T<10^{-4}$, then the prior
primitives $P_{n}$ are used as the guess.  For the PRAD method, that
simple guess is modified by first finding $P$ from $U_{q,i}$
given in Eq.~(\ref{Uaeq2}) that describes some ``optically thin''
limit for the primitive solution as if there were no source term.
Then, the optical depth $\tau$ is computed from $P_n$, and if
$\tau<2/3$, then that ``optically thin limit'' primitive is set as the
guess.  Otherwise, the prior primitive $P^n$ corresponding to $U_q^n$
is used.  For the PMHD method, similarly, $P$ is obtained from an MHD
inversion of $U_{q,i}$ (using MHD inversion tolerance of
$10^{-9}$) and used if $\tau<2/3$, otherwise $P^n$ is used.  For the
PMHD method, the ``stage'' guess for $\ug$ is further modified as an
inversion from the conserved entropy $S u^t$ per unit conserved mass
$\rhorest u^t_{\rm gas}$ to give a specific entropy $s_{\rm gas}$ that together with the
previous density $\rho_0$ is used to obtain an estimate of $\ug$.
This helps avoid issues with stepping when the initial guess for $\ug$
is too small and leads to a guess for the 4-force that is large.

For the first step and after each step, the error is computed using
the TYPE2 limiter when computing the radiation inversion.  If the
limiter was activated because the radiation energy was negative, then
the error excludes the energy term (or entropy term for the
entropy-based methods) in the error function.  The TYPE2 limiter then
ensures the radiation momentum is unchanged, so momentum can continue
to be monitored.  If this leads to an inversion error, then the step
is backed-up half-way until it succeeds for $20$ attempts.

Then, we perform pre-step checks.  We check if the error is already
small, and if so break before computing the Jacobian or step.  We also
check to see (for steps beyond the $5$th full 4D type step) that the
error is dropping fast enough (dropping by $0.5$) compared to the
average of $3$ steps starting from $5$ steps to $2$ steps ago.  Each
of the total and iterated errors must satisfy this requirement or the
iterations are stopped to save computational expense when no
improvements are being made in the error.
Another pre-step check is to see if the error is repeatedly rising,
by checking whether the error has risen (instead of dropped)
$5$ times starting after the $5$th full 4D type step.  Once
the error has risen that many times, the iterations are dropped
to avoid computational expense.

The Newton step is then taken for either $U$ or $P$ using
Eq.~(\ref{eq:newton}).

Then we perform post-step checks.  If $\ug<0$ for PMHD,UMHD methods or
$\bar{E}<0$ for URAD,PRAD methods, then each $\ug$ or $\bar{E}$ are
set to $0.5$ of their absolute magnitudes.  This helps do a one-sided
bisection down to small values.  If this modification occurs $2$ times
for the ``staged'' momentum steps or ``staged'' energy steps for the
energy-based method or $4$ times for the ``staged'' momentum steps or
``staged'' energy steps for the entropy-based methods, then the prior
value is held and the next ``stage'' in the stages approach is attempted.
Another post-step check is we count to see if $\ug<0$ for the
URAD,PRAD methods more than $3$ times.  In that case, we assume
another method or the entropy-based method are more appropriate and
stop the iterations.

For each attempt, the iterations are stopped and Newton method aborted if any
of the following occurs:
\begin{enumerate}
\item the ``total normalized error'' $e_T$ from Eq.~(\ref{totalerror}), falls below a tolerance of ${\rm tol}=10^{-12}$ for the PMHD,UMHD methods or ${\rm
  tol}=10^{-9}$ for the URAD,PRAD methods.
\item the residual $\left|\vec{x}^{(k+1)}/\vec{x}^{(k)} - 1\right|$ falls below
$10\epsilon_m$.
\item a maximum number of Newton steps is hit
(chosen independently for each method given each of their performance issues).
\end{enumerate}
In any case, for a given attempt, a final error is
computed based on the last step, and the best solution (with the
smallest total error) is used as the solution.  Lastly, when using the
PMHD, UMHD, or ENTROPYUMHD methods, if that best solution used the
TYPE2 limiter and encountered a negative radiation energy density, then
the radiative inversion is recomputed using the BASIC limiter.  This
ensures that while we chose the best solution possible, the actual
solution used reduces the radiation primitives to $\bar{E}\sim 0$ to
avoid run-away energy gains.

In any case, if the attempted tolerance is not met, an error of
$e_T<10^{-9}$ is considered acceptable regardless of the attempted error
and no further attempts at using other methods are made.  An error of
$e_T>10^{-9}$ is considered not quite acceptable, in which case the next
method in line is attempted to seek a smaller error.  The solution
with the lowest error over all steps and methods used is taken as the
energy-based solution.

\subsection{Backup solvers/solutions when Implicit Energy solver Fails}
\label{s.backup}

If the energy-based method fails to obtain the required error or if
the energy-based solution gives $\ug<0$, then the entropy-based
methods are attempted in the same sequence as the energy-based methods
(the ENTROPYUMHD method is currently not used).  As with the
energy-based method, the entropy solution with the lowest error over
all steps and methods used is taken as the entropy-based solution.

If the entropy method fails and the energy method gives $\ug<0$ or if
both energy and entropy methods fail, then we consider using the cold
MHD equations. Let $t^t_\mu = T^t_\mu + \rhorest u^t_{\rm gas}$, then the cold MHD
solver is only attempted if $\ug<0.1 \rho_0 |\uvec^i \uvec_i|$, and
$|t^t_t t^{tt}| < 0.1 |t^t_i t^{ti}|$, and $|R^t_t R^{tt}| < 0.1
|R^t_i R^{ti}|$.  Only the PMHD method is used.  The final value of
$\ug$ is spatially averaged over neighbors that had good energy or
entropy inversions, or averaged over all bad neighbors if no good
neighbors exist.

The stages, entropy, and especially cold backup methods are used
rarely, but they help avoid lack of inversion.

No solution is considered to be when the total error is $e_T>10^{-7}$, as
experimenting shows that beginning around $10^{-4}$ or so, such a
large error can imply a static solution with no changes when evolution
should occur.  When none of energy, entropy, or cold MHD methods meet
this tolerance, then diffusive back-up methods are used.  In this very
rare case where none of these find an acceptable solution, then the
radiative source term is temporarily updated explicitly using backup
inversion methods as in HARM for the gas quantities, but then the full
set of primitives is spatially averaged over neighbors that had a good
inversion.  If no good neighbors exist for a point, then averages are
performed over all neighbors, but this happens in none of the tests
considered in this paper.  Note that using static values (rather than
averages) as a final backup method can lead to catastrophic evolution
because then (e.g.) the induction equation (that must evolve the
magnetic field in order to preserve the solenoidal constraint) is an
simple differential equation for a constant $v^i$ that gives an
exponentially growing $B^j$.

In cases where a solution is acceptable but has too small $\rho_0, \ug$,
we use the numerical floor approach described in \citet{2012MNRAS.423.3083M}.
The only constraint on $\bar{E}$ is that it is forced to be positive
by setting it to $10^{-150}$ if it was non-positive.

In our experimentation with explicit sub-cycling as an alternative/backup to
the implicit inversion, some conditions to use sub-cycle methods were
experimented with, but nothing was found to be generally applicable in
all regimes.  So sub-cycling was completely abandoned in favor of the
implicit method.

\section{Test problems}
\label{s.tests}

We consider several radiative tests to ensure the numerical method is
accurate, robust, and fast.  Our goal is not to have exceptionally
sharp discontinuities or noise-less solutions by tweaking the
numerical method used for each test, but rather our goal is to use a
high-resolution interpolation scheme and ensure the noise or issues
are manageable so we understand how the code would operate when used to
study general problems involving magnetized accretion flows around
black holes.  So we accept some noise and study under what extreme
situations that noise appears.

Non-radiative tests were performed in \citet{2003ApJ...589..444G} and
so are not considered in this paper.  Most of the radiative tests are
based on test problems from \citet{2013MNRAS.429.3533S,
  2013arXiv1311.5900S}.  We extend this set by considering a larger
physical parameter space in linear wave convergence tests
(section~\ref{RADWAVE}), MHD radiative Bondi flow (section
\ref{RADBONDIMAG}), and for the first time present a fully 3D
radiation GRMHD simulation of a disk-jet accretion system.

We used fixed numerical parameters for all tests and the fiducial 3D
simulation in section~\ref{sec:fiducial}, but all tests use LAXF
except the double shadow test (see section~\ref{RADDBLSHADOW}) uses
HLL (we discuss the minor LAXF issues with this test).  We use a
maximum radiative Lorentz factor of $\gamma_{\rm rad,\rm max} = 100$
(or twice larger than the injected beam's value for tests that inject
higher $\gamma_{\rm rad}$), 3rd order Runge-Kutta method (RK3), LAXF
flux, PPM reconstruction, and Courant factor $C=0.49999$.  Note that
using a lower-order reconstruction like MINM or MC leads to more
diffusion and does not stress the method as much as using PPM, while
PPM can lead to some additional grid-scale artifacts.  Also, we use a
3rd order Runge-Kutta with $C\sim 1/2$ in order to more generally
handle cases where the gas temperature is low, giving a relatively low
internal energy density compared to the kinetic or magnetic energies.
None of these tests exhibit any implicit solver failures.

\subsection{Radiation modified MHD linear waves in 1D Cartesian Minkowski}
\label{RADWAVE}

\begin{figure*}
\includegraphics[width=0.9\textwidth]{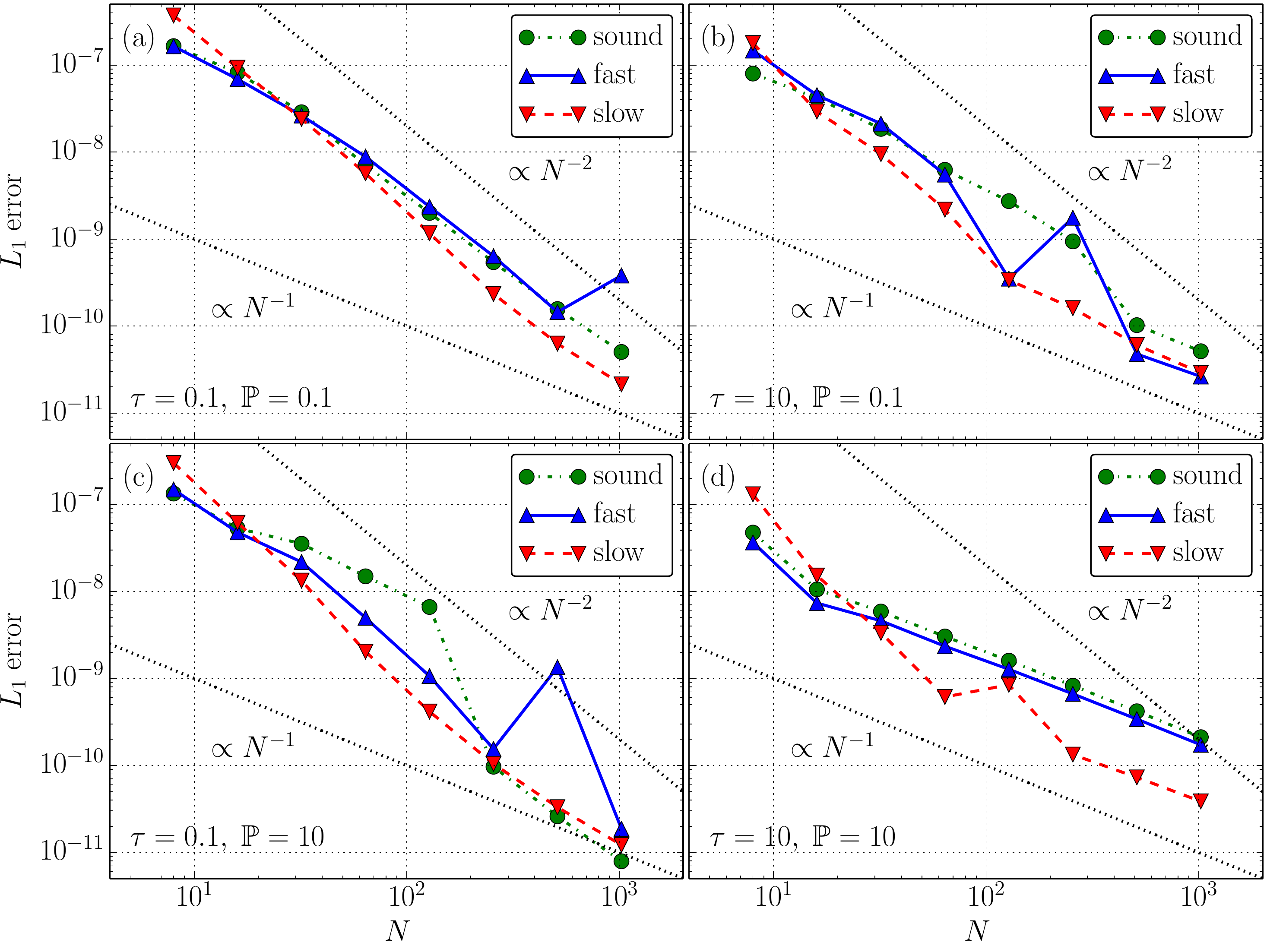}
\caption{Spatial convergence rate of the radiation modified MHD linear
  waves. Similar to Table~\ref{t.waves}, the four panels in this table
  show eigenmodes of linear waves: top row of panels [panels (a) and (b)] shows
  gas-dominated and bottom row of panels [panels (c) and (d)]
  radiation-dominated systems; left column of panels [panels (a) and (c)] shows
  optically-thin and right column of panels [panels (b) and (d)]
  optically-thick systems. Numerical results for sound waves are shown
  with green circles, fast waves with blue triangles, and slow waves
  with red inverted triangles. The black dotted lines show quadratic
  and linear convergence.  Summarizing, the code converges at the
  expected rate, although PPM leads to some non-monotonic convergence
  that does not occur with MINM or MC limiters.}
  \label{f.rmhdl1}
\end{figure*}

First, we test the accuracy with which our numerical scheme propagates linear
MHD waves in the presence of radiation. As waves propagate
through the gas, their interactions with the photon field feed back into
the gas and modify the nature of the perturbations themselves relative to the
non-radiative case. Making sure that this interaction is correctly
captured in the numerical scheme and that the numerical solution
converges to the analytical solution at the expected order is a stringent test of the
numerical method. We will consider sound, slow, and fast waves.
We do not consider Alfv\'en waves because they are less affected by
radiative effects \citep{2012ApJS..199...14J}.

For each of the tests, we initialize a single eigenmode, of form
\begin{equation}
q^i = {\it Re}\left[q^i_a + \delta q^i e^{i (\omega t - k
  x)}\right],  \label{eq:qi}
\end{equation}
with eigenvectors $\delta q^i$ given in
Table~\ref{t.waves} and computed using the method described
in \citet{2013arXiv1311.5900S}.  Here ${\it Re\,}(\dots)$ indicates the
real part of a variable. The ambient background medium is uniform,
of density
$\rho_a=1$ and sound speed $c_{\rm s,a} = 0.1$. This corresponds to
gas internal energy, $e_{\rm gas,a}=\rho[\Gamma(\Gamma-1) c_{\rm
  s,a}^{-2}-\Gamma]^{-1}=0.009137055837563452$, and the thermal
pressure of the ambient gas, $p_{\rm gas,a} = (\Gamma-1)e_{\rm gas}$,
where we choose $\Gamma = 5/3$.
The ambient medium is at rest ($v_a^x =
v_a^y = 0$), and the radiation flux in the
fluid frame vanishes $\widehat F^x_a=\widehat F^y_a=0$.
For sound waves we set all magnetic field components to zero, whereas
for slow and fast waves, we set $B_a^x = B_a^y = B_0 =
0.10075854437197568$ (see Table~\ref{t.waves} and
\citealt{2013arXiv1311.5900S}).

We carry out the simulations on a 1D domain, $0\le x \le 1$, however,
we allow for velocities and magnetic fields in the $y-$direction. We use periodic boundary conditions in the $x$-direction. We
consider a wavenumber $k = 2\pi$ such that one wavelength fits in the
domain. We set the absorption opacity to zero, $\kappa_{\rm abs} = 0$.  We
vary the scattering opacity given by the optical depth of the domain,
$\kappa_{\rm es}=\tau$, and set the radiation pressure via a
dimensionless parameter, $\mathbb P = p_{\rm rad,a}/p_{\rm gas,a}$
(see Table~\ref{t.waves}).

After one period of the wave, i.e., $P = 2\pi/{\it
  Re\,}(\omega)$, the numerical solution, $\rho_0^j(t=P)$, deviates from
the analytic solution, $\rho^j_{\rm 0,A}(t=P)$, which is given by
eq.~\eqref{eq:qi}.  We measure this deviation using $L_1$ norm:
\begin{equation}
  \label{eq:1}
  L_1 = N^{-1}\sum_j\left| \rho_0^j - \rho^j_{0,A}\right|,
\end{equation}
where the summation is carried out over all of the $N$ grid cells.
Figure~\ref{f.rmhdl1} shows convergence of our scheme in four panels, with each panel
demonstrating the convergence of the eigenmodes given in the
corresponding panel in Table~\ref{t.waves}.  As seen in
Fig.~\ref{f.rmhdl1}(a),(c), in the optically-thin case, $\tau = 0.1$,
the numerical solution converges to the analytic solution at 2nd
order, as expected. (Note that PPM leads to some non-monotonic
convergence that does not occur with MINM or MC limiters.)  As is
clear from Fig.~\ref{f.rmhdl1}(b), higher optical depth simulations converge at 2nd
order at low and moderate resolutions and switch over to 1st order convergence
at high resolutions. Thus, the
radiation component of the code, which is treated implicitly and
converges at 1st order, affects
the overall convergence rate at $\tau\gg1$. As expected,
Fig.~\ref{f.rmhdl1}(d) demonstrates that
radiation-dominated optically-thick case converges at first order.

\begin{table*}
\begin{minipage}{\textwidth}
\caption{Eigenmodes of linear
  waves.  Top row, panels (a) and (b), shows gas-dominated and
  bottom row, panels (c) and (d), radiation-dominated systems; left
  column, panels (a) and (c), shows
  optically-thin and right column, panels (b) and (d),
  optically-thick systems. Other parameters: $\rho_a=1$, $e_{\rm gas,a}=0.009137055837563452$, $u_a^x=u_a^y=0$, $\widehat F^x_a=\widehat F^y_a=0$,
$B^x_a=B^y_a=B_0$, $\kappa_{\rm abs}=0$, $k=2\pi$,
$\Gamma=5/3$, and ${\mathbb P} \equiv p_{\rm rad,a}/p_{\rm gas,a}$, $\kappa_{\rm es}=\tau$ as given in the Table. Here we choose $B_0 = 0$ for sound waves and $B_0 =
0.10075854437197568$ for fast and slow waves. Perturbations are of the form
$q^i = {\it Re\,}[q^i_a + \delta q^i e^{i (\omega t - k x)}]$.
}
\label{t.waves}
\begin{tabular}{clr@{$\,$}c@{$\,$}l}
\hline
(a) & \multicolumn{4}{|c|}{Optically-thin, gas-dominated ($\tau = 0.1$, ${\mathbb P} = 0.1$)} \\
\hline
\parbox[t]{2mm}{\multirow{9}{*}{\rotatebox[origin=c]{90}{sound wave}}}& $\delta\rhorest$        &                 1e-06 & $+$ & 0                    $i$ \\
& $\delta \ug$            &  1.51556529080798e-08 & $+$ & 7.69692971953054e-10 $i$ \\
& $\delta u^x$            &  9.97992249118626e-08 & $+$ & 2.55207217592872e-09 $i$ \\
& $\delta u^y$            &                     0 & $+$ & 0                    $i$ \\
& $\delta B^y$            &                     0 & $+$ & 0                    $i$ \\
& $\delta \widehat E$     &  1.33147769911346e-13 & $+$ & 3.6001746512389e-11  $i$ \\
& $\delta \widehat F^x$   & -2.52471264862269e-10 & $+$ & 7.40040781015203e-11 $i$ \\
& $\delta \widehat F^y$   &                     0 & $+$ & 0                    $i$ \\
& $\omega$                &     0.627057023634126 & $+$ & 0.0160351423986572   $i$ \\
\hline
\parbox[t]{2mm}{\multirow{9}{*}{\rotatebox[origin=c]{90}{fast wave}}}& $\delta\rhorest$        &                 1e-06 & $+$ & 0                    $i$ \\
& $\delta \ug$            &   1.5198360895975e-08 & $+$ & 4.81575290993662e-10 $i$ \\
& $\delta u^x$            &  1.60251314293283e-07 & $+$ & 7.2383120050772e-10  $i$ \\
& $\delta u^y$            & -9.79544263084857e-08 & $+$ & 9.83678950177998e-10 $i$ \\
& $\delta B^y$            &  1.62343664101617e-07 & $-$ & 8.96662164240542e-10 $i$ \\
& $\delta \widehat E$     &  1.48421181882934e-12 & $+$ & 6.06322316295508e-11 $i$ \\
& $\delta \widehat F^x$   & -3.95432710842345e-10 & $+$ & 8.51051304663626e-11 $i$ \\
& $\delta \widehat F^y$   &  2.36679521545993e-10 & $+$ & 2.11182386936598e-11 $i$ \\
& $\omega$                &      1.00688870342377 & $+$ & 0.00454796556390826  $i$ \\
\hline
\parbox[t]{2mm}{\multirow{9}{*}{\rotatebox[origin=c]{90}{slow wave}}}& $\delta\rhorest$        &                 1e-06 & $+$ & 0                    $i$ \\
& $\delta \ug$            &  1.50174235106495e-08 & $+$ & 1.22298943455801e-09 $i$ \\
& $\delta u^x$            &  6.15332754996702e-08 & $+$ & 1.83139801648519e-09 $i$ \\
& $\delta u^y$            &  9.89772118301622e-08 & $+$ & 6.54185791061938e-09 $i$ \\
& $\delta B^y$            & -6.14882091832378e-08 & $-$ & 5.88315338295397e-09 $i$ \\
& $\delta \widehat E$     &   1.9170283012363e-13 & $+$ & 2.18721458210053e-11 $i$ \\
& $\delta \widehat F^x$   & -1.65180532693438e-10 & $+$ & 7.1752041819694e-11  $i$ \\
& $\delta \widehat F^y$   & -2.23678888272661e-10 & $-$ & 7.43141463935117e-11 $i$ \\
& $\omega$                &     0.386624972522161 & $+$ & 0.0115070131087776   $i$ \\

\\[-0.1cm]
\hline
(c) & \multicolumn{4}{|c|}{Optically-thin, radiation-dominated ($\tau = 0.1$, ${\mathbb P} = 10$)} \\
\hline
\parbox[t]{2mm}{\multirow{9}{*}{\rotatebox[origin=c]{90}{sound wave}}}& $\delta\rhorest$        &                 1e-06 & $+$ & 0                    $i$ \\
& $\delta \ug$            &  9.14134312414906e-09 & $+$ & 3.83221342644378e-10 $i$ \\
& $\delta u^x$            &  7.74804632510917e-08 & $+$ & 3.58319335378836e-09 $i$ \\
& $\delta u^y$            &                     0 & $+$ & 0                    $i$ \\
& $\delta B^y$            &                     0 & $+$ & 0                    $i$ \\
& $\delta \widehat E$     & -1.52193074553572e-10 & $+$ & 9.38040606030636e-10 $i$ \\
& $\delta \widehat F^x$   & -1.93666448665796e-08 & $-$ & 7.93046885686635e-10 $i$ \\
& $\delta \widehat F^y$   &                     0 & $+$ & 0                    $i$ \\
& $\omega$                &     0.486824108292728 & $+$ & 0.0225138678333065   $i$ \\
\hline
\parbox[t]{2mm}{\multirow{9}{*}{\rotatebox[origin=c]{90}{fast wave}}}& $\delta\rhorest$        &                 1e-06 & $+$ & 0                    $i$ \\
& $\delta \ug$            &  9.20593402786723e-09 & $+$ & 7.43788681605899e-10 $i$ \\
& $\delta u^x$            &  1.51647542124337e-07 & $+$ & 2.92430743208669e-09 $i$ \\
& $\delta u^y$            & -1.11471565905267e-07 & $+$ & 6.59617536694267e-10 $i$ \\
& $\delta B^y$            &  1.74787152941627e-07 & $-$ & 1.8658034881622e-09  $i$ \\
& $\delta \widehat E$     & -4.70212889643857e-10 & $+$ & 1.98234059740446e-09 $i$ \\
& $\delta \widehat F^x$   & -3.79422297680303e-08 & $-$ & 3.18097469382448e-10 $i$ \\
& $\delta \widehat F^y$   &  2.69349129873962e-08 & $+$ & 2.6704669357673e-09  $i$ \\
& $\omega$                &     0.952829608545529 & $+$ & 0.0183739654909631   $i$ \\
\hline
\parbox[t]{2mm}{\multirow{9}{*}{\rotatebox[origin=c]{90}{slow wave}}}& $\delta\rhorest$        &                 1e-06 & $+$ & 0                    $i$ \\
& $\delta \ug$            &  9.13449746980553e-09 & $+$ & 2.48194479126941e-10 $i$ \\
& $\delta u^x$            &  5.01905313372291e-08 & $+$ & 2.38661835159163e-09 $i$ \\
& $\delta u^y$            &  6.76529059165483e-08 & $+$ & 3.82639444942364e-09 $i$ \\
& $\delta B^y$            & -3.51141271790158e-08 & $-$ & 1.22066297927614e-09 $i$ \\
& $\delta \widehat E$     & -7.35475223430181e-11 & $+$ & 6.00744732838449e-10 $i$ \\
& $\delta \widehat F^x$   & -1.25407400097346e-08 & $-$ & 5.53621772791961e-10 $i$ \\
& $\delta \widehat F^y$   & -1.48948162136833e-08 & $-$ & 5.73105396778065e-09 $i$ \\
& $\omega$                &     0.315356409057614 & $+$ & 0.0149955653605657   $i$ \\

\end{tabular}
\hfill
\begin{tabular}{clr@{$\,$}c@{$\,$}l}
\hline
(b) & \multicolumn{4}{|c|}{Optically-thick, gas-dominated ($\tau = 10$, ${\mathbb P} = 0.1$)} \\
\hline
\parbox[t]{2mm}{\multirow{9}{*}{\rotatebox[origin=c]{90}{sound wave}}}& $\delta\rhorest$        &                 1e-06 & $+$ & 0                    $i$ \\
& $\delta \ug$            &   1.1797669003418e-08 & $+$ & 3.04292104805925e-09 $i$ \\
& $\delta u^x$            &  9.29098565575498e-08 & $+$ & 1.44382411133922e-08 $i$ \\
& $\delta u^y$            &                     0 & $+$ & 0                    $i$ \\
& $\delta B^y$            &                     0 & $+$ & 0                    $i$ \\
& $\delta \widehat E$     &  1.98197717210158e-09 & $+$ & 2.2064547519989e-09  $i$ \\
& $\delta \widehat F^x$   & -4.36777061885618e-10 & $+$ & 4.31621443724302e-10 $i$ \\
& $\delta \widehat F^y$   &                     0 & $+$ & 0                    $i$ \\
& $\omega$                &      0.58376984561456 & $+$ & 0.0907181444251821   $i$ \\
\hline
\parbox[t]{2mm}{\multirow{9}{*}{\rotatebox[origin=c]{90}{fast wave}}}& $\delta\rhorest$        &                 1e-06 & $+$ & 0                    $i$ \\
& $\delta \ug$            &  1.31055121995324e-08 & $+$ & 2.26907894858283e-09 $i$ \\
& $\delta u^x$            &  1.59214919060003e-07 & $+$ & 4.26730946717188e-09 $i$ \\
& $\delta u^y$            & -9.86565964237949e-08 & $+$ & 6.11642411231371e-09 $i$ \\
& $\delta B^y$            &  1.63044500663243e-07 & $-$ & 5.5401556994765e-09  $i$ \\
& $\delta \widehat E$     &  2.95346370713069e-09 & $+$ & 1.59680781762658e-09 $i$ \\
& $\delta \widehat F^x$   &  -2.7219589023768e-10 & $+$ & 6.08653555066593e-10 $i$ \\
& $\delta \widehat F^y$   &  3.23862842131416e-12 & $+$ & 2.38270732468968e-11 $i$ \\
& $\omega$                &       1.0003768401216 & $+$ & 0.0268122961453227   $i$ \\
\hline
\parbox[t]{2mm}{\multirow{9}{*}{\rotatebox[origin=c]{90}{slow wave}}}& $\delta\rhorest$        &                 1e-06 & $+$ & 0                    $i$ \\
& $\delta \ug$            &  1.03536412109072e-08 & $+$ & 2.54899319130145e-09 $i$ \\
& $\delta u^x$            &  5.51070747399867e-08 & $+$ & 6.81771795916397e-09 $i$ \\
& $\delta u^y$            &  7.68347744278899e-08 & $+$ & 1.80695857270459e-08 $i$ \\
& $\delta B^y$            & -4.16352105336264e-08 & $-$ & 1.54220614899084e-08 $i$ \\
& $\delta \widehat E$     &   9.0589203896316e-10 & $+$ & 1.85948699094183e-09 $i$ \\
& $\delta \widehat F^x$   & -3.83040910798238e-10 & $+$ & 1.9926795423209e-10  $i$ \\
& $\delta \widehat F^y$   &  2.11405825132413e-12 & $-$ & 6.39415393177625e-12 $i$ \\
& $\omega$                &     0.346247962327932 & $+$ & 0.0428369853095135   $i$ \\

\\[-0.1cm]
\hline
(d) & \multicolumn{4}{|c|}{Optically-thick, radiation-dominated ($\tau = 10$, ${\mathbb P} = 10$)} \\
\hline
\parbox[t]{2mm}{\multirow{9}{*}{\rotatebox[origin=c]{90}{sound wave}}}& $\delta\rhorest$        &                 1e-06 & $+$ & 0                    $i$ \\
& $\delta \ug$            &  1.17069780348943e-08 & $+$ & 1.88153271018629e-09 $i$ \\
& $\delta u^x$            &  2.66250797919881e-07 & $+$ & 6.33514446524509e-08 $i$ \\
& $\delta u^y$            &                     0 & $+$ & 0                    $i$ \\
& $\delta B^y$            &                     0 & $+$ & 0                    $i$ \\
& $\delta \widehat E$     &  2.05419184448571e-07 & $+$ & 1.49858618430197e-07 $i$ \\
& $\delta \widehat F^x$   & -2.07308153187279e-08 & $+$ & 3.77555645793647e-08 $i$ \\
& $\delta \widehat F^y$   &                     0 & $+$ & 0                    $i$ \\
& $\omega$                &      1.67290310151504 & $+$ & 0.39804886622888     $i$ \\
\hline
\parbox[t]{2mm}{\multirow{9}{*}{\rotatebox[origin=c]{90}{fast wave}}}& $\delta\rhorest$        &                 1e-06 & $+$ & 0                    $i$ \\
& $\delta \ug$            &  1.17305399804545e-08 & $+$ & 1.71290104013922e-09 $i$ \\
& $\delta u^x$            &  2.78499110985032e-07 & $+$ & 5.23803931682287e-08 $i$ \\
& $\delta u^y$            & -2.81093153546097e-08 & $+$ & 6.25587500197671e-09 $i$ \\
& $\delta B^y$            &  1.10169648551894e-07 & $-$ & 4.0333708502353e-09  $i$ \\
& $\delta \widehat E$     &  2.07294118408597e-07 & $+$ & 1.36363627261037e-07 $i$ \\
& $\delta \widehat F^x$   & -1.83330848150565e-08 & $+$ & 3.63663817465466e-08 $i$ \\
& $\delta \widehat F^y$   &  2.67581446787218e-10 & $+$ & 1.242717958826e-09   $i$ \\
& $\omega$                &      1.74986152220373 & $+$ & 0.329115716738904    $i$ \\
\hline
\parbox[t]{2mm}{\multirow{9}{*}{\rotatebox[origin=c]{90}{slow wave}}}& $\delta\rhorest$        &                 1e-06 & $+$ & 0                    $i$ \\
& $\delta \ug$            &  9.46188670634028e-09 & $+$ & 1.21375747602521e-09 $i$ \\
& $\delta u^x$            &  8.34268734887456e-08 & $+$ & 1.20828776295457e-08 $i$ \\
& $\delta u^y$            &   1.1363325579508e-07 & $+$ & 2.72697239558014e-07 $i$ \\
& $\delta B^y$            & -8.03822945939874e-08 & $-$ & 3.03114251603687e-07 $i$ \\
& $\delta \widehat E$     &  2.59666249422664e-08 & $+$ & 9.67891091325844e-08 $i$ \\
& $\delta \widehat F^x$   & -1.98262867256072e-08 & $+$ & 5.47609651018276e-09 $i$ \\
& $\delta \widehat F^y$   &   3.6607454416002e-09 & $-$ & 1.14749752402299e-09 $i$ \\
& $\omega$                &     0.524186505728416 & $+$ & 0.0759189591904107   $i$ \\

\end{tabular}
\end{minipage}
\end{table*}

\subsection{HD Radiative Shocks in 1D Cartesian Minkowski}
\label{RADTUBE}

For our next test, we set up a number of radiative shock tube problems
as described in \cite{farrisetal08} and \cite{roedigetal12}.  The
system begins with gas in two different states (left and right),
separated by a membrane. The membrane is removed at $t=0$ and the
system is allowed to evolve. The left- and right-states
of all the tests except test No.~5 are set up in such a way that the
shock asymptotically becomes stationary \citep[see Appendix C
  of][]{farrisetal08}.

Table~\ref{t.radtube} lists the parameters describing the initial
states of seven test problems that we have simulated. The scattering
opacity in all the tests is set to zero, so $\kappa_{\rm
  tot}=\kappa_{\rm abs}$.  The value of the radiative constant
$\sigma_{\rm rad}=a_{\rm rad}c/4$ in code units is given in the
table. All the tests were solved on a grid of $800$ uniformly spaced
points and evolved till $t=300$ for all tests except No. 5 that is run
till $t=13$.  The M1 closure was used, while prior work used the
Eddington approximation, but this only leads to minor differences
right at the shock in the fluid-frame radiative fluxes.

\begin{table*}
\begin{minipage}{2\columnwidth}
\caption{Radiative shock tubes}
\label{t.radtube}
\begin{tabular}{@{}lcccccccccccccc}
\hline
Test & & & & & \multicolumn{4}{c}{Left state:}&  &\multicolumn{4}{c}{Right state:} \\
No. & $\Gamma$ &  $\sigma_{\rm rad}$ & $\kappa_{\rm abs}/\rhorest$ &  &$\rhorest$ & $\pg$ & $u^x$ & $\widehat E$ &  &  $\rhorest$ & $\pg$ & $u^x$ & $\widehat E$  \\
\hline
1 & 5/3 & $3.085\cdot10^9$ &0.4 & & 1.0 & $3.0\times 10^{-5}$ & 0.015 & $1.0\times10^{-8}$ & & 2.4 & $1.61\times10^{-4}$&$6.25\times10^{-3}$&$2.51\times10^{-7}$\\
2 & 5/3 & $1.953\cdot10^4$ &0.2 & & 1.0 & $4.0\times 10^{-3}$ & 0.25  & $2.0\times10^{-5}$ & & 3.11& $4.512\times10^{-2}$&$8.04\times10^{-2}$&$3.46\times10^{-3}$\\
3a& 2   & $3.858\cdot10^{-8}$&0.3&&  1.0 & $6.0\times 10^{1}$ & 10.0  & $2.0   $            & & 8.0& $2.34\times10^{3}$&$1.25$&$1.14\times10^{3}$\\
3b& 2   & $3.858\cdot10^{-8}$&25.0&&  1.0 & $6.0\times 10^{1}$ & 10.0  & $2.0   $            & & 8.0& $2.34\times10^{3}$&$1.25$&$1.14\times10^{3}$\\
4a& 5/3 & $3.470\cdot10^{7}$&0.08&&  1.0 & $6.0\times 10^{-3}$ & 0.69  & $0.18   $            & & 3.65& $3.59\times10^{-2}$&$0.189$&$1.3$\\
4b& 5/3 & $3.470\cdot10^{7}$&0.7&&  1.0 & $6.0\times 10^{-3}$ & 0.69  & $0.18   $            & & 3.65& $3.59\times10^{-2}$&$0.189$&$1.3$\\
5& 2   & $3.858\cdot10^{-8}$&1000.0&&  1.0 & $6.0\times 10^{1}$ & 1.25  & $2.0   $            & & 1.0& $6.0\times 10^{1}$&$1.10$&$2.0$\\
\hline
\end{tabular}
\end{minipage}
\end{table*}

\begin{figure}
  \centering\hspace{-.15in}\vspace{-.15in}
\includegraphics[width=.9\columnwidth,angle=0]{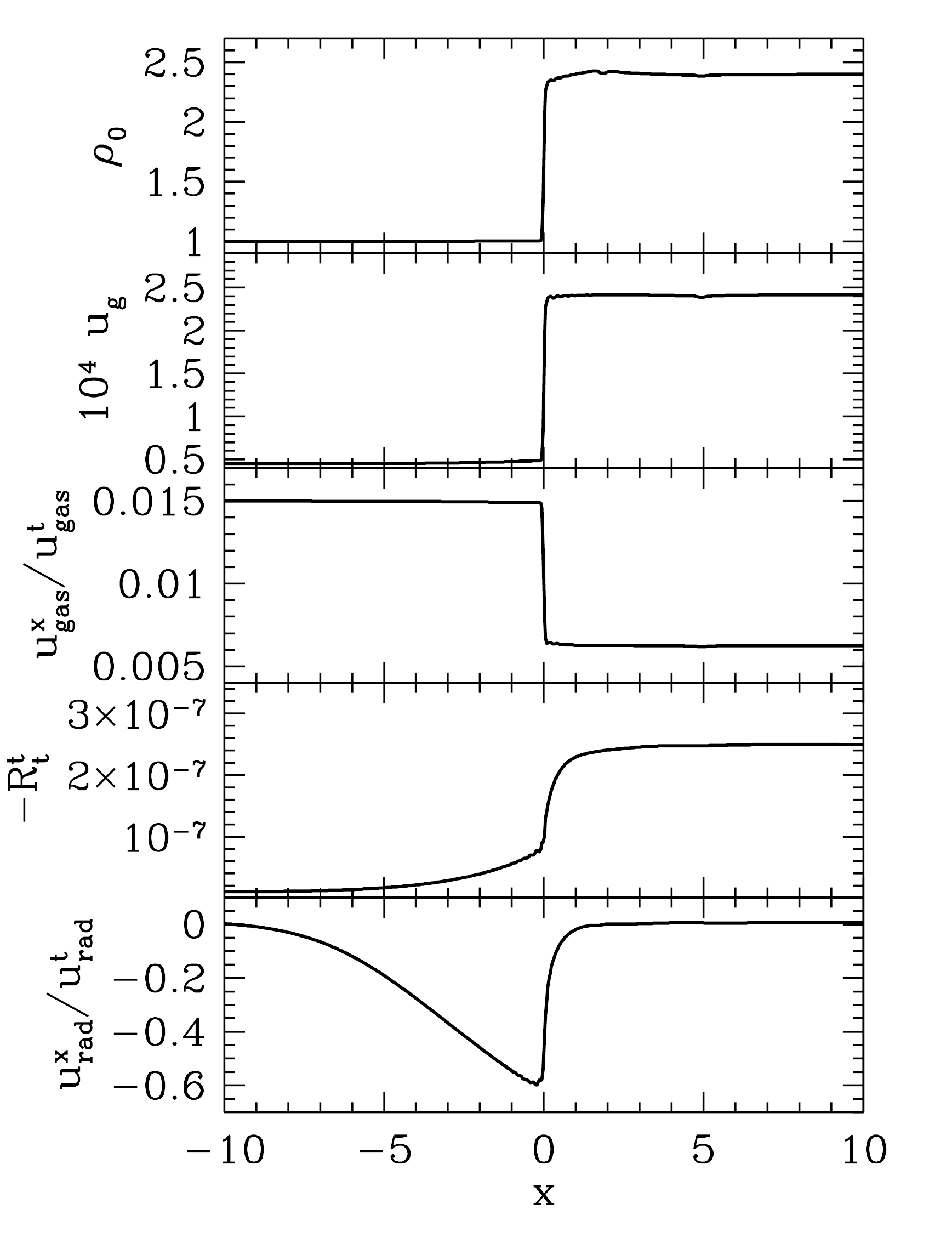}
\caption{Results obtained for radiative shock tube test No.~1. From
  top to bottom, the panels show the profiles rest-mass density
  ($\rhorest$), gas internal energy density ($\ug$), gas 3-velocity
  ($u^x_{\rm gas}/u^t_{\rm gas}$), lab-frame radiative energy density
  ($-R^t_t$), and radiation 3-velocity ($u^x_{\rm rad}/u^t_{\rm
    rad}$).  The profiles match the analytical solution, except for a
  slight bump in density near the shock.}
  \label{f.radtube1}
\end{figure}

Fig.~\ref{f.radtube1} shows the numerical solution for radiative shock
tube problem No.~1, which corresponds to a non-relativistic strong
shock. This plot can be compared to the corresponding figures and
analytical solutions provided elsewhere
\citep{farrisetal08,zanottietal11, fragileetal12}. The agreement is
good, except for a slight smoothing of the numerical profiles at the
position of the shock (see the bottom panel).  Also, the shock shows
some oscillations that lead to a mild bump to the right of the shock
in rest-mass density by the end of the simulation.  Lower-order MINM
and even DONOR cell reconstruction, any other Runge-Kutta method, and
both LAXF and HLL lead to the same/similar bump, so PPM or RK3 or
LAXF/HLL are not the origin of the bump.

%oooooooooooooooooooooooooooooooooooooooooooooooooooooooooooooo
%oooooooooooooooooooooooooooooooooooooooooooooooooooooooooooooo

\begin{figure}
  \centering\hspace{-.15in}\vspace{-.15in}
\includegraphics[width=.9\columnwidth,angle=0]{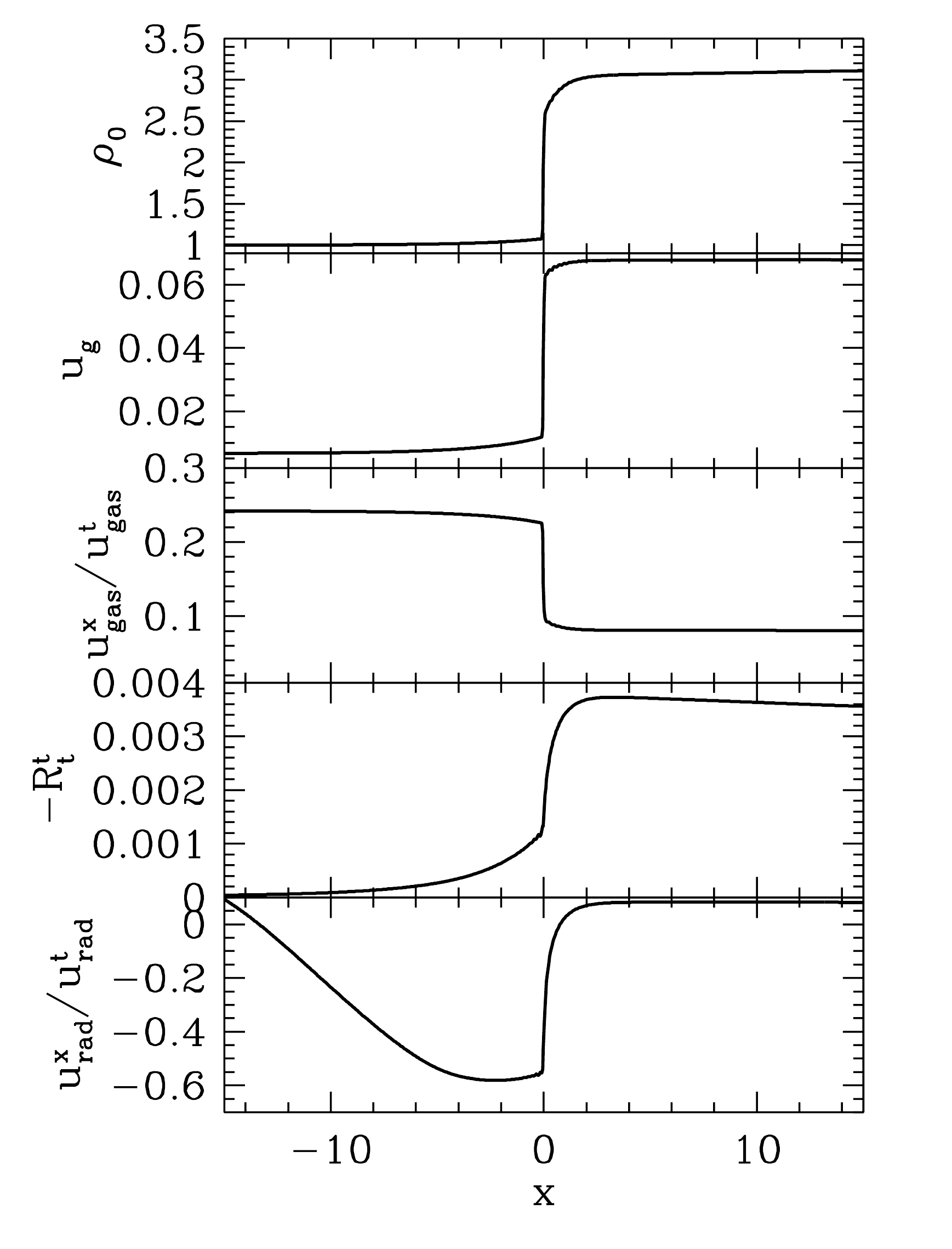}
\caption{Same as Fig.~\ref{f.radtube1} but for radiative shock tube
  test No.~2.  The profiles match the analytical solution.}
  \label{f.radtube2}
\end{figure}

Fig.~\ref{f.radtube2} shows results for radiative shock tube test
No.~2, which corresponds to a mildly relativistic strong shock. Again,
the agreement between the numerical and semi-analytical
\citep{farrisetal08} profiles is good.

%oooooooooooooooooooooooooooooooooooooooooooooooooooooooooooooo
%oooooooooooooooooooooooooooooooooooooooooooooooooooooooooooooo

\begin{figure}
  \centering\hspace{-.15in}\vspace{-.15in}
\includegraphics[width=.9\columnwidth,angle=0]{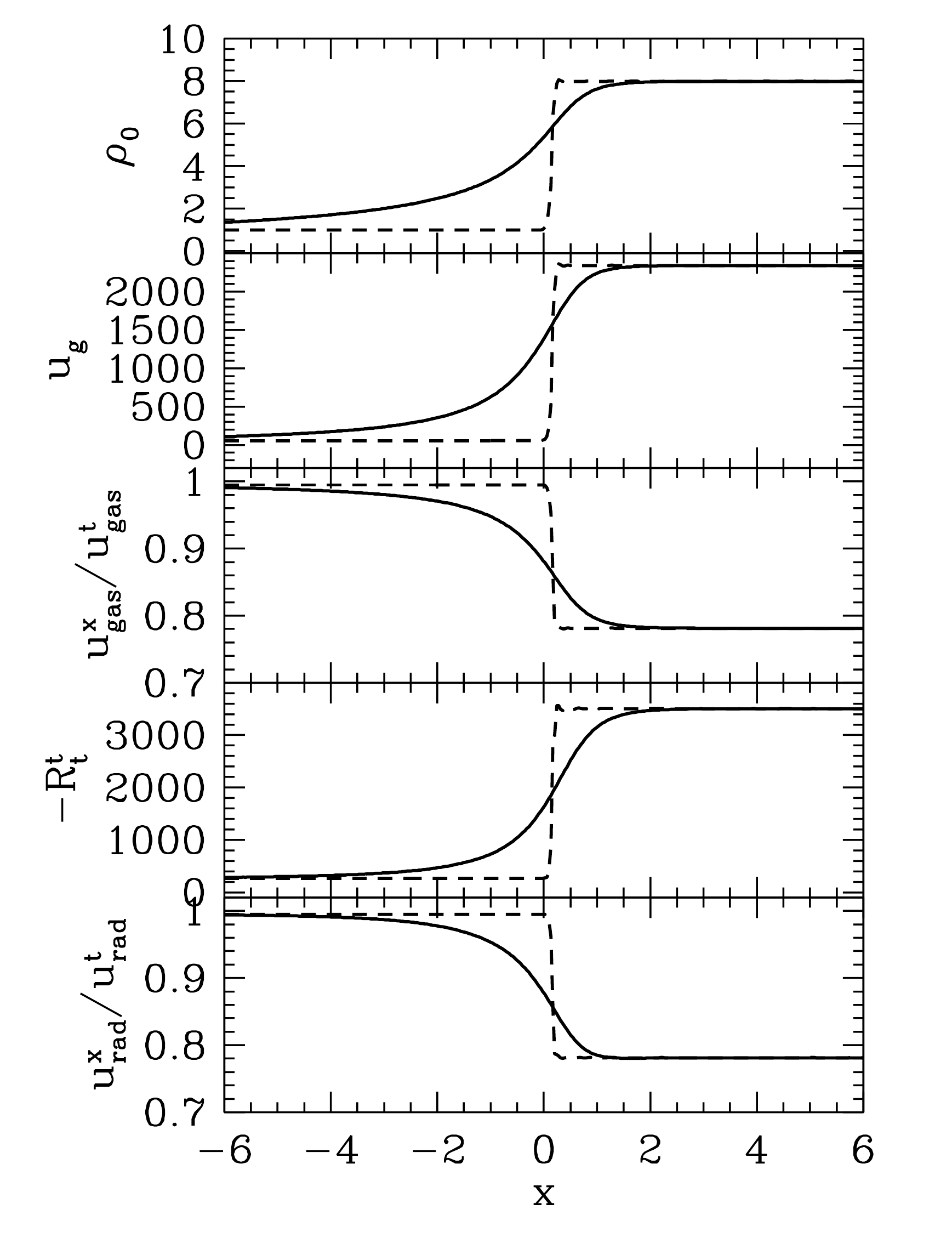}
\caption{Similar to Fig.~\ref{f.radtube1} but showing results for
  radiative shock tube tests No. 3a (solid) and No. 3b (dashed).  The
  profiles match the analytical solution.}
  \label{f.radtube3}
\end{figure}

Fig.~\ref{f.radtube3} shows results corresponding to radiative shock
tube tests No.~3a and 3b. These are strongly relativistic shocks with
upstream $u^x=10$. Test No.~3a corresponds to shock tube test 3 of
\cite{farrisetal08}, while test 3b is the optically thick version of
the same test which was proposed and solved by
\cite{roedigetal12}. These two tests verify that the code is able to
resolve a highly relativistic wave in two very different optical depth
limits. In both cases, the numerical solution reaches a steady state
and closely follows the corresponding semi-analytical solution as
presented in \citep{farrisetal08,2013MNRAS.429.3533S}.  The acceptable
and normal amount of mild oscillations near the shock appear because
we use the high-order PPM reconstruction and the flattener is only
moderately efficient at reducing the order of spatial interpolation
near shocks.

%oooooooooooooooooooooooooooooooooooooooooooooooooooooooooooooo
%oooooooooooooooooooooooooooooooooooooooooooooooooooooooooooooo

\begin{figure}
  \centering\hspace{-.15in}\vspace{-.15in}
\includegraphics[width=.9\columnwidth,angle=0]{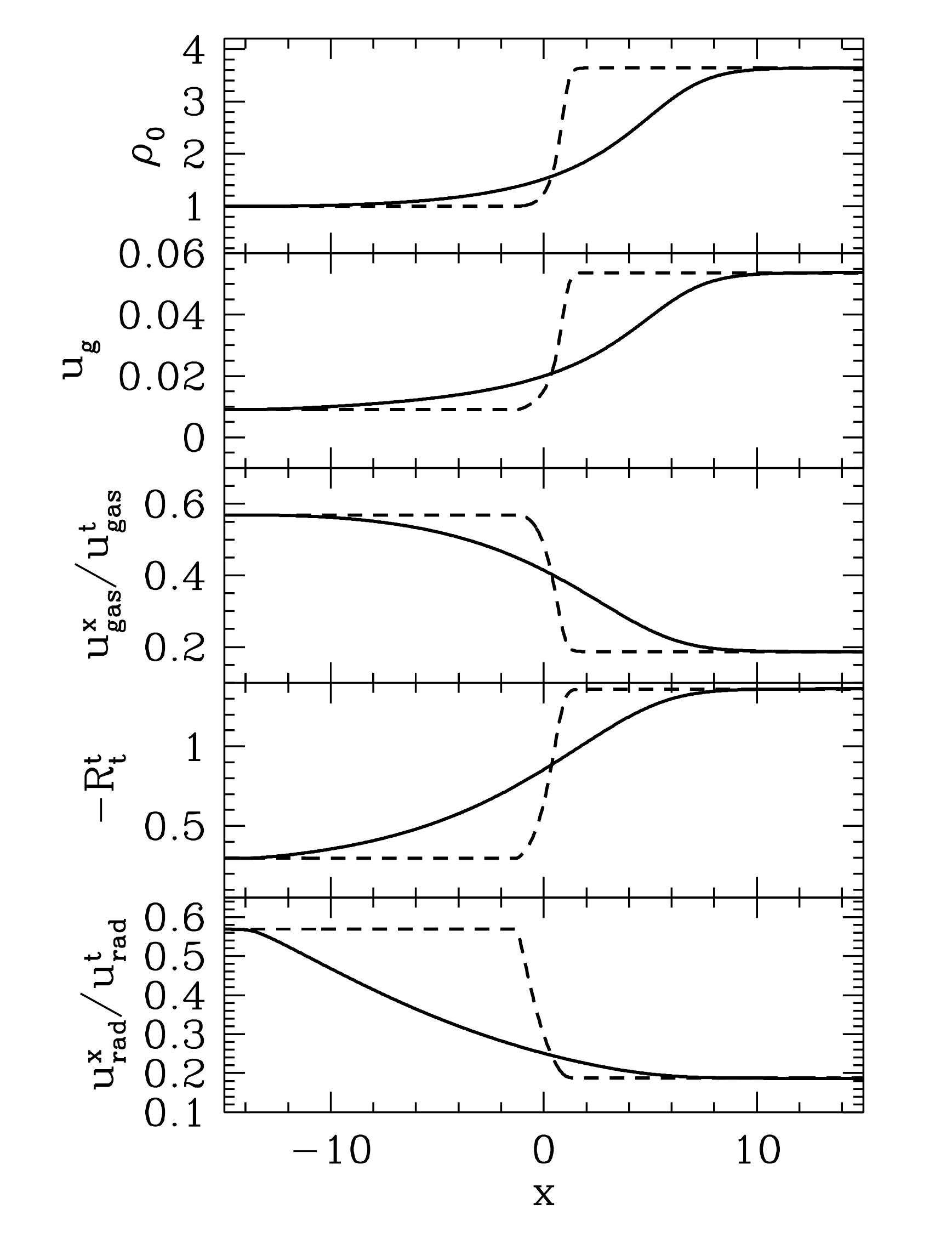}
\caption{Similar to Fig.~\ref{f.radtube1} but showing results for
  radiative shock tube tests No. 4a (solid) and No. 4b (dashed).  The
  profiles match the analytical solution.}
  \label{f.radtube4}
\end{figure}

Fig.~\ref{f.radtube4} shows results for radiative shock tube tests
No.~4a and 4b. These tests correspond to radiation pressure dominated
mildly relativistic waves. Test~4b is the optically thick version of
test~4a that was proposed by \cite{roedigetal12}. In both tests, the
numerical solution reaches a stationary state and agrees well with the
semi-analytical solution.  The opacity coefficient $\kappa_{\rm abs}$
in tests 3b and 4b are the maximum values that the scheme by
\citet{roedigetal12} could handle while remaining stable. The
algorithm implemented in \harmrad\ has no such limitation.

%oooooooooooooooooooooooooooooooooooooooooooooooooooooooooooooo
%oooooooooooooooooooooooooooooooooooooooooooooooooooooooooooooo

\begin{figure}
  \centering\hspace{-.15in}\vspace{-.15in}
\includegraphics[width=.9\columnwidth,angle=0]{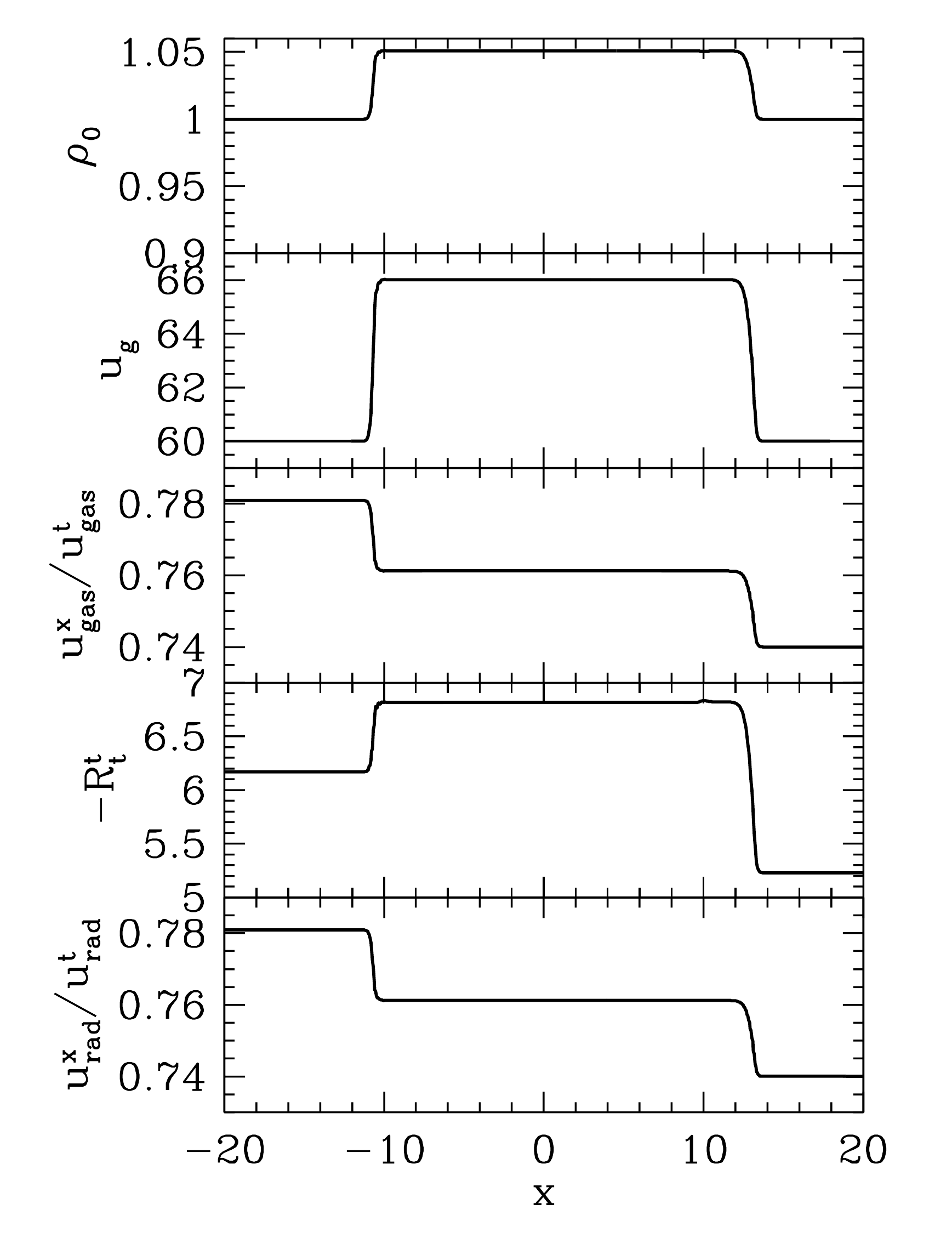}
\caption{Same as Fig.~\ref{f.radtube1} but for radiative shock tube
  test No.~5. There is no analytical solution available for this problem,
but it agrees with prior work.}
  \label{f.radtube5}
\end{figure}

Fig.~\ref{f.radtube5} corresponds to radiative shock tube test
No.~5.  This is the only test that does not asymptote to a stationary
solution. This test was proposed and solved by \citet{roedigetal12}
and represents an optically thick flow with mildly relativistic
velocities. The left- and right-states are identical except that they
have different velocities. As a result, two shock waves propagate in
opposite directions. This test does not have an analytical
solution. However, by comparing our numerical solution with that
presented in \cite{roedigetal12}, we confirm that our scheme
performs well.

%oooooooooooooooooooooooooooooooooooooooooooooooooooooooooooooo
%oooooooooooooooooooooooooooooooooooooooooooooooooooooooooooooo

\subsection{Optically Thin Radiative Pulse in 3D Cartesian Minkowski}
\label{RADPULSE3D}

We now test the ability of our scheme to handle the evolution of a
radiation pulse in the optically thin limit.  We set up a Gaussian
distribution of radiative energy density at the center of a 3D
Cartesian coordinate system. The pulse radiative temperature is set
according to, \be T_{\rm
  rad}=\left(\frac{E}{4\sigma}\right)^{1/4}=T_a\left(1+100
e^{-(x^2+y^2+z^2)/w^2}\right), \ee with $T_a=10^6$, $w=5.0$.  The
value of $a_{\rm rad}\approx 8.77\times 10^{-12}$. We assume zero
absorption opacity ($\kappa_{\rm abs}=0$) and scattering opacity ($\kappa_{\rm
  es}=0$). The background fluid field has constant density $\rhorest=1$
and temperature $T=T_a$. We solve the problem in three dimensions on a
coarse Cartesian grid of $32x32x32$ and $50x50x50$ cells (showing only
the $50x50x50$ result).

\begin{figure*}
  \centering\hspace{-.15in}\vspace{-.15in}
   \subfigure{\includegraphics[height=.343\textwidth,angle=0]{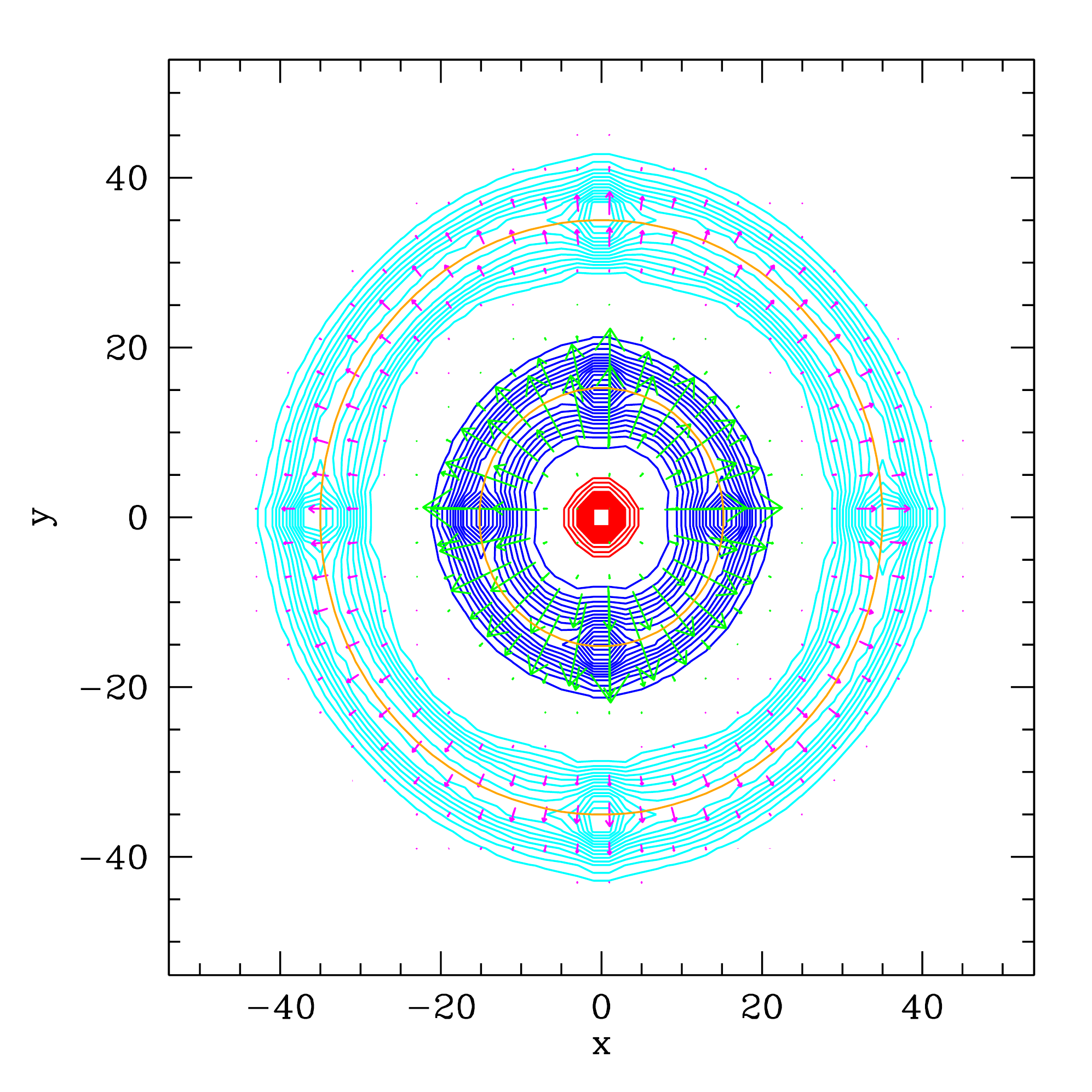}}
   \subfigure{\includegraphics[height=.343\textwidth,angle=0]{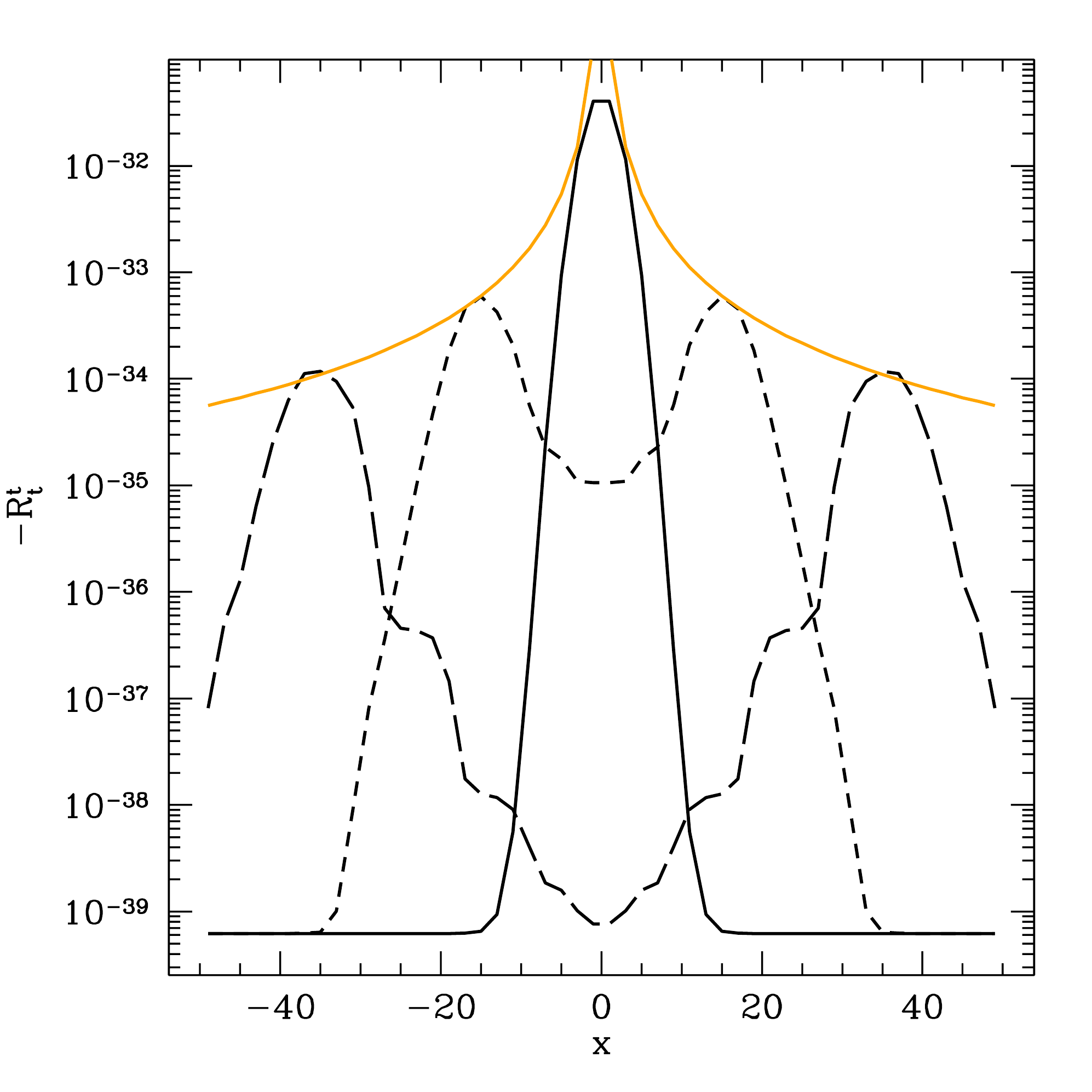}}\\
   \caption{Profiles of the lab-frame radiative energy density
     ($-R^t_t$) for the optically thin radiative pulse test described
     in section~\ref{RADPULSE3D}. The left panel shows the
     distribution in the $x-y$ plane at $z=0$ at times $t=0$ (red,
     $15$ contours from $6.24\times 10^{-40}$ to $4.038\times
     10^{-32}$), $t=15.21$ (blue, $15$ contours from $6.24\times
     10^{-40}$ to $5.963\times 10^{-34}$), and $t=35.01$ (cyan, $15$
     contours from $6.24\times 10^{-40}$ to $1.174\times 10^{-34}$)
     with lab-frame energy flux ($-R^i_t$) shown for $t=15.21$ (green
     vectors) and $t=35.01$ (magenta vectors).  The orange circles
     correspond to the initial pulse spreading at the speed of light
     from $x=y=z=0$, and the centroid of the pulse distribution
     matches well with this position for both evolved times.  The
     right panel shows the same times at $t=0$ (solid), $t=15.21$
     (dashed), and $t=35.01$ (long-dashed) in the $y=z=0$ line.  The
     pulse should decay as $1/x^2$, which is shown as an orange line,
     and there is a reasonable match.}
   \label{f.pulsethin}
 \end{figure*}

The initial pulse in radiative energy density is expected to spread
isotropically with the speed of light (optically thin medium) and to
decrease inversely proportionally to the square of radius (energy
conservation). Such behavior is visible in Fig.~\ref{f.pulsethin}
showing the radiative energy distribution in the $z=0$ plane (left
panel) and its cross-section along $y=z=0$ (right panel). The orange
circles in the left panel show the expected size of the pulse. It is
clear that the propagation speed of the pulse is consistent. This
problem was solved on a relatively coarse Cartesian grid, and this
results in deviations from the perfectly spherical shape.  Also, the
PPM scheme uses a stencil size of $\pm 4$ cells, so one should have an
initial radiative distribution mostly within $\gg 8$ cells to avoid
grid-induced artifacts.  The solution becomes more isotropic at larger
resolution or when using a spherical grid. The right panel in
Fig.~\ref{f.pulsethin} shows the profiles of the energy density along
the $x$-axis, which follows the expected rate of energy decrease with
increasing distance from the center.

 %oooooooooooooooooooooooooooooooooooooooooooooooooooooooooooooo
% %oooooooooooooooooooooooooooooooooooooooooooooooooooooooooooooo

\subsection{Optically Thick Radiative Pulse in 1D Cartesian Minkowski}
\label{RADPULSEPLANAR}

To test the optically thick limit we choose to set up a similar pulse
but this time planar instead of a point-like, i.e., according to, \be
T_{\rm rad}=\left(\frac{E}{4\sigma}\right)^{1/4}=T_a\left(1+100
e^{-x^2/w^2}\right).  \ee This time we set the scattering opacity to
$\kappa_{\rm es}=10^{3}$ and solve the problem as one-dimensional on
$100$ grid points distributed uniformly between $x=-50$ and $x=50$
with periodic boundary conditions in $y$ and $z$. The total optical
depths per cell and per pulse are therefore $\tau=10^3$ and
$\tau=10^4$, respectively.  The value of $a_{\rm rad}\approx
8.77\times 10^{-12}$ in code units and the absorption opacity is zero.

In the optically thick limit the evolution of such a system is
described by a diffusion equation with diffusion coefficient
$D=1/(3\kappa_{\rm tot})$.  An initially Gaussian pulse of radiative
internal energy will diffusive as
\begin{equation}\label{diffsol}
-R^t_t(t) = A \exp\left(\frac{-x^2}{4D(t+t_0)}\right) \left(\frac{t+t_0}{t_0}\right)^{-n/2} ,
\end{equation}
for $n=1$ dimensions, $t_0\approx 4800$ and $A\approx 5.49\times 10^{-32}$.

\begin{figure}
  \centering\hspace{-.15in}\vspace{-.15in}
\includegraphics[width=.95\columnwidth,angle=0]{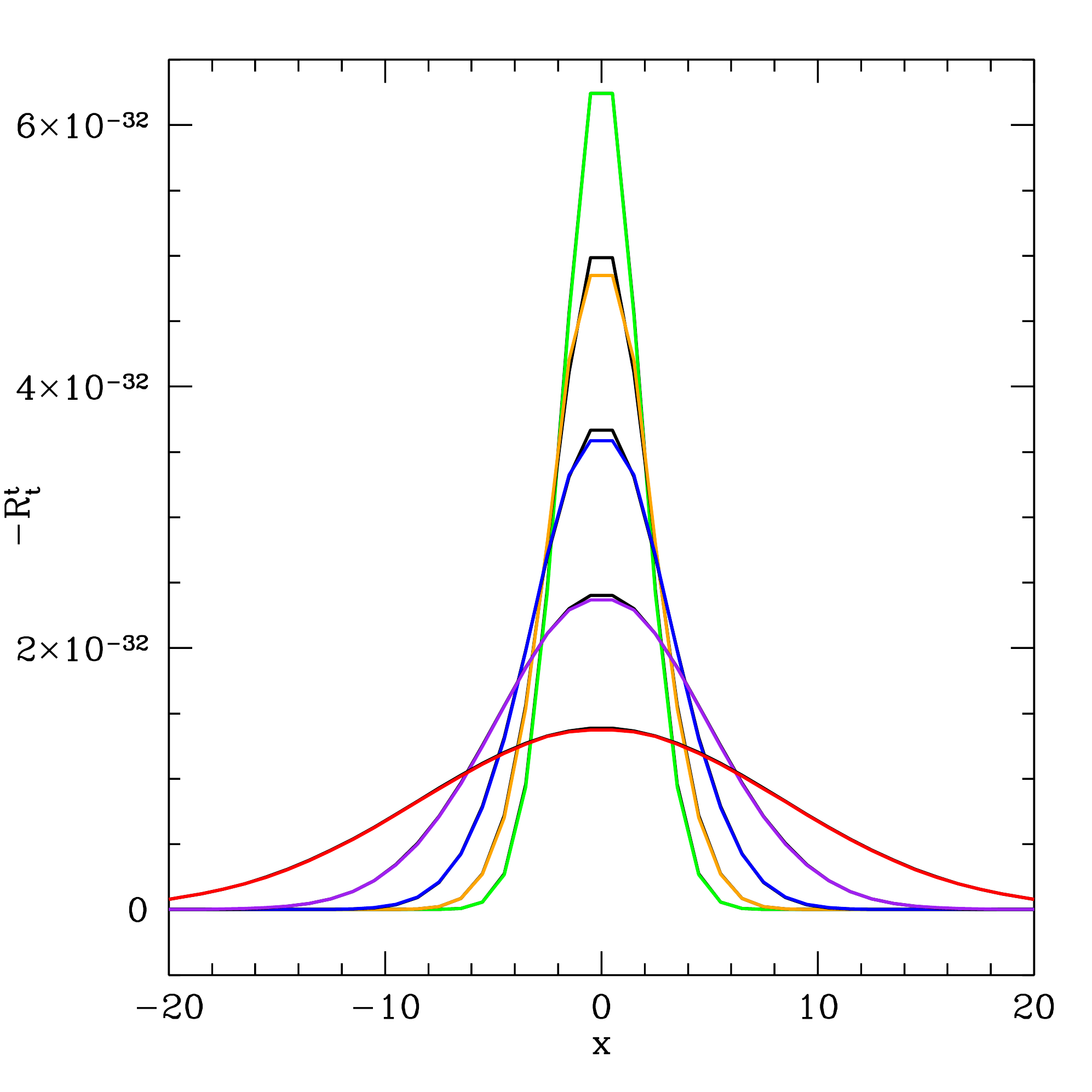}
\caption{The radiative energy density for the optically thick pulse
  described in section~\ref{RADPULSEPLANAR}. The colored lines show
  times $t=0$ (green), $t=2951.05$ (orange), $t=9867.95$ (blue),
  $t=29877.1$ (purple), and $t=10^5$ (red).  Behind each solution is a
  black line for the analytical solution from Eq.~\ref{diffsol} (which
  overlaps at $t=0$).  The analytical and numerical solutions agree
  very well in this diffusion regime at high optical depth.}
  \label{f.pulsethick}
\end{figure}

In Fig.~\ref{f.pulsethick} we plot profiles of the radiative energy at
various moments and compare them to the analytical solution given by
Eq.~\ref{diffsol}. The numerical solution for the central two points
diffuses slightly faster due to the additional numerical dissipation
introduced by the scheme.  At later times this difference becomes
insignificant.

A code's speed can be sensitive to this high optical depth case,
depending upon the way the initial guess is chosen in the implicit
solver as well as how the wavespeeds are determined.  \harmrad\ only
takes $200$ steps with $2.6$ average energy-based iterations using the
URAD scheme without stages, and despite the large timesteps, the
Newton method does not need to reject any implicit Newton steps as
could happen for large steps when conserved quantities step out of
bounds without a physical primitive inversion.

%oooooooooooooooooooooooooooooooooooooooooooooooooooooooooooooo
%oooooooooooooooooooooooooooooooooooooooooooooooooooooooooooooo

\subsection{Single Beam of Light in 2D Cartesian Minkowski}
\label{RADBEAMFLAT}

\begin{figure*}
  \centering\hspace{-.15in}\vspace{-.15in}
\subfigure{\includegraphics[width=.35\textwidth,angle=0]{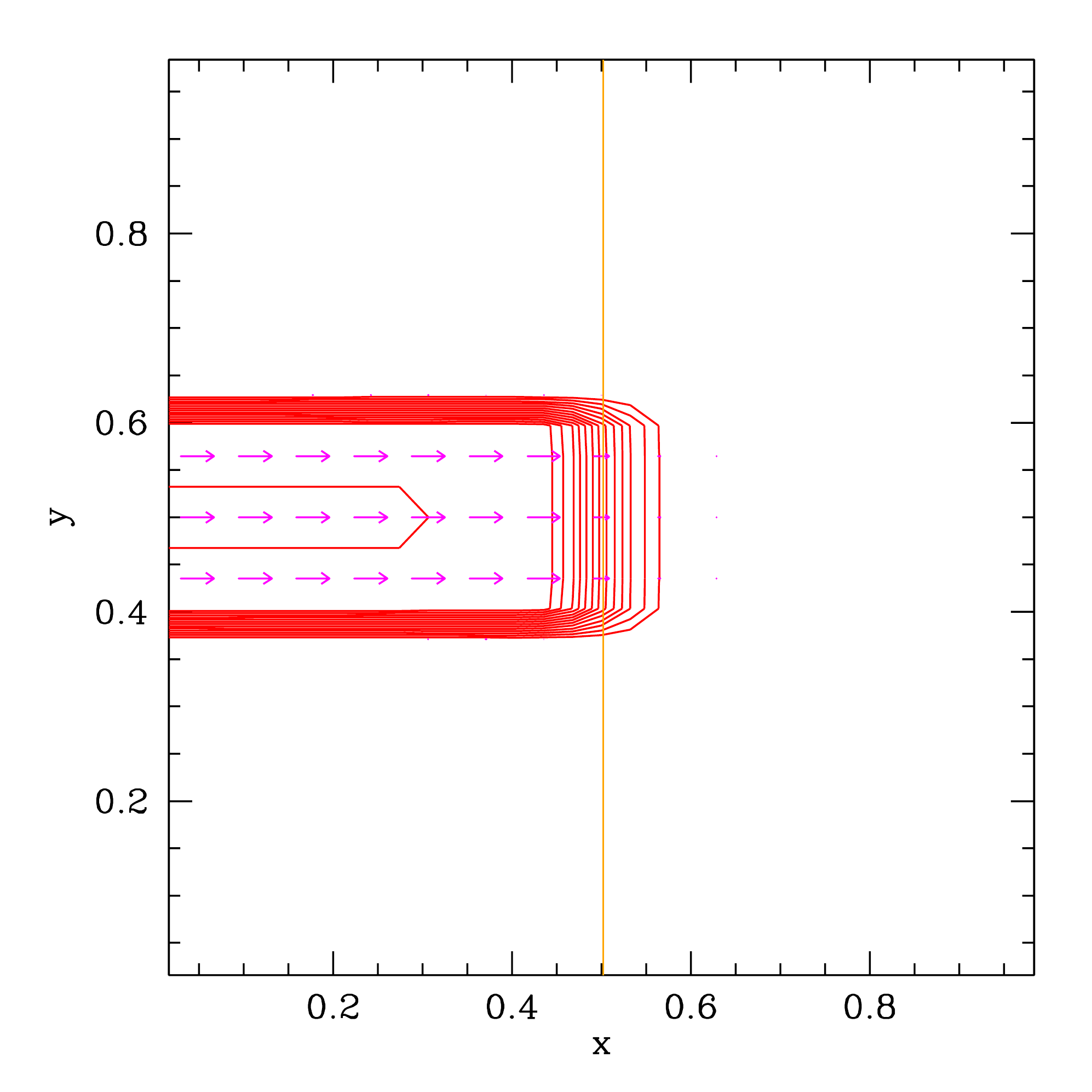}}
\subfigure{\includegraphics[width=.35\textwidth,angle=0]{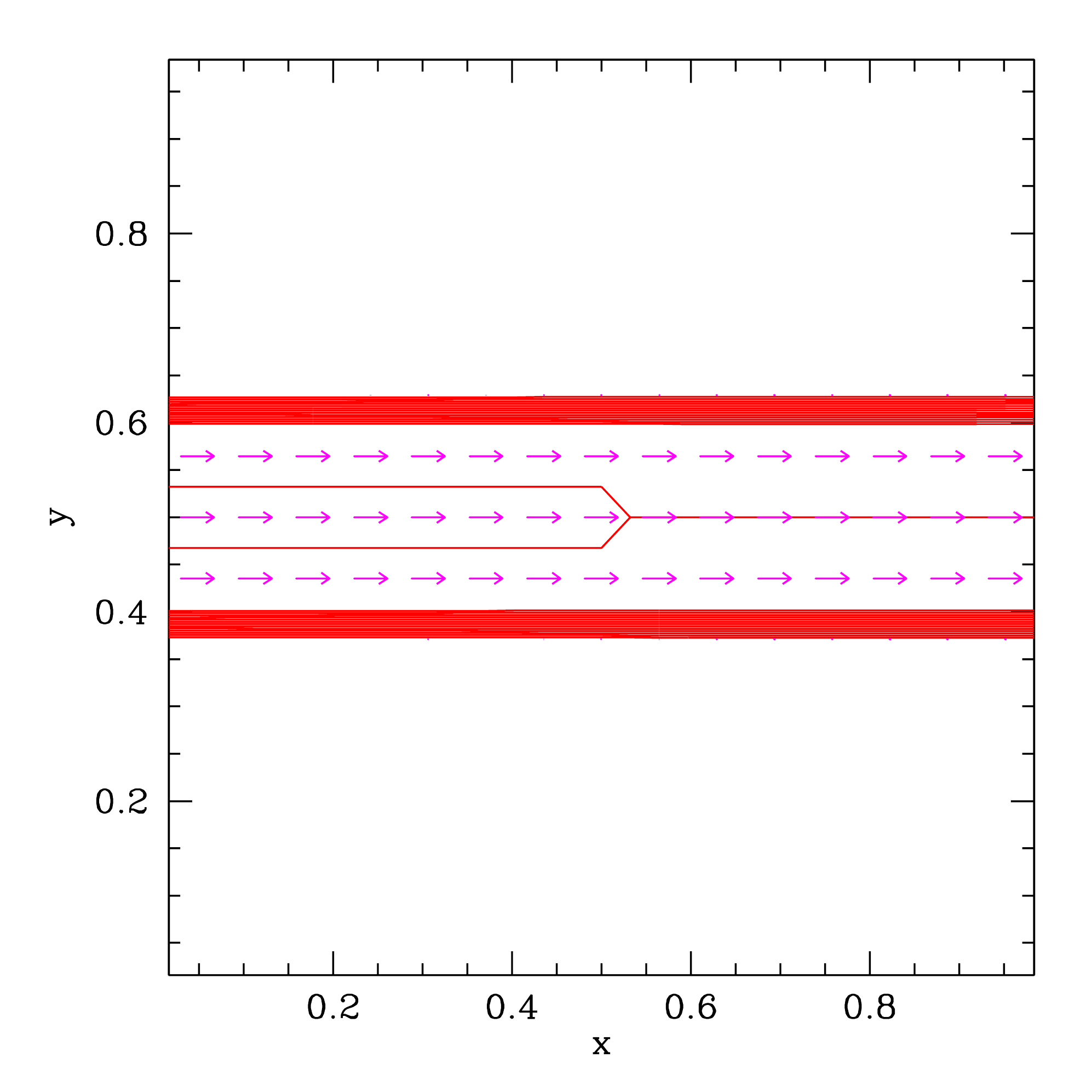}}
\caption{Results for single beam of light in Minkowski for beam
  injected with $\gamma_{\rm rad}\approx 500$ in
  section~\ref{RADBEAMFLAT}.  Left panel shows $t=0.5023$ with $15$
  red contours for $-R^t_t$ from $0.3666$ to $3.333\times 10^7$ and
  $R^i_t$ as magenta vectors scaled by $5\times 10^{-8}$ shown every
  other grid point.  Orange vertical line indicating the light travel
  distance since injection.  Right panel shows $t=10$ with $15$ red
  contours for $-R^t_t$ from $0.494$ to $3.333\times 10^7$ and $R^i_t$
  as magenta vectors scaled by $5\times 10^{-8}$ shown every other
  grid point. The beam travels out as expected.}
  \label{f.radbeamflat}
\end{figure*}

Fig.~\ref{f.radbeamflat} show the results for an injected single beam
of light with a top-hat distribution.  The gas and radiation are
decoupled by neglecting absorptions and scatterings ($\kappa_{\rm
  abs}=\kappa_{\rm tot}=0$). The grid is two-dimensional in the $x-y$
plane with $31$ points distributed uniformly from $0$ to $1$ in each
dimension.  All initial values for primitive quantities are negligibly
small.  The ideal gas constant is set to $\gamma=4/3$.  Outflow
boundary conditions are used on all borders, except the region covered
by the beam from $y=0.4$ to $y=0.6$, where we set the injected
lab-frame radiation energy density to be $100$ times the ambient
value, and we set the radiative 3-velocity to be to be $v^x_{\rm rad}
= 0.99998$ or $\gamma_{\rm rad} \approx 500$.  The value of $a_{\rm
  rad}\approx 1.18\times 10^{17}$, and the maximum radiative gamma
allowed is $\gamma_{\rm rad}=1000$. Note that for $\gamma_{\rm
  rad}\gtrsim 2000$, the beam edge can appear mildly unstable with
PPM, but there is no significant disruption and the beam moves at the
correct speed.  Higher $\gamma_{\rm rad}$ are achievable with MINM
limiter at the cost of resolution/accuracy in more general
simulations.

%oooooooooooooooooooooooooooooooooooooooooooooooooooooooooooooo
%oooooooooooooooooooooooooooooooooooooooooooooooooooooooooooooo

\subsection{Single Shadow in 2D Cartesian Minkowski}
\label{RADSHADOW}

Here we test the ability of the M1 closure scheme, as incorporated in
\harmrad, to resolve shadows. We set up a blob of dense, optically
thick gas in flat space-time, surrounded by an optically thin medium,
and we illuminate this system.

We start with a single source of light imposed on the left boundary. We
solve the problem in two dimensions on a $100\times 50$ grid, with the density
blob distribution set to be
\be
\rhorest = \rho_a+(\rho_b-\rho_a)\,e^{-\sqrt{x^2+y^2}/w^2},
\ee
where $\rho_a=10^{-4}$, $\rho_b=10^3$ and $w=0.22$. The gas
temperature is adjusted so as to give constant pressure throughout
the domain,
\be
T = T_a\frac{\rho_a}{\rhorest} ,
\ee
with $\gamma=1.4$.

The initial radiative energy density is set to the local thermal
equilibrium value, and the initial velocities and radiative fluxes are
zero. We apply periodic boundary conditions at the top and bottom and
outflow boundary conditions at the right border of the domain. At the
left border we have the external source of light, which we specify
with $E_L=4\sigma T_L^4$, $F^x=0.99999E_L$, $T_L=100T_a$. All other
quantities are set to match the ambient gas. We assume
$\kappa_{\rm abs}=\kappa_{\rm tot}=\rhorest$.  The value of $a_{\rm rad}\approx 351.37$ in
code units, and the maximum radiative gamma allowed is $\gamma_{\rm
  rad}=447.215$.

\begin{figure*}
  \centering\hspace{-.15in}\vspace{-.15in}
\subfigure{\includegraphics[width=.35\textwidth,angle=0]{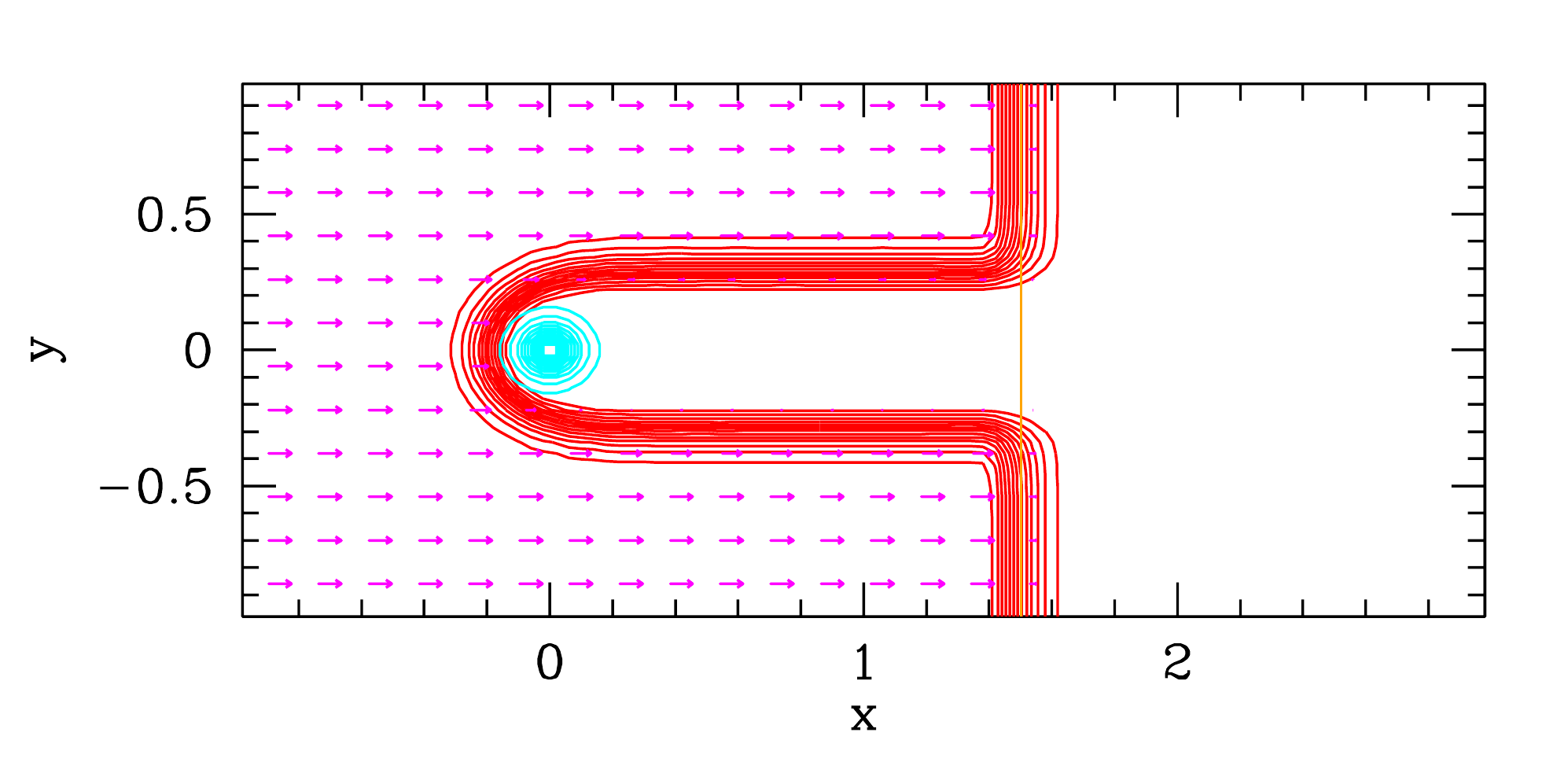}}
\subfigure{\includegraphics[width=.35\textwidth,angle=0]{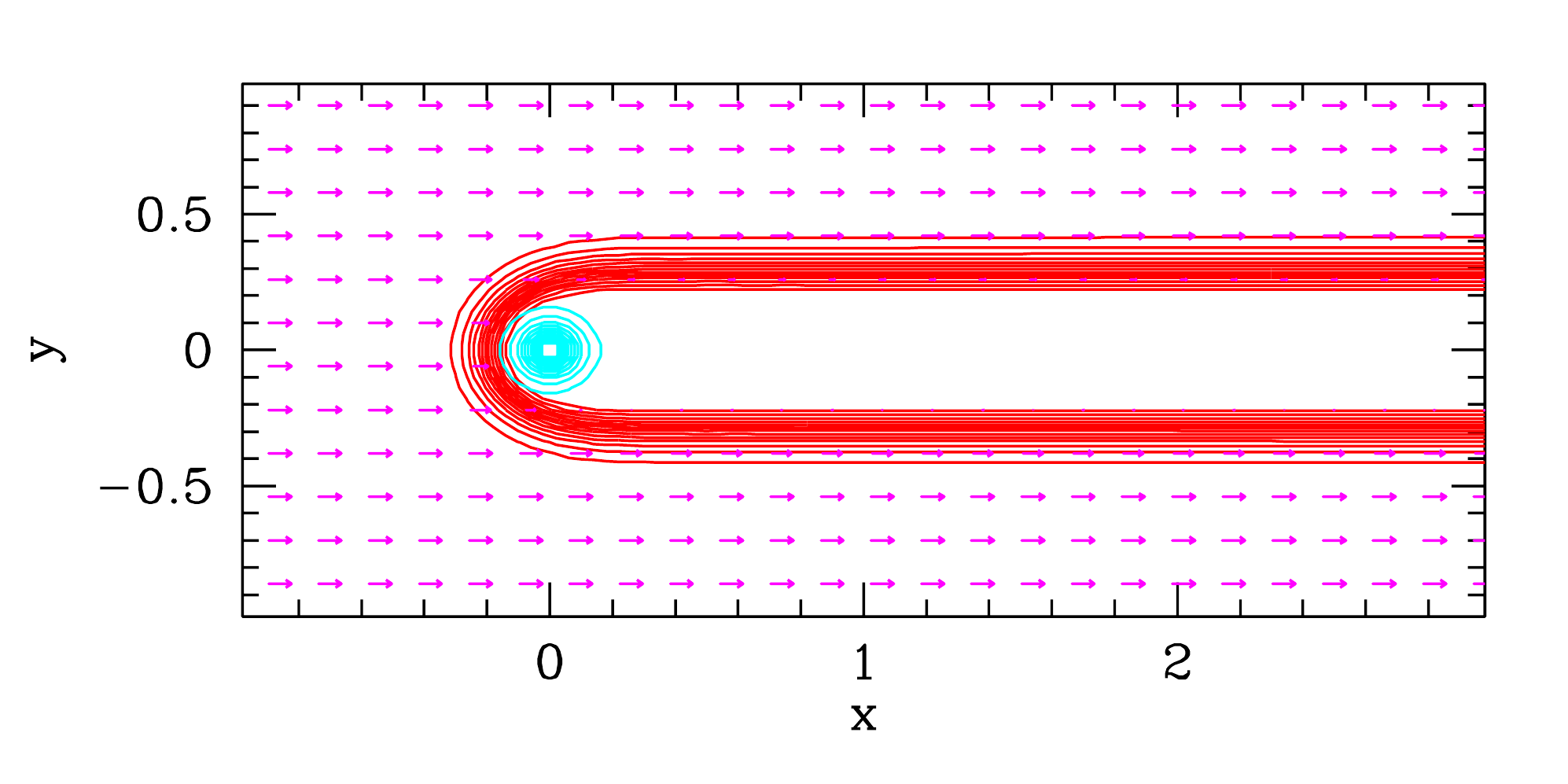}}
\caption{Results for shadow in Minkowski for beam injected with
  $\gamma_{\rm rad}\approx 158$ in section~\ref{RADSHADOW}.  Left
  panel shows $t=2.502$ with $15$ red contours for $-R^t_t$ from
  $9\times 10^{-25}$ to $2.5\times 10^{-14}$ and $R^i_t$ as magenta
  vectors scaled by $2\times 10^{12}$ shown every 4th grid point.
  Orange vertical line indicating the light travel distance since
  injection.  Right panel shows $t=10$ with $15$ red contours for
  $-R^t_t$ from $2.085\times 10^{-26}$ to $2.5\times 10^{-14}$ and
  $R^i_t$ as magenta vectors scaled by $2\times 10^{12}$ shown every
  4th grid point.  In both panels the massive blob is shown as
  $\rhorest$ with $15$ cyan contours from $0.0001$ to $557.4$.  The
  beam travels out as expected with no beam instability or disruption.
  As required, the shadow is sharply defined with negligible radiative
  flux behind the blob of mass.}
  \label{f.shadow}
\end{figure*}

Fig.~\ref{f.shadow} shows the results at $t=10$. By this time, the
initial radiation wave has passed through the domain and the system
has reached a stationary state.  The M1 closure is designed to keep
flux moving parallel to itself in optically thin regions for $F\approx
E$. As a result, a strong shadow develops behind the optically thick
blob.

Note that the beam is stable up to a choice of $\gamma_{\rm
  rad}\gtrsim 2000$ for the beam injected.  However, the radiation
behind the blob has a few lack of implicit inversion solutions beyond
$\gamma_{\rm rad}\sim 500$, although this only affects the radiative
velocity where the radiative energy density is negligible.

%oooooooooooooooooooooooooooooooooooooooooooooooooooooooooooooo
%oooooooooooooooooooooooooooooooooooooooooooooooooooooooooooooo

\subsection{Double Shadow in 2D Cartesian Minkowski}
\label{RADDBLSHADOW}

We also consider a two-beam test problem similar to the one described
in \citet{jiangetal12}. We set up similar initial conditions for gas
and radiation as in the single beam shadow test.  This time, however,
we set up a reflection symmetry at the lower boundary ($y=0$) and we
impose an inclined (lab-frame $F^x_0=0.93E_0$, $F^y_0=-0.37E_0$ with
$\gamma_{\rm rad}$ limited as stated) beam on the upper boundary and
on the part ($y>0.3$) of the left boundary.  As a result, the domain
is effectively lit by two self-crossing beams of light.

\begin{figure*}
  \centering\hspace{-.15in}\vspace{-.15in}
\includegraphics[width=0.6\textwidth,angle=0]{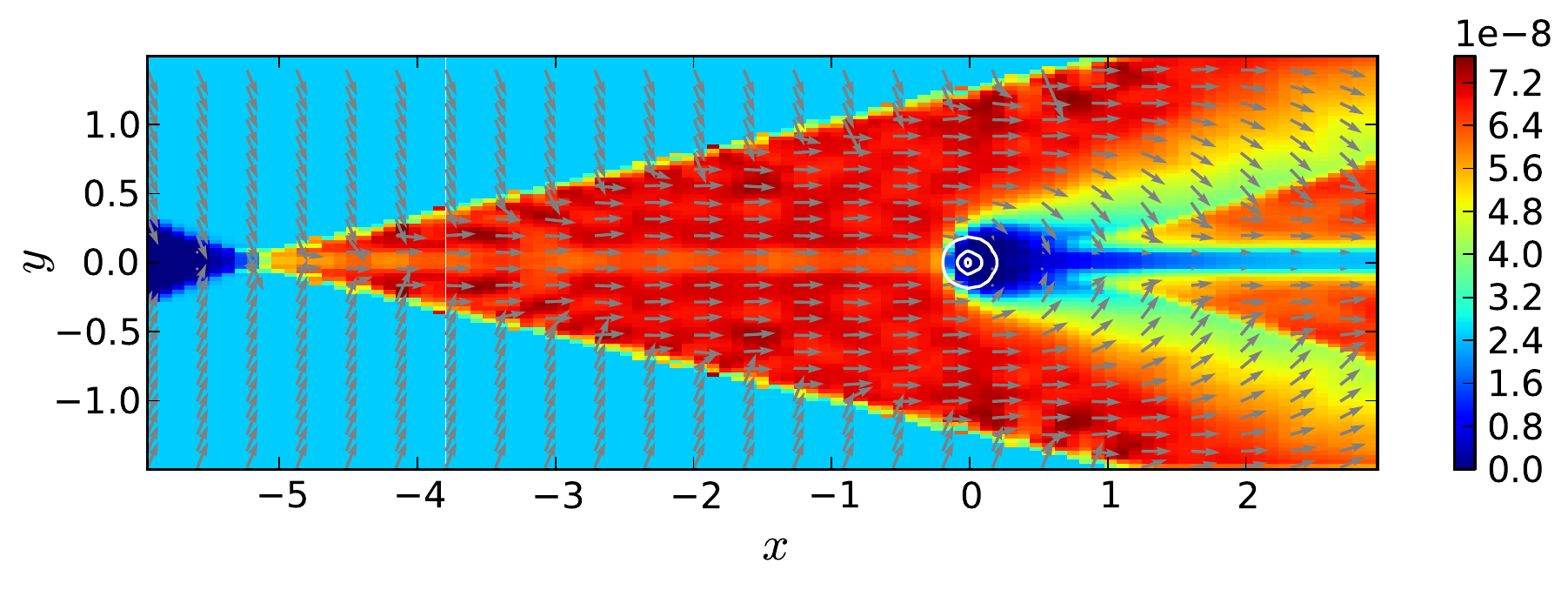}
\caption{Results for double shadow test in Minkowski for beam injected
  with $\gamma_{\rm rad}\approx 22$ at an angle, which then reflects
  and interacts with the original beam before casting a shadow due to
  the blob in section~\ref{RADDBLSHADOW}.  Shows lab-frame $-R^t_t$ at
  $t=20$ as color (with legend), and lab-frame $-R^i_t$ as vectors.
  White contours show three logarithmic rest-mass contours from
  $\rhorest=100$ to $\rhorest=650$.  The radiative intersection is
  sharply defined, but PPM with HLL (as shown) generates slight
  oscillations that drives weak waves into the beam.}
  \label{f.shadow2}
\end{figure*}

We plot the result of a numerical simulation in
Fig.~\ref{f.shadow2}. In the region near the left top, where the beams
do not overlap, the direction of the flux follows the imposed boundary
condition. In the region of the overlap the radiative energy density
increases twice ($E=2E_0$) while the flux becomes equivalent to the
superposition of the beam-intrinsic fluxes, i.e., it is purely
horizontal and its $x$-component equals
$F^x=2F^x_0=1.86E_0=0.93E$. The clump of optically thick gas is,
therefore, effectively illuminated by a purely horizontal beam.
Unlike in the planar beam case, there are regions of the partial
shadow (penumbra) resulting from these perpendicular photons allowed
by the closure when $F^x<E$. The region of the total shadow (umbra) is
therefore limited by the edges of the penumbra and follows the
expected shape (compare Fig.~11 in \citet{jiangetal12}) to a good
accuracy.  The M1 closure, however, produces an extra narrow
horizontal shadow along the $x$-axis that should not be present.

For Fig.~\ref{f.shadow2} we set the beam's $\gamma_{\rm rad}\approx
22$, because with PPM any higher values lead to mild oscillations
driven away from the stationary inclined radiative edge into the rest
of the beam.  Even at $\gamma_{\rm rad}\approx 500$ these oscillations
are only at the $\le 20\%$ level and are proportional to the beam's
$(\gamma_{\rm rad})^{1/2}$, but it is visually obvious in such plots.
We plan to continue to improve PPM and HLL/LAXF (designed for fluid
shocks, not radiative jumps) to work better at radiative
discontinuities that are stationary and not aligned with the grid.  No
such oscillations appear with MINM that is significantly more
diffusive for general applications.  Much weaker oscillations $<2\%$
for $\gamma_{\rm rad}\approx 2000$) appear when using HLL, but for
general applications (e.g. highly magnetized or nearly force-free
flows near a rotating black hole), we have found HLL to also be
unstable (apparently due to the conflict between the causal one-sided
flux solution and the acausal centered reconstruction stencil).

%oooooooooooooooooooooooooooooooooooooooooooooooooooooooooooooo
%oooooooooooooooooooooooooooooooooooooooooooooooooooooooooooooo

This test shows limits of the M1 closure approach but at the same time
stresses the fact that, in principle, it does not limit specific
intensity to one particular direction (assuming only its symmetry with
respect to the flux). It performs much better than the Eddington
approximation, but in the case of multiple sources of light it must be
used with caution.

\subsection{Static Radiative Atmosphere in 1D Spherical Polar Minkowski}
\label{RADATM}

An important aspect of radiation in accretion disks is momentum
transfer between radiation and gas, such as the balance expressed by
the Eddington luminosity described in section~\ref{sec_intro}.  To
validate the treatment of gas-radiation momentum exchange, we consider
a static atmosphere which is in equilibrium under the action of
gravity, a gas pressure gradient, and the radiation force. We take the
optically thin limit and assume that gas-radiation interactions occur
only through a scattering coefficient, i.e., $\kappa_{\rm abs}=0$,
$\kappa_{\rm tot}=\kappa_{\rm es}$.  We consider a polytropic
atmosphere with equation of state $\pg=K\rhorest^\Gamma$.

An analytical solution can be obtained for this model problem.  For a
polytropic equation of state and $\kappa_{\rm abs}=0$, there is no
energy equation, and the radial component of the momentum equation can
be used to find the solution. In the non-relativistic limit ($r\gg2$),
assuming stationarity ($\partial_t=0$) and zero velocity ($v^i=0$),
the radial momentum equation takes the form
\be\label{eq.atm1}
\frac1\rhorest\pder{p}{r}=-\frac{1-f}{r^2},
\ee
where
\be
f=\kappa_{\rm es}F_{\rm in}r_{\rm in}^2.
\ee
Here $F_{\rm in}$ is the radiative flux imposed as a boundary
condition at the bottom of the atmosphere, $r=r_{\rm in}$, and $f$
gives the ratio of the radiative to gravitational (or geometrical)
forces; $f=1$ corresponds to the Eddington limit, where the luminosity
is $L_{\rm Edd}=4\pi/\kappa_{\rm es}$ and the radiative flux is
$F_{\rm in} = F_{\rm Edd} = 1/\kappa_{\rm es}r_{\rm in}^2$. Since
radiative energy must be conserved, in the stationary state the flux
must satisfy $F=F_{\rm in}r_{\rm in}^2/r^2$ (non-relativistic limit).

The solution to Eq.~(\ref{eq.atm1}) is
\be\label{eq:rho}
\rhorest=\rho_a\equiv\left[\frac{(\Gamma-1)}{\Gamma
  K}\left(C+\frac{1-f}r\right)\right]^{\frac1{\Gamma-1}},
\ee
where
\be\label{eq:f}
C=\frac{\Gamma K}{(\Gamma-1)}\rho_{0,\rm in}^{\Gamma-1} -
\frac{1-f}{r_{\rm in}},
\ee
and $\rho_{0,\rm in}$ is the assumed rest-mass density
at $r=r_{\rm in}$. The entropy constant $K$ is calculated at the
bottom of the atmosphere from the assumed gas temperature $T_{\rm in}$.

We set up a uniform spherical polar grid in Boyer-Lindquist
coordinates with only $40$ points between $r=10^6$ and $1.4\times10^6$
gravitational radii with $\theta$ spanning only $1$ cell from
$0.99\pi/2$ to $1.01\pi/2$.  We scaled all quantities to physical
units assuming $M=1M_\odot$ and $\kappa_{\rm es}=0.4\rhorest\, {\rm
  cm^{-1}}$. At the innermost radius we set $\rho_{0,\rm
  in}=10^{-15}\, {\rm g\,cm^{-3}}$ (optically thin atmosphere) and
$T_{\rm in}=10^6 \,\rm K$. All the velocities were initially zero and
the radiative energy density $E=F_{\rm in}/0.99999$. Initial values of
the gas density and temperature in the domain and in the ghost cells
were assigned based on the analytical solution. We ran four models
corresponding to four luminosities: $10^{-10}$, $0.1$, $0.5$ and
$1.0\, L_{\rm Edd}$.  Each model was run up to a time $t=2\times10^9
M$, which is sufficient to reach relaxed steady state for these
optically thin atmospheres.

%oooooooooooooooooooooooooooooooooooooooooooooooooooooooooooooo
\begin{figure}
  \centering
\includegraphics[height=.6\textwidth,angle=0]{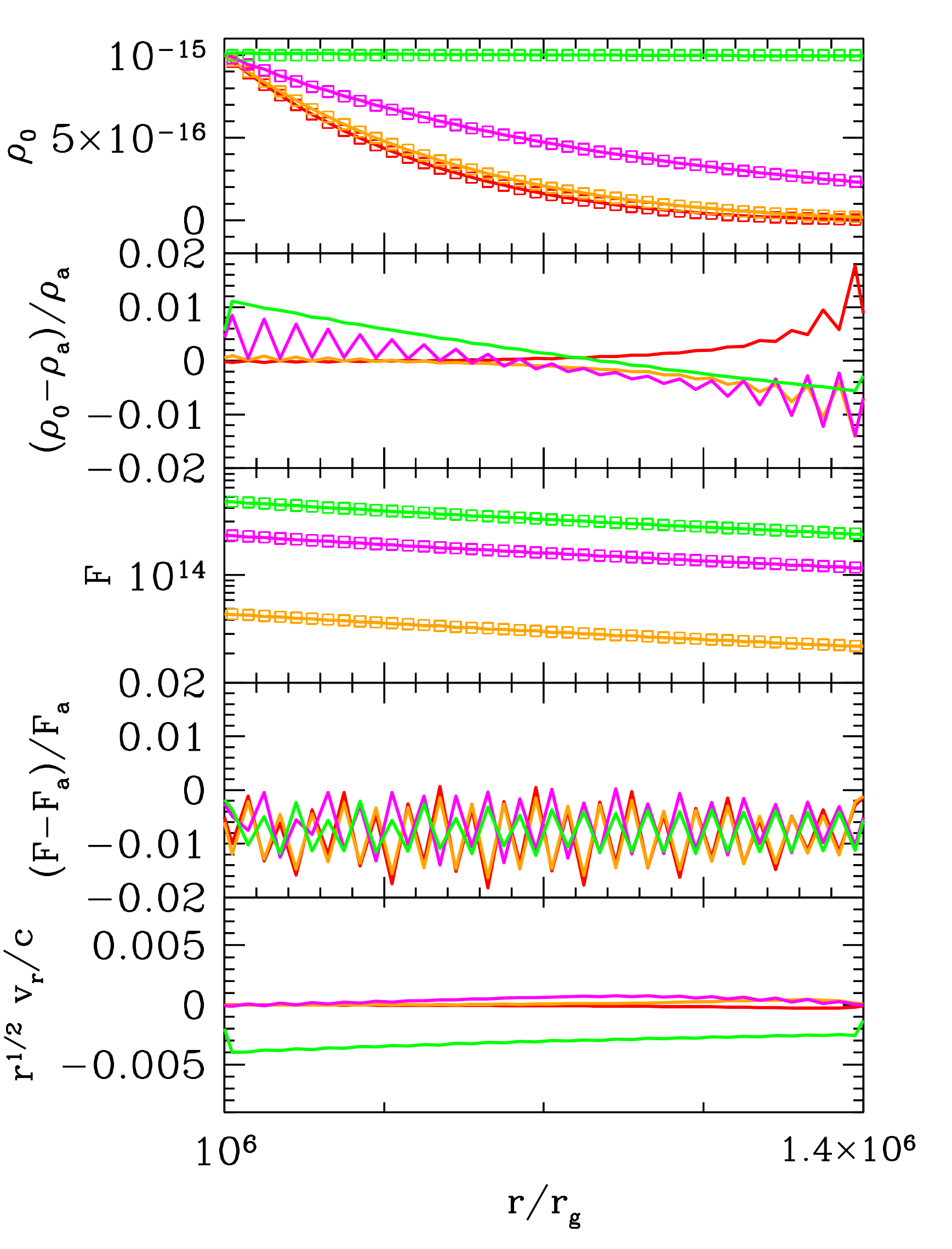}
\caption{Results obtained with the static atmosphere test in
  section~\ref{RADATM}. Numerically determined profiles and residuals
  between the numerical and analytical solutions are plotted for the
  density (top panel), radial lab-frame flux (middle panel), and
  radial velocity (bottom panel, residuals only). Colors denote the
  Eddington ratio of the flux boundary condition $F_{\rm in}$ at the
  bottom of the atmosphere: $F_{\rm in} = 10^{-10} F_{\rm Edd}$ (red),
  $0.1 F_{\rm Edd}$ (orange), $0.5 F_{\rm Edd}$ (magenta) and $1.0
  F_{\rm Edd}$ (green). Squares correspond to the numerical solutions
  and lines show the analytical profiles (equation~\ref{eq.atm1}).
  There is agreement at the percent level between the numerical and
  analytical solutions.}
  \label{f.radatm}
\end{figure}
%oooooooooooooooooooooooooooooooooooooooooooooooooooooooooooooo

Fig.~\ref{f.radatm} shows the results.  For the top panel, the
higher the luminosity, the flatter is the density profile, indicating
the effect of the outward force due to radiation. For the particular
case of the Eddington luminosity, the density should be perfectly
constant, reflecting the fact that the gravitational force is exactly
balanced by radiation and no pressure gradient in required.  Even at
this low resolution, \harmrad\ properly handles gas-radiation momentum
exchange as shown by looking at the residuals in the 2nd panel, where
fractional deviations in the density are below $\lesssim 2\%$.  Much
of the error is because of the finite $\gamma_{\rm rad}\approx 158$
rather than $v=c$ (i.e. a smaller $F_{\rm in}/E\sim 0.99$ gives about
$3$ times larger errors), the PPM reconstruction, what form of the
primitives one interpolates (e.g. interpolating more constant
quantities leads to much lower error for the $F_{\rm in}=1.0 F_{\rm
  Edd}$ case), and small relativistic corrections for the
non-relativistic solution.

The middle panel in Fig.~\ref{f.radatm} shows our results for the
radial radiative flux.  Once again, the models behave very well and
the agreement with the analytical solution is excellent.  Finally, the
bottom panel shows the residual radial velocities ($v_r/c$). These are
of the order of $10^{-5}$ (they should be zero), and appear to be
mostly driven by slight inconsistencies near the boundaries for
reasons similar to the reasons given for the density deviations.  Use
of MINM or HLL vs. LAXF or the entropy equations does not improve
these errors, but the errors do decrease with increasing resolution,
but only to first order due the errors being introduced by the
discontinuities at the inner and outer boundaries.

\subsection{Beam of light in 2D spherical polar for $a/M=0$ Black Hole}
\label{RADBEAM2D}

To test the performance of the code for radiation in strong
gravitational field, we study propagation of a beam of light in the
Schwarzschild metric.  The technical aspects of these results are
qualitatively similar for our (not shown) tests of beams in spherical
polar Minkowski, so we only consider the curved space-time case.

We consider three models, in each of which a beam of light is emitted
in the azimuthal direction at a different radius. We decouple gas and
radiation by neglecting absorptions and scatterings ($\kappa_{\rm
  abs}=\kappa_{\rm tot}=0$). We run the models on a two-dimensional
grid with only $30$ points distributed uniformly in $r$ between
$r_{\rm in}$ and $r_{\rm out}$ (see Table~\ref{t.beam} for values) and
only $60$ points distributed uniformly in azimuthal angle $\phi$
between $\phi=0$ and $\pi/2$.  Initially, we assign negligibly small
values for all primitive quantities, including the radiation energy
density and flux. We use outflow boundary conditions on all borders
except the region covered by the beam at the equatorial plane (see the
range of $r_{\rm beam}$ in Table~\ref{t.beam}), where we set the
radiation temperature to $T_{\rm beam}=10^{10}=1000T_a$ and the
lab-frame flux to $F^\phi=0.9999E$.  Here $T_a$ is the initial gas and
radiation temperature of the ambient medium.  We always stop the
simulation when the beam reaches the outer boundary and show that
result.  This corresponds to $t=9$, $t=8.5$, $t=16.5$ for Model 1,2,3,
respectively.

\begin{table}
\caption{Model parameters for the light beam tests}
\label{t.beam}
\centering\begin{tabular}{@{}lccc}
\hline
 Model & $r_{\rm beam}$ & $r_{\rm in}$ & $r_{\rm out}$\\
\hline
1 & $3.0\pm0.1$ & 2.6 & 3.5 \\
2 & $6.0\pm0.2$ & 5.5 & 7.5 \\
3 & $16.0\pm0.5$ & 14.5 & 20.5 \\
\hline
\end{tabular}
 \end{table}

\begin{figure*}
  \centering\hspace{-.15in}\vspace{-.15in}
  \subfigure{\includegraphics[width=.35\textwidth,angle=0]{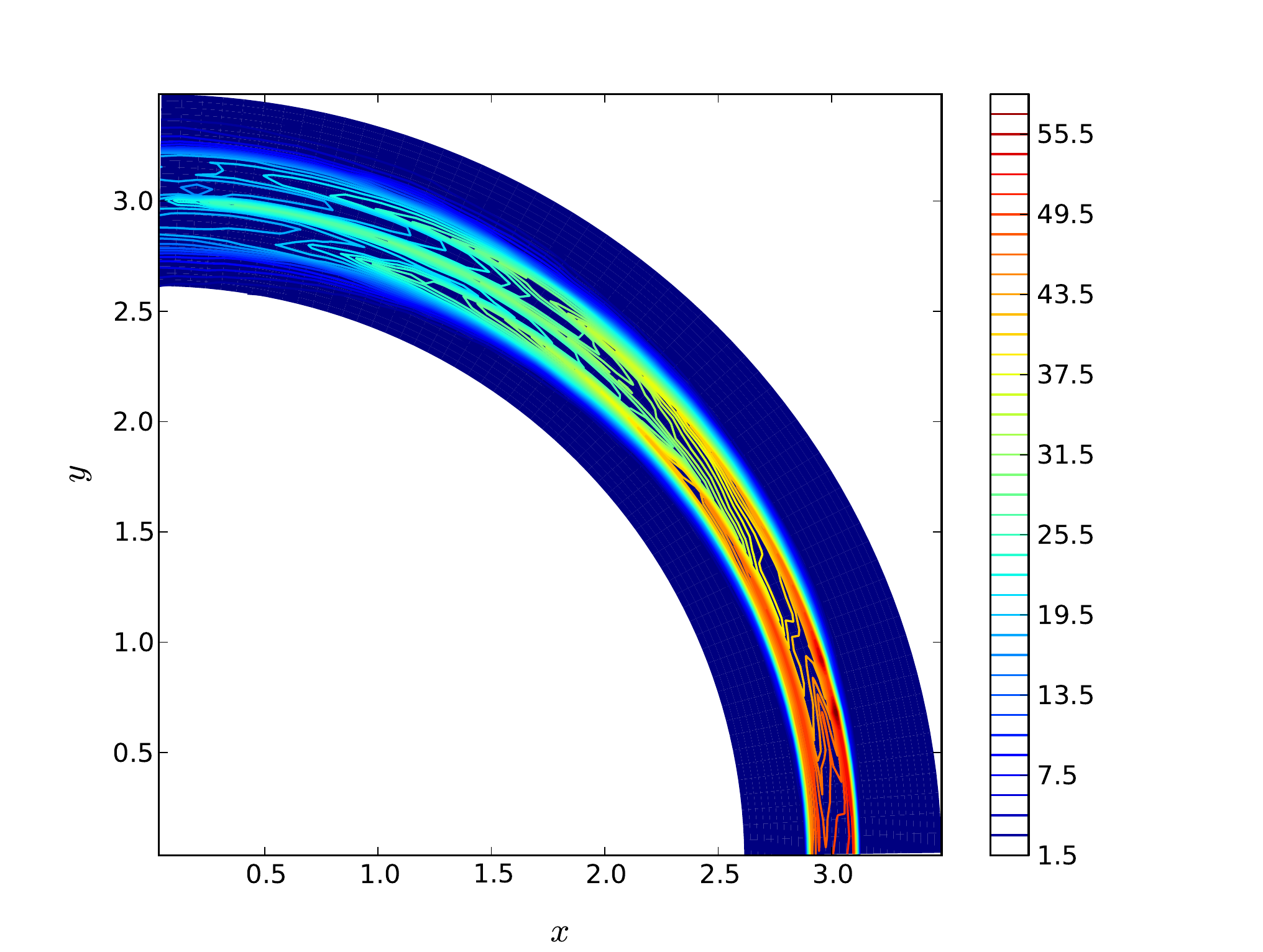}}\hspace{-.1in}
  \subfigure{\includegraphics[width=.35\textwidth,angle=0]{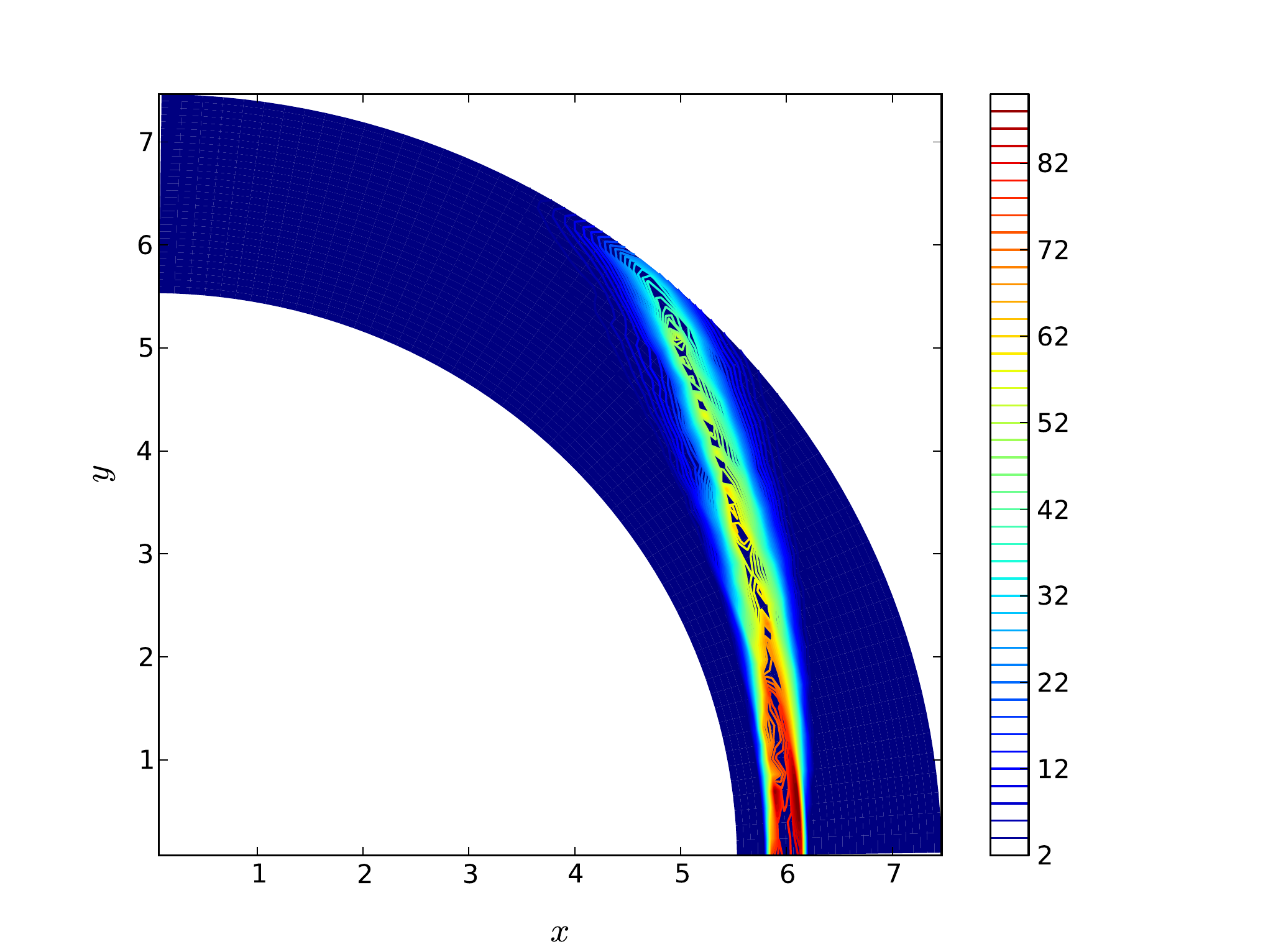}}\hspace{-.2in}
  \subfigure{\includegraphics[width=.35\textwidth,angle=0]{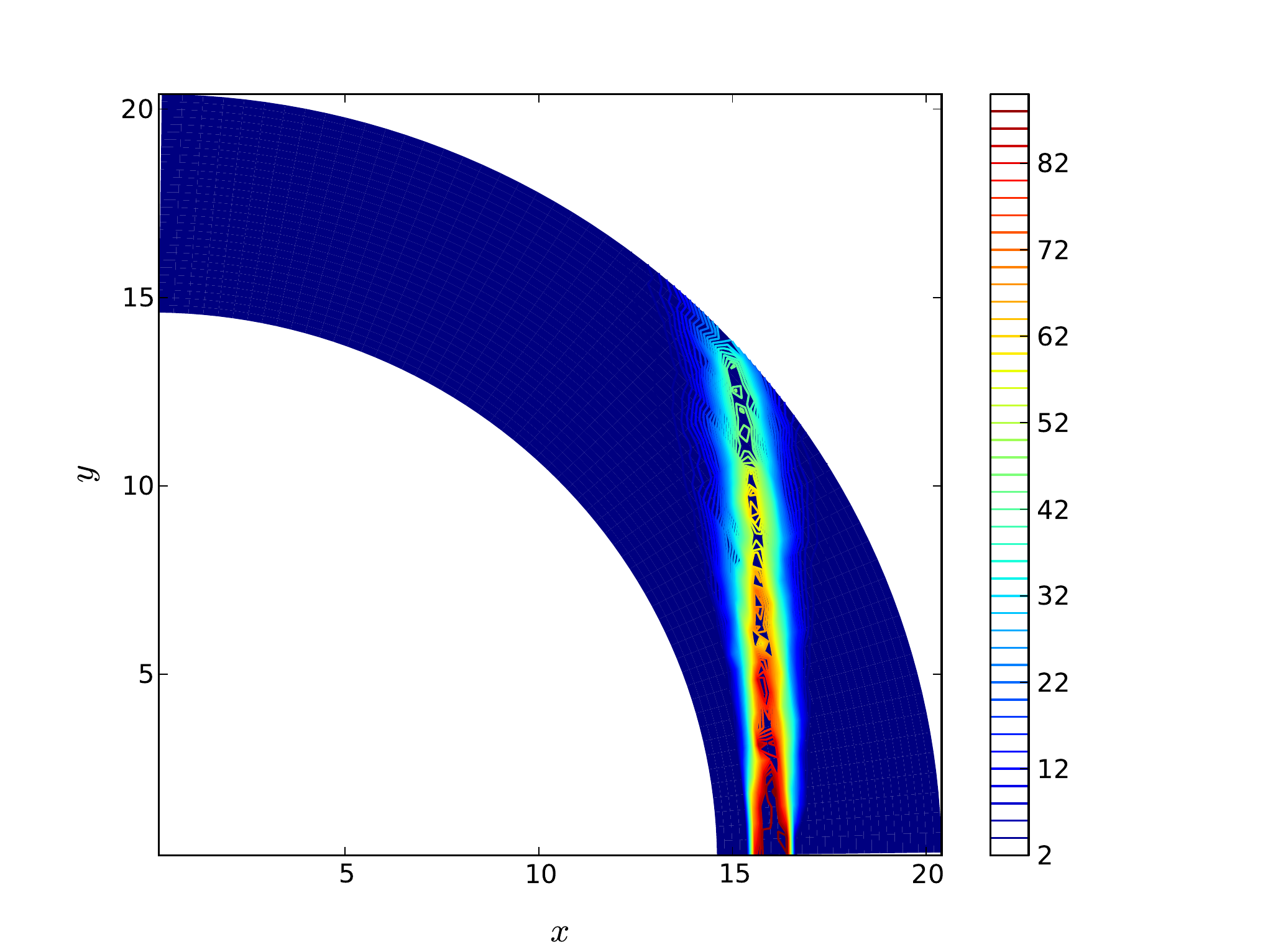}}
\caption{Results for Model 1 (left panel), Model 2 (middle), Model 3
  (right), involving light beams propagating near a Schwarzschild BH
  (see Table~\ref{t.beam} for model details) described in
  section~\ref{RADBEAM2D}. The BH is at $r=0$ (i.e., $x=y=0$) with
  horizon $r=2$. The beams are introduced via a boundary condition on
  the $x$-axis. The beams initially move vertically, i.e., in the
  azimuthal direction, in the lab-frame. Contours indicates the
  lab-frame radiation energy density.  The beams travel as expected in
  curved space-time.}
  \label{f.radbeam}
\end{figure*}

The panels in Fig.~\ref{f.radbeam} show the results for the three beam
models. Consider the right panel, which corresponds to Model 3
(Table~\ref{t.beam}) with the beam centered at $r_{\rm beam}=16$. At
such a large radius we do not expect significant bending of photon
geodesics and this is indeed the case --- the beam is only slightly
bent towards the BH. We also expect the beam to be tightly confined,
i.e., it should propagate with a nearly constant width. However, the
numerical solution shows some artificial broadening (but much less
than MINM used in Koral ; \citealt{2013MNRAS.429.3533S}).

The middle panel in Fig.~\ref{f.radbeam} shows Model 2, where the
beam is centered at the marginally stable orbit: $r_{\rm beam}=6$. At
this radius, photon geodesics are significantly deviated by gravity,
resulting in strong curvature in the beam. The numerical beam follows
the correct trajectory.

Finally, the left panel in Fig.~\ref{f.radbeam} shows Model 1, where
the center of the beam is exactly at the photon orbit: $r_{\rm
  beam}=3$.  An azimuthally oriented ray at this radius is expected to
orbit around the BH at a constant $r$. This is seen clearly in the
numerical solution.  Some of the diffusion seen is due to the need to
use $F^\phi=0.9999E$ with PPM, while some is physical broadening due
to photons emitted inside the $r=3$ curve bending inwards while those
emitted outside $r=3$ bend outward and head towards radial infinity.

A value of $F^\phi=0.99999E$ (giving $\gamma_{\rm rad}\approx 220$)
leads to some disruptions of the beam when using PPM and LAXF.  Using
HLL or decreasing $F^\phi$ improves stability.  No such disruptions
occur with MINM even at $F^\phi=0.99999E$.  Even with PPM and
$F^\phi=0.9999E$ (giving $\gamma_{\rm rad}\approx 70$), some
disruptions can eventually occur where the incoming beam interacts
with reflections (due to the simplified outflow boundary conditions)
off the outer boundary where the beam contacts.  As our goal is to
test the on-grid behavior (not advanced boundary conditions), we
always stop the simulation when the beam reaches the outer boundary.
We plan to improve PPM's behavior to add a bit of diffusion to keep
such beam's more uniform, but the behavior non-uniformity shown is due
to the beam being roughly the size of the PPM stencil size and PPM can
exaggerate features on such unresolved scales.  Note that use of HLL
does not improve the solution compared to the solution shown in
Fig.~\ref{f.radbeam}.

%oooooooooooooooooooooooooooooooooooooooooooooooooooooooooooooo
%oooooooooooooooooooooooooooooooooooooooooooooooooooooooooooooo

\subsection{Radiative spherical (Bondi) accretion in 1D spherical polar for $a/M=0$ Black Hole}
\label{RADBONDI}

Our next test problem considers radiative spherical accretion onto a
non-rotating BH. This problem has been studied in the past by
\cite{1983bhwd.book.....S,vitello84} and \cite{nobilietal91} and more
recently by \cite{roedigetal12} and \cite{fragileetal12}. We follow
\cite{fragileetal12} in the setup of our simulations to facilitate
comparison with their results.  As in their work, we consider Thomson
scattering and thermal bremsstrahlung, which give the following
opacity coefficients,
\bea
\hspace{.7in}\kappa_{\rm abs}&=&1.7\times10^{-25} T^{-7/2} m_p^{-2} \rhorest\, {\rm cm^{-1}},\\
\hspace{.7in}\kappa_{\rm tot}&=&\kappa_{\rm abs}+0.4 \rhorest\, {\rm cm^{-1}},
\eea where $\rhorest$ is in ${\rm g\,cm^{-3}}$ and $m_p$ is the mass of
the proton. Our numerical grid spans from
$r_{\rm in}$ to $r_{\rm out}=2\times 10^4$ and is resolved by 512 grid points spaced logarithmically following $r = R_0 + \exp(x^{(1)})$.
We assume a BH mass of $3 M_\odot$.
For the initial state, we choose the mass accretion rate $\Mdot$
(see Table~\ref{t.bondi} for values) and set the density profile accordingly,
\be\label{eq.bondi1} \rhorest=-\frac{\Mdot}{4\pi r^2 u^r}, \ee where the
radial velocity $u^r$ is equal to its free fall value
$u^r=-\sqrt{2/r}$. The gas temperature is given by
\be\label{eq.bondi2} T=T_{\rm out}\left(\frac\rhorest{\rho_{0,\rm
    out}}\right)^{\Gamma-1}, \ee where $T_{\rm out}$ is the
temperature at the outer radius and $\Gamma$ is the adiabatic
index. The latter is calculated from the radiation to gas pressure ratio $f_p=p_{\rm rad}/p_{\rm gas}$ of the initial state (Table~\ref{t.bondi}),
\be \Gamma=1+\frac13\left(\frac{2+2f_{\rm
    p}}{1+2f_{\rm p}}\right).  \ee The radiative energy density is set to
$E=3 f_{\rm p} p_{\rm gas}$.

The numerical simulations are run in one (radial) dimension. The
primitive quantities at the outer boundary are fixed at their initial
values, as described above. At the inner boundary we apply outflow
boundary conditions.  We could apply special extrapolating and interpolating
radial dependencies, but we avoid changing the boundary conditions to keep
the results applicable to more general simulations.
Table~\ref{t.bondi} lists the parameter values we used corresponding
to five models.  The first model, E1T6, is characterized by the lowest
mass accretion rate and is designed to highlight the ability of our
scheme to handle optically thin media. The other four models are
identical to simulations described in \cite{fragileetal12}.

\begin{table}
\caption{Model parameters and results for radiative spherical accretion tests}
\label{t.bondi}
\centering\begin{tabular}{@{}lcccc}
\hline
 &  &  & & Measured \\
 Model & $\Mdot c^2/L_{\rm Edd}$ & $T_{\rm out} [K]$ & $f_{\rm p}=\frac{p_{\rm rad}}{p_{gas}}$&$L/L_{\rm Edd}$\\
\hline
E1T6 & 1.0 & $10^6$&$1.2\times10^{-4}$ & $2.8\times10^{-8}$ \\
E10T5 & 10.0 & $10^5$&$1.2\times10^{-7}$& $1.3\times10^{-6}$ \\
E10T6 & 10.0 & $10^6$&$1.2\times10^{-4}$& $3.4\times10^{-6}$ \\
E10T7 & 10.0 & $10^7$&$1.2\times10^{-1}$& $7.5\times10^{-6}$ \\
E100T6 & 100.0 & $10^6$&$1.2\times10^{-4}$& $1.0\times10^{-4}$ \\
\hline
\end{tabular}\\
Model names and parameters after \cite{fragileetal12}.
 \end{table}

For all models, we choose a grid such that $r=R_0+\exp(x_1)$ for some
uniform grid $x_1$.  We can set, e.g., $r_{\rm in}=1.9$ for
Kerr-Schild coordinates, while $r_{\rm in}=2.5$ also works for
Boyer-Lindquist coordinates.  We show results for Kerr-Schild
coordinates using Kerr-Schild coordinates with $r_{\rm in}=2.5$ and
$R_0=2.2$ for all models except E10T5 where we choose $r_{\rm in}=1.9$
and $R_0=1.85$ in order to place more cells near the BH to ensure the
energy equation evolve remains accurate for $\ug$ despite the small
$T_{\rm gas}$.  The models are run till $t=3.13\times 10^4$, $t=1.674\times
10^4$, $t=1.21\times 10^5$, $t=1.485\times 10^5$, and $t=1.351\times
10^5$ for the models E1T6, E10T5, E10T6, E10T7, and E100T6,
respectively.

\begin{figure}
  \centering\hspace{-.15in}\vspace{-.15in}
\includegraphics[width=1.0\columnwidth,angle=0]{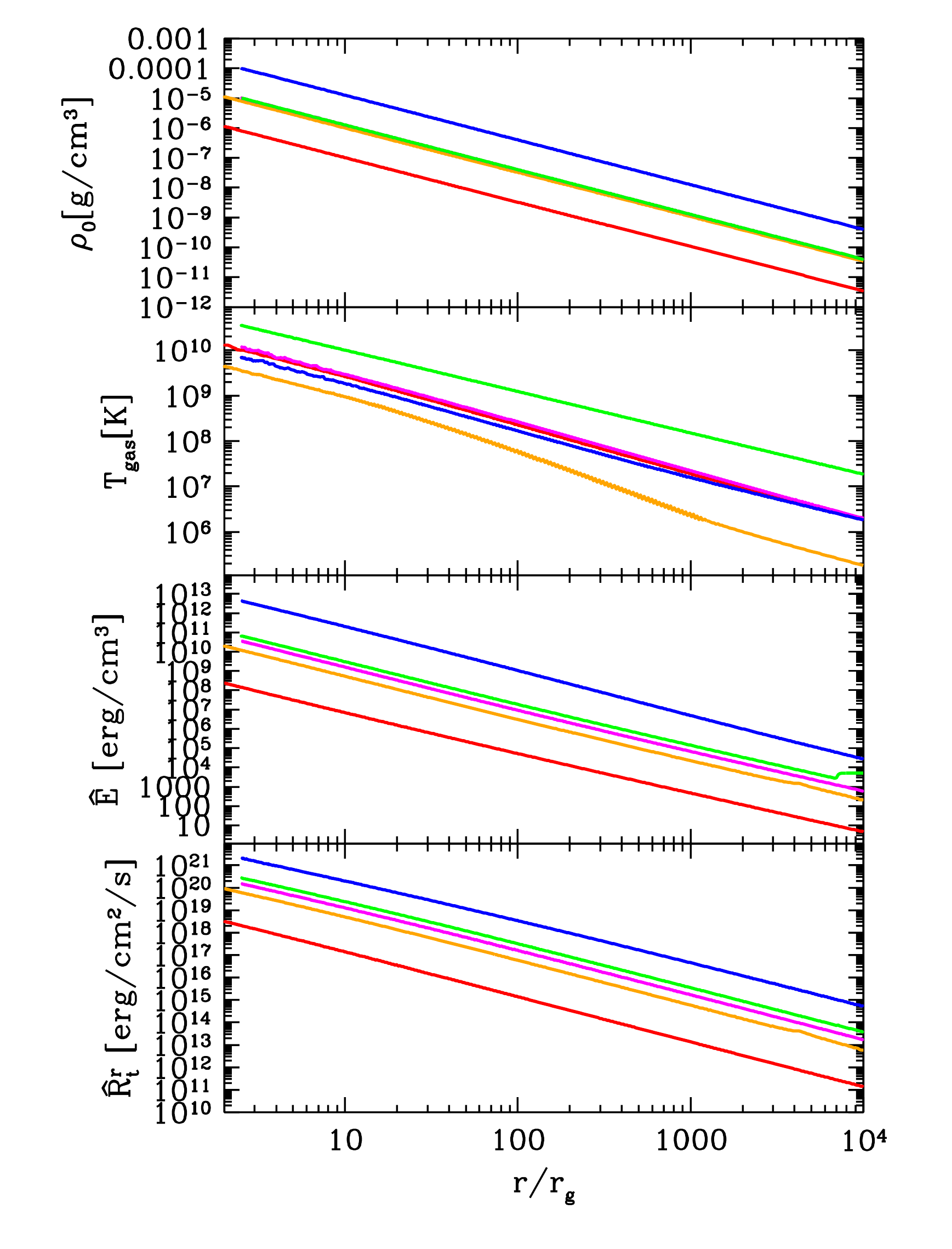}
\caption{Numerical results obtained with \harmrad\ for five models of
  spherical Bondi accretion with radiation.  Parameters of the models
  are given in Table~\ref{t.bondi}.  The panels show density (top),
  gas temperature (2nd), fluid-frame radiative energy density (3rd),
  and fluid-frame radiative radial flux (bottom).  Lines correspond to
  models E1T6 (red), E10T5 (orange), E10T6 (magenta), E10T7 (green),
  and E100T6 (blue).  The results show that the code handles well both
  optically-thin and optically-thick regions for a large radial
  dynamic range in curved space-time.}
  \label{f.bondi}
\end{figure}

Fig.~\ref{f.bondi} shows the numerical solutions obtained with
\harmrad. The top panel presents profiles of density, which follow
the initial profile (equation~\ref{eq.bondi1}) throughout the
simulation.

The 2nd panel of Fig.~\ref{f.bondi} shows the gas temperature. For
all but the coldest model, E10T5, the temperature follows
equation~(\ref{eq.bondi2}). In the case of model E10T5, the gas is
hotter than the analytical result.  This is because of gas-radiation
coupling which heats up the gas as it approaches the BH (the
analytical solution assumes that there is no interaction).  Some
models show mild oscillations in the temperature (or $\ug$), which is
due to the large dynamic range in radius when using PPM and the energy
equation due to the differences between point and average quantities
due to the non-linearity of the energy equation.  The noise in $\ug$
is especially pronounced in model E10T5.  A higher-order scheme like
WHAM \citep{2007MNRAS.379..469T} or a scheme that interpolates
conserved quantities (e.g. MP5 in Koral) can improve the temperature
behavior.  Also, use of the entropy equations in \harmrad\ avoids all
such oscillations, and the entropy equations need much lower
resolution to achieve the same accuracy as the energy equation.  Also,
the noise in temperature (or $\ug$) with the energy equation is less
as one increases resolution or focuses resolution toward the BH by
using $R_0$ closer to $R_{\rm in}$.  Note that lowering the implicit
solver tolerance (say always requiring $e_T<10^{-13}$) for the energy
method does not avoid the noise.

The 3rd and 4th panels in Fig.~\ref{f.bondi} show radial profiles of
the fluid-frame radiative energy density and fluid frame radial energy
flux for the five models.  Both quantities follow roughly an $r^{-2}$
scaling, reflecting the fact that in steady-state (barring redshift
factors) the luminosity is equal to $4\pi Fr^{2}$ and should be
conserved.  The glitches at large radius are due to the finite run
time of those particular models, as the radiative variables are still
evolving towards equilibrium.

%OOOOOOOOOOOOOOOOOOOOOOOOOOOOOOOOOOOOOOOOOOOOOOOOOOOOOOOOOOOOOO
%oooooooooooooooooooooooooooooooooooooooooooooooooooooooooooooo

Because the flux in these models is non-negligible compared to the
energy density (e.g., $F\approx 0.9E$ for the E10 family of models),
the Eddington closure scheme does not work very well, especially at
low optical depth.  For instance, \cite{fragileetal12} used Eddington
closure and obtained unphysical noise or breaks in their profiles of
radiative quantities (see their Figure 5) in all models with $\Mdot <
300 L_{\rm Edd}$. This just reflects the fact that their closure
cannot handle optically thin media.  Our algorithm uses the M1 closure
scheme and has no problems with either optically thick or thin
regimes. To emphasize this point, we have solved an additional model,
E1T6, in which the accretion rate is an order of magnitude lower than
the smallest rate considered by \cite{fragileetal12}. \harmrad\ works
fine for this model, and can, in fact, handle even more extreme
configurations, both at lower and higher accretion rates.

For direct comparison of our results with those reported in
\citet{fragileetal12}, we have calculated for all our models the
luminosities,
\be
L=4\pi Fr^2,
\ee
emerging at radius $r=7700$ (which is within $10\%$ of the value
at $r\sim 1360$.

\begin{figure}
  \centering\hspace{-.15in}\vspace{-.15in}
\includegraphics[width=1.0\columnwidth,angle=0]{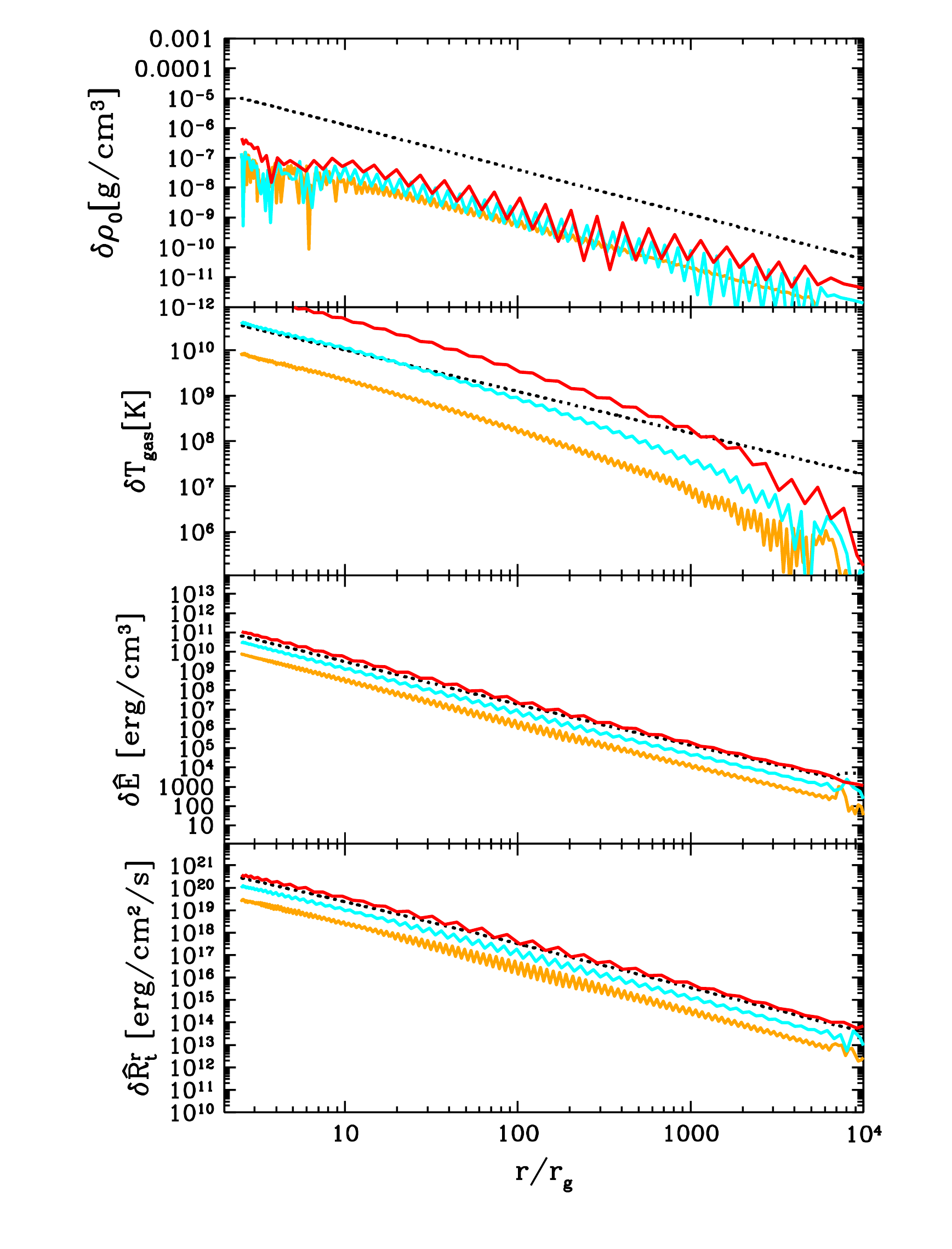}
\caption{Convergence test for 1D radiative Bondi test (similar
  quantities plotted as in Fig.~(\ref{f.bondi})).  Here we only
  consider model E10T7 with radial resolutions of $64$ (solid red),
  $128$ (solid cyan), $256$ (solid orange), and $512$ (dotted black)
  (as used in Fig.~(\ref{f.bondi}, where this model is shown as green
  lines in that other figure)).  Model with resolution $512$ is
  directly plotted for reference (dotted black), while remaining
  models at lower resolutions are plotted as an absolute difference
  away from the $512$ resolution model in order to show the relative
  error.  Shows convergence to correct solution to 2nd order, with
  most sensitive dependence in temperature.}
  \label{f.bondiconv}
\end{figure}

Fig.~\ref{f.bondiconv} shows model E10T7 for different radial
resolutions ($64$, $128$, $256$, $512$) to show convergence behavior.
The energy equations fail for $32$ cells for such a large radial
dynamic range, where clearly visible oscillations appear with $64$
cells.  The error is dominated by the temperature, which converges to
2nd order.  E.g., for the first radial cell, the relative differences
in gas temperature between models are $0.2466$, $1.194$, and $5.125$
for $256$, $128$, and $64$, respectively, which is a drop by a factor
of $4$ for each increase in resolution by a factor of $2$.  Use of the
entropy equation instead of the energy equation reduces the
temperature error substantially, with the same convergence rate at
that lower error.

%OOOOOOOOOOOOOOOOOOOOOOOOOOOOOOOOOOOOOOOOOOOOOOOOOOOOOOOOOOOOOO
%oooooooooooooooooooooooooooooooooooooooooooooooooooooooooooooo

\subsection{MHD Radiative Bondi Flow in 2D spherical polar with $a/M=0$ BH}
\label{RADBONDIMAG}

\def\Lone{{\mathcal{L}}_1}

Our next test involves the same radiative Bondi problem as in
section~(\ref{RADBONDI}) for model E10T7, but now we include a strong
monopolar magnetic field with vector potential component
$A_\phi\propto (1-\cos(\theta))$.  We choose $b^2/\ug\approx 812$
(giving $b^2/\rhorest\approx 4.9$) at the horizon at $r=2M$.  The flow
is along the magnetic field, so all magnetic forces cancel exactly.
This is a difficult test, however, because numerically the magnetic
terms cancel only to truncation error, and the small value of $\ug$
has to be recovered from the total energy equation dominated by
truncation errors in the magnetic field.  This causes problems at high
magnetic field strengths, unless one uses the entropy evolution
equation that involves lower-order velocity terms.  For entropy
evolution, we get highly accurate results even if we choose
$b^2/\ug\sim 10^5$ and even higher, at relatively lower resolutions.

We now also solve the problem in Kerr-Schild coordinates in full 2D
spherical polar coordinates with resolution $N_r\times N_\theta =
512\times 16$ with angular span from $\theta=0$ to $\pi$ and using
$R_{\rm in}=1.9$, $R_{\rm out}=2\times 10^4$, and $R_0=1.88$ chosen to
allow the energy evolution method to accurately evolve this scenario.
This appears to simply be a one-dimensional test, but for HARM it is
actually two dimensional.  Although the pressure is independent of the
Boyer-Lindquist coordinate $\theta$, the $\theta$ acceleration does
not vanish identically due to round-off error.  This is because
pressure enters the momentum equations through a flux ($-\del_\theta
(p \sin\theta)$ in the Newtonian limit) and a source term ($p
\cos\theta$ in the Newtonian limit).  Analytically these terms cancel;
numerically they produce an acceleration that is of order the
truncation error for the original HARM \citep{2003ApJ...589..444G} and
round-off error for HARM.  This test also exercises many terms in the
code because in Kerr-Schild coordinates only three of the ten
independent components of the metric are zero.

\begin{figure}
  \centering\hspace{-.15in}\vspace{-.15in}
\includegraphics[width=1.0\columnwidth,angle=0]{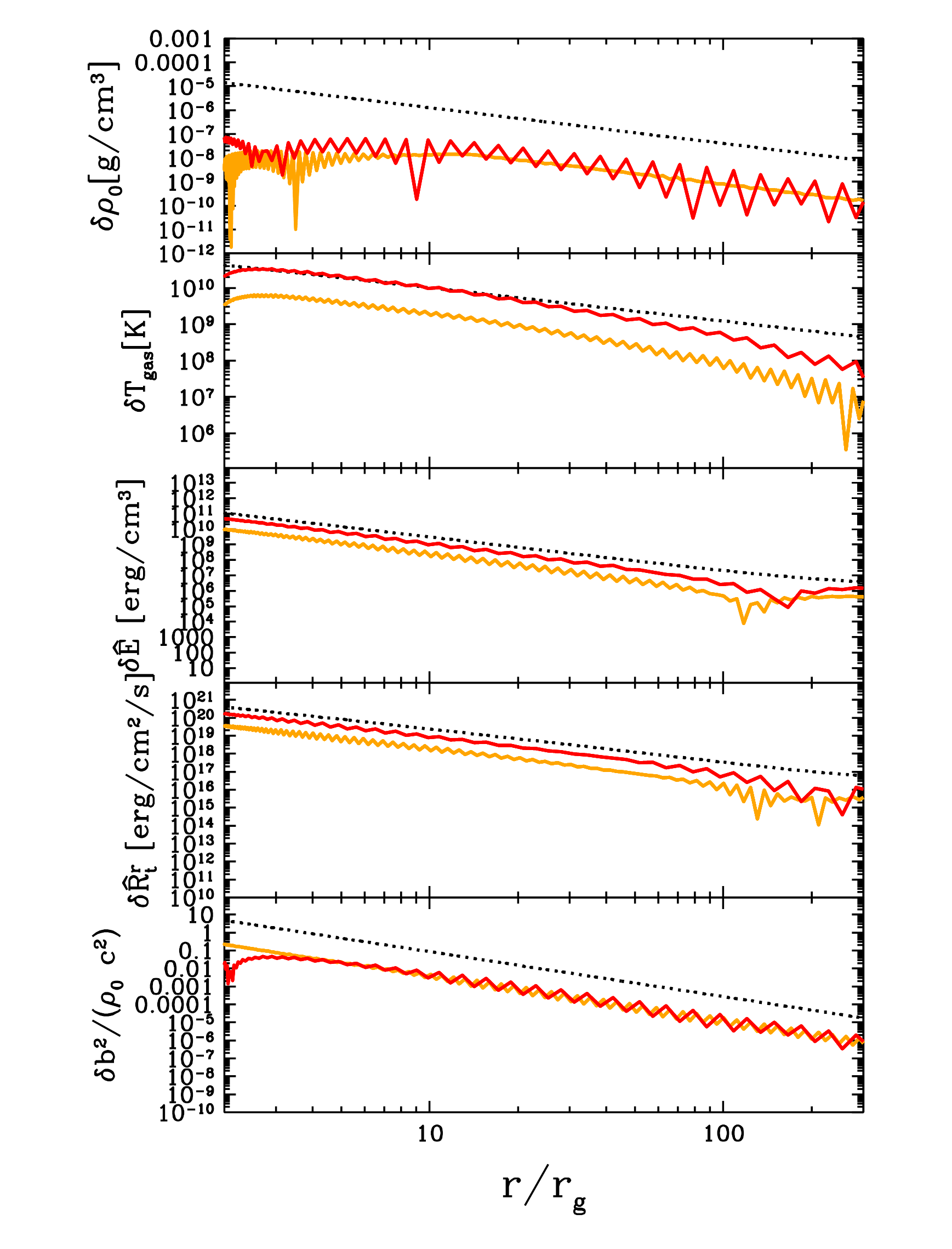}
\caption{Convergence test for model E10T7 for a relativistically
  strong radial magnetic field with spherical Bondi accretion with
  radiation in 2D (until here, our Bondi tests were in 1D). The panels
  show density (top), gas temperature (2nd), fluid-frame radiative
  energy density (3rd), fluid-frame radiative radial flux (4th), and
  $b^2/(\rhorest c^2)$ (bottom).  The $512$ resolution model is
  plotted directly (dotted black) for reference, while the $256$ model
  (orange) and $128$ model (red) are plotted as the differences from
  the $512$ model to show convergence.  The models have $16$ angular
  cells, and all are shown in the difference. This results show that
  the code handles well the case of 2D relativistically strong
  magnetic fields in radiative flows, the angular direction introduces
  no anisotropy, and the solution converges to 2nd order in the
  temperature that dominates the error.}
  \label{f.magbondi}
\end{figure}

Fig.~\ref{f.magbondi} shows the model E10T7 in 2D and with the
magnetic field.  We show resolutions: high ($512\times 16$), medium
($256\times 16$), and low ($128\times 16$).  While the outer
radius is $2\times 10^4$, we only evolved for a finite
time and so only show out to $r\sim 300r_g$.

All resolutions do well, except for a radial resolution of $128$ for
which the temperature starts to deviate at smaller radii.  For even
smaller radial resolutions of $64$ and $32$, the implicit solver
fails.  This occurs because of the energy equation's limits over such
a large dynamic range.  An evolution with the entropy equations has no
such limits on resolution and the temperature behaves accurately even
at low resolutions.  Over the span in equilibrium, the simulations
converge to at least 2nd order in space in temperature, the quantity
that dominates the error.  E.g., for any of the first several radial
cells, the relative errors are $0.04264$ and $0.319$ for the $256$ and
$128$ models, respectively.

%OOOOOOOOOOOOOOOOOOOOOOOOOOOOOOOOOOOOOOOOOOOOOOOOOOOOOOOOOOOOOO
%oooooooooooooooooooooooooooooooooooooooooooooooooooooooooooooo

%%% Local Variables:
%%% mode: latex
%%% TeX-master: "ms"
%%% End:

\section{Fiducial Fully 3D GR Radiative MHD Disk Model}
\label{sec:fiducial}

As a test of the scheme to handle astrophysically realistic
situations, we consider magnetized radiative accretion onto a rotating
black hole.  This model with black hole mass of $\MBH=10M_{\odot}$ and
$a/M=0.9375$ has nominal thin disk efficiency $\eta_{\rm NT} \approx
17.91\%$, so that $\Mdotedd \approx 7.8\times 10^{18}$g/s (see
Eq.~\ref{Ledddef}).  Then rescaling $\Mdotedd$ by only $c$, $G$, and
$M$, we normalize other quantities by letting $\rho_{\rm Edd}\approx
1.2\times 10^{-4}$g, $u_{\rm Edd}\approx 1.07\times 10^{17}$erg, and
$b_{\rm Edd}\approx 3.2\times 10^8$G.

We use the same numerical parameters as all prior tests given in
section~\ref{s.tests}.  The implicit solver takes an average of 3.0
iterations per complete solution (including all unused attempts),
where $99.8\%$ are from the PMHD method, $0.2\%$ are from the URAD
method, and $0.002\%$ are from the PRAD method.  Among all used
solutions, $99.7\%$ come from the energy-momentum equations, $0.3\%$
come from the entropy-momentum equations, $0\%$ come from the cold MHD
equations, and $0.0004\%$ have no solution such that diffusive backups
are used.  Nearly $100\%$ of those unacceptable solutions occur inside
the horizon near the polar axis.  As compared to an otherwise
identical non-radiative simulation, this simulation is about 2-3 times
slower per core, while it operates at slightly higher parallel
efficiency because there are more per-core operations between any
cross-core Message Passing Interface (MPI) operations.

\subsection{Initial Mass Distribution: Polish doughnut and Atmosphere}

We set up analytical equilibrium
torii \citep[Polish doughnuts,][]{abramowiczetal78} in the
Kerr ($a/M=0.9375$) metric as initial conditions. For the analytical model, we
assume a constant specific angular momentum, $\ell =
-u_\phi/u_t=$\, constant. From the condition $u^\mu
u_\mu=-1$, it follows that \be u_t^{-2}=-g^{tt}+2\ell
g^{t\phi}-\ell^2g^{\phi\phi}.  \ee We choose the specific
internal energy at the inner edge of the torus, $u_{t,\rm in}$, which
determines the radius of the inner edge of the torus, and we then
calculate the fluid enthalpy, $h=\rhorest+\ug+\pg$
\citep[e.g.,][]{hawleyetal84a}, \be h=\frac{u_{t,\rm in}}{u_t}.  \ee
Using an equation of state $p=\kappa_1 \rho^\Gamma$ (where the constant
$\kappa_1$ determines the entropy of the torus gas), we obtain
\bea
\hspace{1in}\rhorest&=&\left[\frac{(h-1)}{\kappa_1}\,\frac{(\Gamma-1)}{\Gamma}\right]^{1/(\Gamma-1)}, \\
\hspace{1in}u&=&\rhorest\,\frac{(h-1)}{\Gamma}.
\eea
We set the initial velocity to $v^r=v^\theta=0$, $v^\phi=u^\phi/u^t$
within the doughnut, and set the gas to the ZAMO velocity outside.
We choose $\Gamma=5/3$, where $\Gamma=4/3$ may lead to somewhat
different results \citep{2004ApJ...611..977M,mm07}.  The specific angular momentum is
set to be $\ell=4.5$, $u_{t,\rm in}=0.9999999$, corresponding to a
torus inner radius $r_{\rm in}\approx 8.5$ and
pressure maximum at $r_{\rm max}\approx 18r_g$.
We set $\rhorest=\rho_{\rm
  max}=1$ at the maximum rest-mass density.  To seed the magneto-rotational instability (MRI), $\ug$ is
perturbed by a factor $1 + F_R(E-0.5)$, where $F_R=0.1$ and $E$ is a
random number from $0$ to $1$.  The torus is surrounded by an
atmosphere with $\rhorest=10^{-4} (r/r_g)^{-2}$, $\ug=10^{-6}
(r/r_g)^{-5/2}$, $\reluvec^i=0$, and $B^i=0$.
The density can consistently drop to zero in the jet that
emerges, so we use a numerical density floor that ensures a maximum of
$b^2/\rhorest = 200$, maximum of $b^2/\ug=10^5$, and maximum of
$\ug/\rhorest=10^{10}$.

\subsection{Initial Magnetic Field}
\label{sec:modelsetupfield}

We consider an initial poloidal field geometry that does not lead to
magnetic flux saturation near the BH, so a magnetically arrested disk
(MAD) or magnetically choked accretion flow (MCAF) does not form.  A
single set of field loop of a single polarity are inserted.  For this
poloidal field geometry, the $\phi$-component of the magnetic vector
potential is
\begin{eqnarray}
\Avpotvec_\phi &\propto& f_1 f_2 ,\\
f_1 &=& |q|^p |r\sin{(\theta)}|^\nu ,\nonumber \\
q &=& u_{\rm g}/u_{\rm g,\rm max} - f_c ,\nonumber \\
f_2 &=& \sin{(\log{(r/S)}/T)} ,\nonumber
\end{eqnarray}
where $f_1$ has $p=1$ and $\nu=2$, $q$ has $f_c=0.2$, $u_{\rm gas,\rm max}$
is the maximum $\ug$, $q=0$ is set if $q<0$, and $f_2$ has
$S=0.5r_{\rm in}$ and $T=0.28$ for the flipping field and $f_2=1$ for
the non-flipping field.

The magnetic field strength is set via the plasma $\beta=\pg/p_b \sim
(2/\Gamma)(c_s/v_a)^2$ where $v_a^2 = b^2/(\rho_0 + \ug + \pg + b^2)$
gives the~\alf~speed $v_a$.  Our model has $\beta_{\rm min}$, the
smallest value of $\beta$ (within the resolved disk region, e.g.,
$r\sim 1000r_g\ll R_{\rm out}$) of $\beta_{\rm min}\approx 100$. An
alternative measure is $\beta_{\rm rat-of-maxes}\equiv p_{\rm gas,\rm
  max}/p_{b,\rm max}\approx 240$, where $p_{\rm gas,\rm max}$ is the
maximum thermal pressure on the domain and $p_{b,\rm max}$ is the
maximum magnetic pressure on the domain.  Another alternative is
$\beta_{\rm rat-of-avg}\equiv p_{\rm gas,\rm avg}/p_{b,\rm avg} =
\langle \pg \rangle/\langle p_b \rangle\approx 3800$.  These $\beta$
are computed with condition $b^2/\rho_0<1$.  Our choices for $\beta$
ensure that $\Qtwo>1$ so the MRI operates, while we push close to
$\Qtwo\sim 1$.

\subsection{Initial Radiation}

The initial solution solves for the hydrostatic equilibrium torus with
$\Gamma=4/3$ (which is the correct value for a radiation pressure
dominated disk). This pressure is then assumed to be the actual total
pressure $p_{\rm tot}$, which is then distributed between gas and
radiation so as to satisfy local thermal equilibrium (LTE,
$\widehat{E}= a_{\rm rad} T^4$) based upon a single temperature
satisfying
\begin{equation}
p_{\rm tot} = \pg + p_{\rm rad} = k_{\rm B} \rhorest T + \frac{1}{3}a_{\rm rad} T^4 .
\end{equation}
Once this initial state has been set, we reset $\Gamma=5/3$ for a
nonrelativistic gas.  Because the gas is very optically thick, the
initial state holds hydro-radiation-static equilibrium to good
accuracy.

The radiative fluxes in each direction $i$ are set as based upon the
flux-limited diffusion approximation
\begin{equation}
F_i = -\frac{c}{\kappa_{\rm tot}} \frac{d p_{\rm rad}}{dx^i} ,
\end{equation}
where the orthonormal gas-fluid frame radiation fluxes are limited to
$|F|<0.7 \widehat{E}$.

\subsection{Numerical Grid}
\label{sec:numsetupgrid}

The uniform spatial coordinates $\xvec^{(i)}$ have resolution
$N_r\times N_\theta \times N_\phi$ active grid cells and $4$ boundary
cells for each of the $6$ boundaries in 3D.  The radial grid of $N_r=256$
cells spans from $R_{\rm in}$ to $R_{\rm out}$ with mapping
\begin{equation}
r(x^{(1)}) = R_0 + \exp{f[\xvec^{(1)}]}
\end{equation}
where $R_0=0.2$ is chosen in this paper.  For $\xvec^{(1)}<x_{\rm
  break}$
\begin{equation}
f[\xvec^{(1)}] = n_0 \xvec^{(1)} ,
\end{equation}
where $x_{\rm break}=\log(r_{\rm break}-R_0)/n_0$ (with $n_0=1$), and
otherwise
\begin{equation}
f[\xvec^{(1)}] = n_0 \xvec^{(1)} + c_2 (\xvec^{(1)} - x_{\rm break})^{n_2} ,
\end{equation}
where $c_2=1$, and $n_2=10$.  The $\xvec^{(1)}$ grid ranges from
$\xvec^{(1)}_s = (\log(R_{\rm in}-R_0))/n_0$ to $\xvec^{(1)}_f$, which
is $\xvec^{(1)}_f=(\log(R_{\rm out}-R_0))/n_0$ if $R_{\rm out}<r_{\rm
  break}$ and otherwise determined iteratively from $R_{\rm out} =
r[\xvec^{(1)}_f]$.  The value of $R_{\rm in}=1.1\approx 0.816r_{\rm
  H}$ is chosen so that there are $6$ active grid cells inside the
outer horizon, while $R_{\rm in}$ is outside the inner horizon.  So
the boundary cells only connect to stencils (each $\pm 4$ cells) that
are inside the horizon, which avoids causal connection between the
inner boundary and the flow outside the horizon.  We set $R_{\rm
  out}=10^4r_g$ and $r_{\rm break}=5\times 10^2r_g$.  The radius
$r_{\rm break}$ is where the grid changes from exponential to
hyper-exponential, which allows the grid to focus on the dynamics at
small radii while avoiding numerical reflections off the outer grid.
Radial boundaries use absorbing conditions.

The $\theta$-grid of $N_\theta=128$ cells spans from $0$ to $\pi$ with
mapping
\begin{equation}
\theta(\xvec^{(2)}) = \Theta_2 + W (\Theta_1 - \Theta_2)
\end{equation}
where $\xvec^{(2)}$ ranges from $0$ to $1$ (i.e. no polar cut-out ; but
see Appendix~\ref{sec:nummethods}).  The first grid mapping function
is given by
\begin{eqnarray}
\Theta_1 &=& T_0 S_2 + T_2 S_0 ,\\
T_2 &=& (\pi/2)(1 + \arctan{(h_2(\xvec^{(m2)}-(1/2)))}/\arctan{(h_2/2)} ) ,\nonumber \\
h_2 &=& h_3 + ((r-r_{sj3})/r_{0j3})^{n_{j1}} ,\nonumber \\
T_0 &=& \pi  \xvec^{(2)} + ((1- h_0)/2)  \sin{(2 \pi  \xvec^{(2)})} ,\nonumber \\
h_0 &=& 2-Q_j (r/r_{1j})^{-n_{j2}( (1/2) + (1/\pi)   \arctan{(r/r_{0j}-r_{sj}/r_{0j})})} ,\nonumber \\
S_2 &=& (1/2) - (1/\pi) \arctan{((r-r_s)/r_0)} ,\nonumber \\
S_0 &=& (1/2) + (1/\pi) \arctan{((r-r_s)/r_0)} \nonumber ,
\end{eqnarray}
where $r_s=40$ and $r_0=20$.  For $h_2$, we set $h_3=0.3$,
$r_{0j3}=20$, $r_{sj3}=0$, and $n_{j1}=1$ so the jet is resolved with
grid lines following $\theta_j\propto r^{-n_{j1}}$.  For $h_0$, we set
$r_{1j}=2.8$, $n_{j2}=1$, $r_{0j}=15$, $r_{sj}=40$, and $Q_j=1.3$.  We
set $\xvec^{(m2)}=\xvec^{(2)}$ unless $\xvec^{(2)}>1$ then
$\xvec^{(m2)}=2-\xvec^{(2)}$ and unless $\xvec^{(2)}<0$ then
$\xvec^{(m2)}=-\xvec^{(2)}$.  $\Theta_1$ focuses on the disk at small
radii and the jet at large radii.  The second mapping function is
\begin{equation}
\Theta_2 =  (\pi/2) (h_\theta (2 \xvec^{(2)}-1)+(1-h_\theta) (2 \xvec^{(2)}-1)^{n_\theta}+1) ,
\end{equation}
where $n_\theta=5$ and $h_\theta=0.15$.  $\Theta_2$ focuses on the
thin inflow near the horizon in poloidal field models, while it also
avoids small $\phi$ polar cells that would limit the time step.  The
interpolation factor is
\begin{equation}
W = (1/2) + (1/\pi)(\arctan{( (r-r_{sj2})/r_{0j2})} ) ,
\end{equation}
where $r_{sj2}=5$ and $r_{0j2}=2$.  The polar axis boundary condition
is transmissive as described in Appendix~\ref{sec:nummethods}.

The $\phi$-grid of $N_\phi=64$ cells spans from $0$ to $2\pi$ with
mapping $\phi(\xvec^{(3)}) = 2\pi \xvec^{(3)}$.  Many of our simulations have
$\xvec^{(3)}$ vary from $0$ to $1$ such that $\Delta\phi = 2\pi$.  This is
a fully 3D (no assumed symmetries) domain.  Periodic boundary
conditions are used in the $\phi$-direction.

We choose a resolution $N_r\times N_\theta\times N_\phi$ that has a
grid aspect ratio of 1:1:1 for most of the inner-radial domain.  This
allows the $\phi$ dimension to be treated equally to the $r-\theta$
dimensions.  The aspect ratio (as volume-averaged within
$|\theta-\pi/2|\le [\theta^d]_t$) is given as $A_r$ for radius $r$.
We measure $A_{r_{\rm H}}\approx 1.1:1:8$, $A_{12}\approx 1:1:3$, and
$A_{14}\approx 1:2:3$, where $r_{\rm H}$ is the horizon.  There are
about $14$ vertical cells resolving the disk at the horizon where the
disk thins-out.

\subsection{Resolving the MRI and Turbulence}
\label{sec:numsetupgridmri}

The MRI is a linear instability with fastest growing wavelength of
\begin{equation}\label{lambdamri}
\lambda_{x,\rm MRI} \approx  2\pi \frac{|v_{x,\rm A}|}{|\Omega_{\rm rot}|} , \\
\end{equation}
for $x=\theta,\phi$, where $|v_{x,\rm A}|=\sqrt{\bvec_x
  \bvec^x/\epsilon}$ is the $x$-directed~\alf~speed, $\epsilon\equiv
b^2 + \rhorest + \ug + \pg$, and $r\Omega_{\rm rot} = v_{\rm rot}$.
$\lambda_{\rm MRI}$ is accurate for $\Omega_{\rm rot}\propto r^{-5/2}$
to $r^{-1}$.  $\Omega_{\rm rot},v_{\rm A}$ are separately
angle-volume-averaged at each $r,t$.

The MRI is resolved for grid cells per wavelength
(Eq.~(\ref{lambdamri})),
\begin{equation}\label{q1mri}
\Qx \equiv \frac{\lambda_{x,\rm MRI}}{\Delta_{x}} ,
\end{equation}
of $\Qx\ge 6$, for $x=\theta,\phi$, where $\Delta_{r} \approx
d\xvec^{(1)} (dr/d\xvec^{(1)})$, $\Delta_{\theta} \approx r
d\xvec^{(2)} (d\theta/d\xvec^{(2)})$, and $\Delta_{\phi} \approx r
\sin\theta d\xvec^{(3)} (d\phi/d\xvec^{(3)})$.  Volume-averaging is
done as with $\Qtwo$, except $v_{x,\rm A}/\Delta_{x}$ and
$|\Omega_{\rm rot}|$ are separately $\theta,\phi$-volume-averaged
before forming $\Qx$.  The $t=0$ values and time-averaged values are
measured at same radii as $\Qtwo$, and we find $\Qone\approx 8$ and
$\Qtwo\approx 2$ near the pressure maximum.

The MRI suppression factor corresponds to the number of MRI
wavelengths across the full disk:
\begin{equation}\label{q2mri}
\Qtwo \equiv \frac{2 r \theta^d}{\lambda_{\theta,\rm MRI}} .
\end{equation}
Wavelengths $\lambda<0.5\lambda_{\theta,\rm MRI}$ are stable, so the
linear MRI is suppressed for $\Qtwo<1/2$ when no unstable wavelengths
fit within the full disk \citep{1998RvMP...70....1B,pp05}.  $\Qtwo$
(or $\Qtwoweak$) uses averaging weight $w=(b^2\rholab)^{1/2}$ (or
$w=\rholab$), condition $\beta>1$, and excludes regions where density
floors are activated. Weight $w=(b^2\rholab)^{1/2}$ is preferred,
because much mass flows in current sheets where the magnetic field
vanishes and yet the MRI is irrelevant.  When computing the averaged
$\Qtwo$, $v_{\rm A}$ and $|\Omega_{\rm rot}|$ are separately
$\theta,\phi$-volume-averaged within $\pm 0.2r$ for each $t,r$.  The
averaged $\Qtwo$ is at most $30\%$ smaller than $\Qtwoweak$.

\subsection{Modes and Correlation Lengths}
\label{sec_cor}

The flow structure is studied via the discrete Fourier transform of
$dq$ (related to quantity $Q$) along $x=r,\theta,\phi$ giving
amplitude $a_p$ for $p=n,l,m$, respectively.  The averaged amplitude
is
\begin{equation}\label{am}
\left\langle \left|a_p\right| \right\rangle \equiv \left\langle \left|\mathcal{F}_p\left(dq\right)\right| \right\rangle \equiv \int_{\rm not\ x} \left| \sum_{k=0}^{N-1}dq~{\rm e}^{\frac{-2\pi i p k}{N-1} } \right| ,
\end{equation}
computed at $r=r_{\rm H},4r_g,8r_g,30r_g$.  The $x$ is one of
$r,\theta,\phi$ and ``not x'' are others (e.g. $\theta,\phi$ for
$x=r$).  The $dq$ is (generally) a function of $x$ on a uniform grid
indexed by $k$ of $N$ cells that span: $\delta r$ equal to $0.75r$
around $r$ for $x=r$, $\pi$ for $x=\theta$, and $2\pi$ for
$x=\phi$. The $N$ is chosen so all structure from the original grid is
resolved, while the span covered allows many modes to be resolved.

For all $x$, $dq \equiv \detg d\xvec^{(1)} d\xvec^{(2)} d\xvec^{(3)}
\delta Q/q_N$.  For $x=r,\theta$, we let $q_N=\int_{\rm\ not\ x} \detg
d\xvec^{(1)} d\xvec^{(2)} d\xvec^{(3)} \langle [Q]_t \rangle$,
$\langle [Q]_t \rangle$ as the time-$\phi$ averaged $Q$, and $\delta Q
= Q - \langle[Q]_t\rangle$.  Using $dq$ removes gradients with
$r,\theta$ so the Fourier transform acts on something closer to
periodic with constant amplitude (see also \citealt{bas11}).  For
$x=\phi$, we let $q_N=1$ and $\delta Q = Q$ because the equations of
motion are $\phi$-ignorable.  For $x=\theta,\phi$, the radial integral
is computed within $\pm 0.1r$.  For $x=r,\theta$, the $\phi$ integral
is over all $2\pi$.  For $x=r,\phi$, the $\theta$ integral is over all
$\pi$.  For all $x$ cases, the $\theta$ range of values uses the
``fdc'' or ``jet'' conditions (respectively called ``Disk'' and
``Jet'',
where these conditional regions are defined via $\phi$-averaged
quantities at each time. Notice we average the mode's absolute
amplitude, because the amplitude of $\langle \delta Q\rangle$
de-resolves power (e.g. $m=1$ out of phase at different $\theta$ gives
$\langle \delta Q\rangle \to 0$ and $a_m\to 0$) and is found to
underestimate small-scale structure.

We also compute the correlation length: $\lambda_{x,\rm cor}=x_{\rm
  cor}-x_0$, where $x_0=0$ for $x=\theta,\phi$ and $x_0$ is the inner
radius of the above given radial span for $x=r$, where $n_{\rm cor} =
\delta r/\lambda_{r,\rm cor}$, $l_{\rm cor}=\pi/\lambda_{\theta,\rm
  cor}$, and $m_{\rm cor} =(2\pi)/\lambda_{\phi,\rm cor}$.  The
Wiener-Khinchin theorem for the auto-correlation gives
\begin{equation}\label{mcor}
\exp(-1) = \frac{\mathcal{F}^{-1}_{x=x_{\rm cor}}[\langle|a_{p>0}|\rangle^2]}{\mathcal{F}^{-1}_{x=x_0}[|\langle a_{p>0}|\rangle^2]} ,
\end{equation}
where $\mathcal{F}^{-1}[\langle |a_{p>0}|\rangle ^2]$ is the inverse
discrete Fourier transform of $\langle |a_p|\rangle^2$ but with
$\langle a_0\rangle$ reset to $0$ (i.e. mean value is excluded).

Turbulence is resolved for grid cells per correlation length
(Eq.~(\ref{mcor})),
\begin{equation}\label{Qmcor}
Q_{p,\rm{}cor} \equiv \frac{\lambda_{x,\rm cor}}{\Delta_{x}} ,
\end{equation}
of $Q_{p,\rm{}cor}\ge 6$, for $x=r,\theta,\phi$ and $p=n,l,m$,
respectively.  Otherwise, modes are numerically damped on a dynamical
timescale (even $Q=5$ would not indicate the mode is marginally
resolved, because numerical noise can keep $Q\approx 5$ at increasing
resolution until finally the mode is actually resolved -- finally
leading to an increasing $Q\ge 6$ with increasing resolution ; as seen
by \citealt{sdgn11}).  Reported $Q_{p,\rm{}cor}$ take $1/\Delta_{x}$
as the number of grid cells covering the span of $\lambda_{x,\rm cor}$
as centered on: middle of $x^{(1)}$ within the used radial span for $x=r$, $\theta=\pi/2$ for
$x=\theta$ for the ``Disk'' and $\theta=0$ for $x=\theta$ for the
``Jet'', and anywhere for $x=\phi$.  For $\Delta\phi<2\pi$,
$\Qthree,Q_{m,\rm{}cor}\ll N_\phi$ is required to avoid truncating the
mode.

\subsection{Diagnostics}
\label{sec:diagnostics}

Diagnostics are computed from snapshots produced every $\sim 4r_g/c$.
For quantities $Q$, averages over space ($\langle Q \rangle$) and time
($[Q]_t$) are performed directly on $Q$ (e.g. on $v_\phi$ rather than
on any intermediate values).  Any flux ratio vs. time with numerator
$F_N$ and denominator $F_D$ ($F_D$ often being mass or magnetic flux)
is computed as $R(t)=\langle F_N(t) \rangle / [\langle F_D
\rangle]_t$.  Time-averages are then computed as $[R]_t$.

\subsubsection{Fluxes and Averages vs. Radius}
\label{sec:avgradquant}

For flux density $F_d$, the flux integral is
\begin{equation}
F(r) \equiv \int dA_{23} F_d ,
\end{equation}
where $dA_{23}=\gdet d\xvec^{(2)} d\xvec^{(3)}$ ($dA_{\theta\phi}$ is
the spherical polar version).  For example, $F_d=\rho_0 \uvec^{(1)}$
gives $F=\Mdot$, the rest-mass accretion rate.  For weight $w$, the
average of $Q$ is
\begin{equation}
Q_w(r) \equiv \langle Q \rangle_w \equiv \frac{\int dA_{\theta\phi} w Q}{\int dA_{\theta\phi} w} ,
\end{equation}
All $\theta,\phi$ angles are integrated over.

\subsubsection{Fluxes and Averages vs. $\theta$}
\label{sec:avgthetaquant}

The flux angular distribution, at any given radius, is
\begin{equation}\label{fluxtheta}
F(\theta) = \int_{\theta'=0}^{\pi/2-|\theta-\pi/2|} dA_{\theta'\phi} F_d + \int^{\pi}_{\theta'=\pi/2+|\theta-\pi/2|} dA_{\theta'\phi} F_d  ,
\end{equation}
which just integrates up from both poles towards the equator, is
symmetric about the equator, and gives the total flux value at
$\theta=\pi/2$.  The average of $Q$ vs. $\theta$ using weight $w$ is
given by
\begin{equation}
Q_w(\theta) = \frac{\int dA_{\theta\phi} w Q}{\int dA_{\theta\phi} w} .
\end{equation}
All $\phi$-angles are integrated over.

\subsubsection{Disk Thickness Measurements}
\label{sec:diskthick1}

The disk's geometric half-angular thickness is given by
\begin{equation}\label{thicknesseq}
\theta^d \equiv \left(\left\langle \left(\theta-\theta_0\right)^2 \right\rangle_{\rholab}\right)^{1/2} ,
\end{equation}
where we integrate over all $\theta$ for each $r,\phi$, and $\theta_0
\equiv \pi/2 + \left\langle \left(\theta-\pi/2\right)
\right\rangle_{\rholab}$ is also integrated over all $\theta$ for each
$r,\phi$, and the final $\theta^d(r)$ is from $\phi$-averaging with no
additional weight or $\detg$ factor.  This way of forming
$\theta^d(r)$ works for slightly tilted thin disks or disordered thick disks.
For a Gaussian distribution in density, this satisfies
$\rholab/(\rholab[\theta=0]) \sim \exp(-\theta^2/(2(\theta^d)^2))$.
For sound speed $c_s=\sqrt{\Gamma \pg/(\rhorest + \ug + \pg)}$, the
thermal half-angular thickness is
\begin{equation}\label{thetateq}
\theta^t_w \equiv \arctan{\left(\frac{\langle c_s\rangle_w}{\langle v_{\rm rot} \rangle_w}\right)} ,
\end{equation}
where $v_{\rm rot}^2 = v_\phi^2 + v_\theta^2$.  For a thin hydrostatic
non-relativistic Keplerian (i.e. $v_{\rm rot}=|v_{\rm K}|$ with
$v_{\rm K}\approx R/(a + R^{3/2})$) Gaussian disk,
$\theta^d=c_s/v_{\rm rot}$ for $c_s$ and $v_{\rm rot}$ at the disk
plane.  Also, $\theta^t_{\rholab}\approx 0.93\theta^d$ for
$\Gamma=4/3$.  Ram pressure forces \citep{2005A&A...433..619B} and
magnetic forces \citep{2011ApJ...737...94C} can cause $\theta^d\ll
\theta^t$. Note that ADAFs have
$\theta^t\gtrsim 1$ \citep{nar94}.

\subsubsection{BH, Disk, Jet, Magnetized Wind, and Entire Wind}
\label{integrations}

Many quantities ($Q$) vs. $r$ or vs. $\theta$ or vs. $\phi$ are
considered for various weights and conditions.  We define the
superscript ``f'' (full flow) case as applies for weight $w=1$ with no
conditions, ``fdc'' (full flow except avoids highly magnetized jet
where numerical floors are activated), ``dc'' (disk plus corona but no
jet) case as applies for $w=1$ with condition $b^2/\rho_0<1$,
``dcden'' (density-weighted average) with $w=\rholab$ and no
conditions, ``$\theta^d$'' (within $1$ disk half-angular thickness)
case with $w=1$ and condition of $|\theta-\theta_0|<\theta^d$, ``eq''
(within 3 cells around the equator) case with $w=1$, and ``jet'' or
``j'' case (jet only) with $w=1$ and the condition that density floors
are activated (see Appendix~\ref{sec:nummethods}).  For quantities
vs. $\theta$ or vs. $\phi$, we radially average within $\pm 0.1r$ at
radius $r$.

Fluxes, described in the next section, have integrals computed for a
variety of (somewhat arbitrary) conditions.  The subscript ``BH'' or
``H'' is for all angles on the horizon.  The subscript ``j'' or
``jet'' is for the ``jet'' with condition $b^2/\rho_0\ge 1$.  When the
jet is measured at a single radius, we use $r=50r_g$ (except the MB09Q
model that uses $r=30r_g$ due to its limited radial range).  The
subscript ``mw'' is for the ``magnetized wind'' with conditions
$b^2/\rho_0<1$ and $\beta<2$ for all fluxes, except for the rest-mass
flux that also has $-(\rhorest+\ug+\pg)\uvec_t/\rhorest>1$ (i.e.
thermo-kinetically unbound).  The ``w'' or ``wind'' subscript is for
the ``entire wind'' with the condition of $b^2/\rhorest<1$ that includes
all of the flow except the jet.  The subscript is ``in'' (``out'') for
the condition $u_r<0$ ($u_r>0$).

\subsubsection{Fluxes of Mass, Energy, and Angular Momentum}
\label{fluxes}

The gas rest-mass flux, specific energy flux, and specific angular
momentum flux are respectively given by
\begin{eqnarray}\label{Dotsmej}
\Mdot &=&  \left|\int\rho_0 \uvec^r dA_{\theta\phi}\right| , \\
\emath \equiv \frac{\dot{E}}{[\Mdot]_t} &=& -\frac{\int (T^r_t+R^r_t) dA_{\theta\phi}}{[\Mdot]_t} , \\
\jmath \equiv \frac{\dot{J}}{[\Mdot]_t} &=& \frac{\int (T^r_\phi+R^r_\phi) dA_{\theta\phi}}{[\Mdot]_t} ,
\end{eqnarray}
as computed in Table~\ref{tbls}.

The net flow efficiency is given by
\begin{equation}\label{eff}
  \eff = \frac{\dot{E}-\Mdot}{[\Mdot]_t} = \frac{\dot{E}^{\rm EM}(r) + \dot{E}^{\rm MAKE}(r) + \dot{E}^{\rm RAD}(r)}{[\dot{M_{\rm H}}]_t} . \\
\end{equation}
Positive values correspond to an extraction of positive energy from
the system at some radius.  One can break-up the efficiency into
contributions from each PAKE, EM, and RAD components to give
$\eta_{\rm PAKE}$, $\eta_{\rm EM}$ and $\eta_{\rm RAD}$ as measured at
various locations (horizon, jet, etc.) or radii.  These $\eff$'s are
computed in Table~\ref{tbls}.

The BH's dimensionless spin-up parameter is
\begin{equation}\label{spinevolve}
s \equiv \frac{d(a/M)}{dt}\frac{M}{[\Mdot]_t}  =  -\jmath - 2\frac{a}{M}(1-\eff) ,
\end{equation}
(computed in Table~\ref{tbls}).  All $\theta$ and $\phi$ angles are
integrated over.  The BH is in ``spin equilibrium'' for $s=0$
\citep{2004ApJ...602..312G}.

\subsubsection{Magnetic Flux}
\label{magneticfluxdiag}

The radial magnetic flux vs. $\theta$ at any radius is
\begin{equation}
\Psi_r(r,\theta) = \int dA_{\theta\phi} \Bvec^r .
\end{equation}
The signed value of the maximum absolute value over all $\theta$
angles (${\rm smaxa}_\theta$) of the magnetic flux is
\begin{equation}
\Psi_{\rm t}(r) \equiv {\rm smaxa}_\theta \Psi_r ,
\end{equation}
and $\Psi_{\rm tH} \equiv \Psi_{\rm t}(r=r_{\rm H})$ is the horizon's
magnetic flux.  The half-hemisphere horizon flux is
\begin{equation}
\Psi_{\rm H} \equiv \Psi_r(r=r_{\rm H}) ,
\end{equation}
as integrated from $\theta=\pi/2$ to $\pi$ (negative compared to the
integral from $\theta=0$ to $\pi/2$). The $\theta$ magnetic flux
vs. radius at angle $\theta$ is
\begin{equation}
\Psi_\theta(r,\theta) = \int_{r_{\rm H}}^{r} \detg d\xvec^{(1)} d\xvec^{(3)} \Bvec^{\xvec^{(2)}} ,
\end{equation}
where the vertical magnetic flux threading the equator is
\begin{equation}
\Psi_{\rm eq}(r) \equiv \Psi_\theta(r,\theta=\pi/2) .
\end{equation}
The total magnetic flux along the equator is
\begin{equation}
\Psi(r) \equiv \Psi_{\rm H} + \Psi_{\rm eq}(r) .
\end{equation}
For all forms of $\Psi$, all $\phi$-angles are integrated over.

The magnetic flux can be normalized in various ways (as computed in
Table~\ref{tbls}).  Normalization by the initial flux at $r_0$ gives
$\Psi(r)/\Psi(r_0)$.  One type of field geometry we will use has
multiple field loops of alternating polarity as a function of radius.
So another normalization is by the initial $i$-th extrema vs. radius,
which gives $\Psi/\Psi_i$ that picks up the extrema in the magnetic
flux over each field loop.  Normalization by the initial value of an
extrema gives $\Psi/\Psi_i(t=0)$.  We also need to form a measure that
indicates how much flux is available to the BH.  So we
consider the normalization by the flux in the disk that is immediately
available to the horizon of the same polarity.  This measure is given
by $\Psi_{\rm H}/\Psi_a$, where $\Psi_a$ is the value where $\Psi(r)$
goes through its first extremum of the same sign of magnetic flux
(i.e. out to the radius with the same polarity of dipolar-like field)
as on the horizon.  If the horizon value is itself an extremum, then
$\Psi_{\rm H}/\Psi_a=1$ implying that the region immediately beyond
the horizon only has opposite polarity field.

The absolute magnetic flux ($\Phi$) is computed similarly to $\Psi$,
except one 1) inserts absolute values around the field (e.g. $\Bvec^r$
and $\Bvec^\theta$ in the integrals); 2) puts absolute values around
the integral ; and 3) divides by $2$ so that a dipolar field has
$|\Psi_{\rm t}|=\Phi$.  For example, $\Phi_r(r,\theta) =
(1/2)\left|\int dA_{\theta\phi} |\Bvec^r| \right|$.  The quantity
$\Phi/\Psi_{\rm t}$ (computed in Table~\ref{tbls}, and which is the
only flux ratio directly time-averaged as $[\Phi/\Psi_{\rm t}]_t$) is
roughly the vector spherical harmonic multipole $l$ of the
$\phi$-component of the magnetic vector potential:
\begin{equation}\label{vpot}
\Avpotvec_\phi=\int_{\theta'=0}^\theta \detg \Bvec^r d\theta'
\end{equation}
as integrated over all $\phi$.  For example, for $l=\{1\ldots 8\}$ one
gets $|\Phi/\Psi_{\rm t}| = 1$, $2$, $2.6$, $3.5$, $4$, $5.6$, $5.7$,
and $6.7$.

The \citet{1999ApJ...522L..57G} model normalization gives
\begin{equation}\label{equpsilon}
\Upsilon \approx 0.7\frac{\Phi_r}{\sqrt{[\Mdot]_t}} ,
\end{equation}
which accounts for $\Phi_r$ being in Heaviside-Lorentz units
\citep{pmntsm10}.  Compared to Gaussian units version of $\phi_{\rm
  H}\equiv \Phi_{\rm H}/\sqrt{\Mdot r_g^2 c}$ defined in
\citet{tnm11}, $\Upsilon\approx 0.2\phi_{\rm H}$.  $\Upsilon_{\rm H}$
and $\Upsilon_j$ are normalized by $\dot{M}_{\rm H}$, $\Upsilon_{\rm
  in}$ by $\dot{M}_{\rm in}$, and $\Upsilon_{\rm mw}$ and
$\Upsilon_{\rm w}$ respectively by $\dot{M}_{\rm mw}$ and
$\dot{M}_{\rm w}$.  $\Upsilon$ is computed in Table~\ref{tbls}.

The field line rotation frequency with respect to the BH spin ($z$)
axis is computed various ways.  We consider $\Omega^a_{\rm F} \equiv
\Fvec_{tr}/\Fvec_{r\phi}$, $\Omega^b_{\rm F} \equiv
\Fvec_{t\theta}/\Fvec_{\theta\phi}$, $\Omega^c_{\rm F} \equiv
|\vvec^\phi| + {\rm sign}[\uvec^r] (v_p/B_p)|\Bvec^\phi|$ with
$v_p=\sqrt{v_r^2 + v_\theta^2}$ and $B_p=\sqrt{B_r^2 + B_\theta^2}$,
and
\begin{equation}\label{omegaeq}
\Omega_{\rm F}\equiv \Omega^d_{\rm F} \equiv \vvec^\phi - \Bvec^\phi \left(\frac{v_r B_r + v_\theta B_\theta}{B_r^2 + B_\theta^2}\right) .
\end{equation}
We also consider $\Omega^e_{\rm
  F}=[|\Fvec_{t\theta}|]_t/[|\Fvec_{\theta\phi}|]_t$.  These
$\Omega_{\rm F}$ are normalized by the BH rotation angular
frequency $\Omega_{\rm H}=a/(2Mr_{\rm H})$.

\subsubsection{Inflow Equilibrium and $\alpha$ Viscosity}
\label{sec_infloweq}

Inflow equilibrium is defined as when the flow is in a complete
quasi-steady-state and the accretion fluxes are constant (apart from
noise) vs. radius and time.  The inflow equilibrium timescale is
\begin{equation}\label{tieofrie}
t_{\rm ie} = N \int_{r_i}^{r_{\rm ie}} dr\left(\frac{-1}{[\langle v_r\rangle_{\rholab}]_t}\right) ,
\end{equation}
for $N$ inflow times from $r=r_{\rm ie}$ and $r_i=12r_g$ to focus on
the more self-similar flow.  $t_{\rm ie}$ is used in
Table~\ref{tbls}, where $r^{\rm dcden}_{\rm i} = r_i$, $r^{\rm
  dcden}_{\rm f}=r_{\rm ie}$ with $N=1$, and $r^{\rm
  dcden}_{\rm o}$ uses $r_{\rm ie}$ with $N=3$.

Viscous theory gives a GR $\alpha$-viscosity estimate for $v_r$ of
$v_{\rm visc}\sim -G\alpha(\theta^d)^2 |v_{\rm rot}|$
\citep{pt74,pmntsm10}, with GR correction $G$ ($\lesssim 1.5$ for
$r\gtrsim 58r_g$) and (not the lapse)
\begin{eqnarray}\label{alphaeq}
\alpha &=& \alpha_{\rm PA} + \alpha_{\rm EN} + \alpha_{\rm M1} + \alpha_{\rm M2} , \\
\alpha_{\rm PA} &\approx& \frac{\rho_0 \delta u_r (\delta \uvec_\phi \sqrt{g^{\phi\phi}})}{p_{\rm tot}} , \nonumber\\
\alpha_{\rm M2} &\approx& -\frac{b_r (\bvec_\phi \sqrt{g^{\phi\phi}})}{p_{\rm tot}} , \nonumber\\
\alpha_{\rm mag} &\approx& -\frac{b_r (\bvec_\phi \sqrt{g^{\phi\phi}})}{p_b} ,\nonumber \\
\alpha_{\rm eff} &\equiv& \frac{v_r}{v_{\rm visc}/\alpha} ,\nonumber
\end{eqnarray}
$\alpha_{\rm eff2} \equiv \alpha_{\rm eff}(|v_{\rm rot}|/|v_{\rm K}|)$,
and (small) $\alpha_{\rm EN} \approx (\ug+\pg)
\delta u_r (\delta \uvec_\phi \sqrt{g^{\phi\phi}})/p_{\rm tot}$ and
$\alpha_{\rm M1} \approx b^2 \delta u_r (\delta\uvec_\phi
\sqrt{g^{\phi\phi}})/p_{\rm tot}$.  Here, $\delta u$ is the deviation
of the velocity from its average (taken over all $\phi$ and over the
time-averaging period). The $\alpha$ (e.g. in Table~\ref{tbls}) is
averaged as follows.  The numerator and denominator are separately
volume averaged in $\theta,\phi$ for each $r$.  Weight $w=1$ with
condition $b^2/\rho_0<1$ gives $\alpha_a$ for the disk+corona, while
$w=\rholab$ gives $\alpha_b$ for the heavy disk.  Notice $\alpha_{\rm
  M2} = \alpha_{\rm mag}/(1+\beta_{\rm mag})$ for some $\beta$ denoted
$\beta_{\rm mag}$, and $\sin(2\theta_b)=\alpha_{\rm mag}$ for tilt
angle $\theta_b$ \citep{2011arXiv1106.4019S}.  These $\alpha$'s are
accurate for $|v|\ll c$ as true for $r\gtrsim 2r_g$ in our models, while
$\alpha_{\rm eff}$ is accurate far outside the inner-most stable
circular orbit (ISCO).

\subsubsection{Optical Depth}
\label{opticaldepth}

The optical depth of the flow is computed in two ways.  One way is as
the optical depth away from the polar axis:
\be\label{taua}
\tau_a\approx \int_0^\theta \gamma \kappa_{\rm tot} \sqrt{g_{\theta\theta}}\, \rm d\theta',
\ee
which assumes the flow is mostly radial, relativistic,
and the region near the polar axis is often optically thin at the radius this is computed.
Another way we compute the optical depth is radially:
\be\label{taub}
\tau_b\approx \int_{r_0}^r \kappa_{\rm tot} \sqrt{g_{rr}}/(2\gamma)\, \rm dr' ,
\ee
which assumes the flow is mostly radial, relativistic, and the region at large
distance is optically thin or has no additional structure that would
affect the optical depth.
The flow's radiative photosphere is then defined as either $\tau_a=1$
or $\tau_b=1$.

The luminosity of the accretion system is computed as the radiative
flux emerging from some chosen radius via
\begin{equation}
L_{\rm rad} = \int dA_{\theta\phi} R^r_t ,
\end{equation}
where we only include those angles where the gas is optically thin
(i.e. only that gas that has $\tau_a<1$).

\subsection{Initial and Evolved Disk Structure}

\begin{figure}
\centering
\includegraphics[width=3.2in,clip]{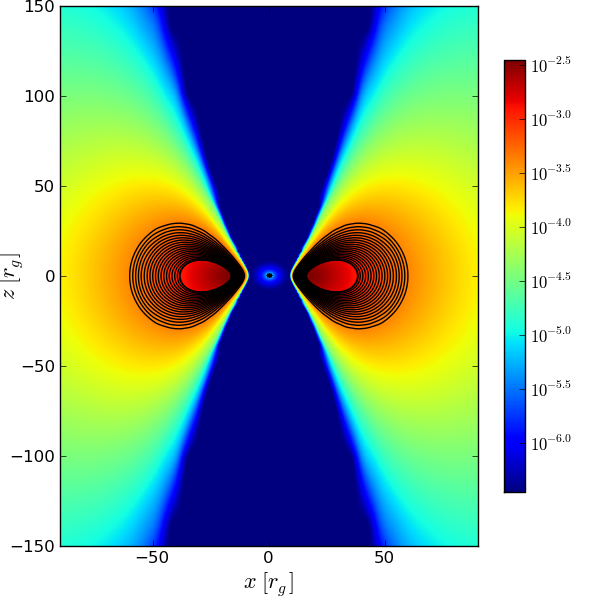}
\caption{The fiducial model's initial ($t=0$) state consists of a
  weakly magnetized geometrically thick torus around a spinning
  ($a/M=0.9375$) BH. $\rho_0$ is shown as color with legend.  Black
  lines show $\Avpotvec_\phi$ (integrated over all $\phi$) composed of
  a single set of field line loops with a single polarity.}
\label{initial3plot}
\end{figure}

\begin{figure}
\centering

\includegraphics[width=3.2in,clip]{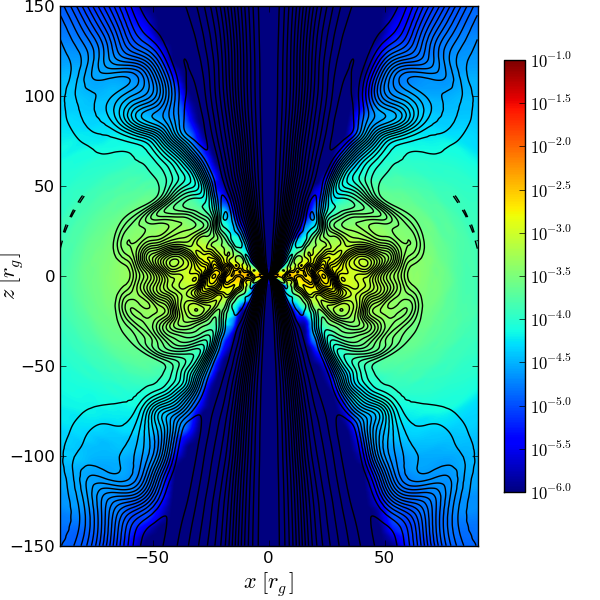}
\caption{The evolved ($t\approx 5600r_g/c$) state of the fiducial
  model (otherwise similar to Fig.~\ref{initial3plot}) consists of
  strongly magnetized gas near the BH that launches a jet.}
\label{middle3plot}
\end{figure}

Fig.~\ref{initial3plot} and Fig.~\ref{middle3plot} show color
plots of $\rhorest$ and field line contours (contours of $\Avpotvec_\phi$
integrated over $\phi$, so is axially symmetric) for the initial and
quasi-steady-state evolved solution, respectively.  The initial
solution consists of a radially extended thick torus within which
a single weak field loop (of single poloidal polarity) is
embedded.  The disk is geometrically thick with $\theta^d\sim 0.4$.

The evolved solution, shown in Fig.~\ref{middle3plot}, show the
simulation when the region within $r\sim 14r_g$ has become
quasi-steady.  The inner part of the poloidal field loop has accreted
onto the black hole.  A plot of the radiation energy density closely
follows that for the rest-mass density.

\begin{figure}
\centering
\includegraphics[width=3.0in,clip]{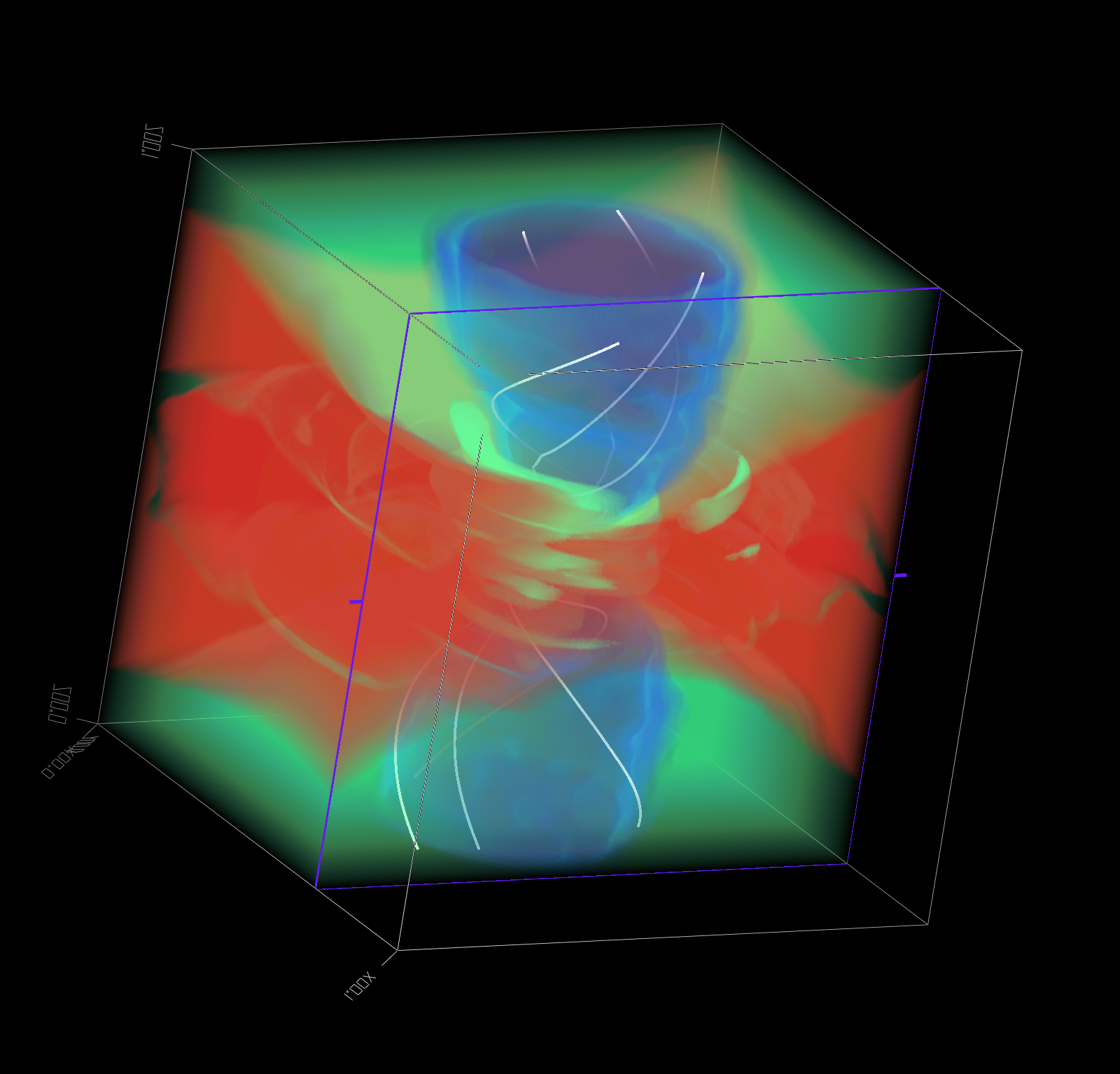}
\caption{Three-dimensional rendering (with clipping near middle, shown
  as purple square) of three main structural elements of the flow from
  near the black hole out to $\pm 10r_g$ in each direction of a
  Cartesian box for the evolved ($t\approx 5600r_g/c$) state of the
  fiducial model.  Figure shows the radiation-dominated disk component
  (orange-red volume rendering), the hot disk-corona gas component
  (green-cyan volume rendering) that comes into equipartition with the
  radiation energy density outside the disk, and the relativistic
  highly magnetized jet component (blue-purple volume rendering with
  white magnetic field lines).}
\label{pic3d}
\end{figure}

Fig.~\ref{pic3d} shows a 3D rendering of the flow's three main
structural elements (hot radiation-dominated component, hot gas, and
relativistic magnetized jet).

\subsection{Overall Time Dependence}\label{sec:timedep}

\begin{figure*}
\centering
\includegraphics[width=6.6in,clip]{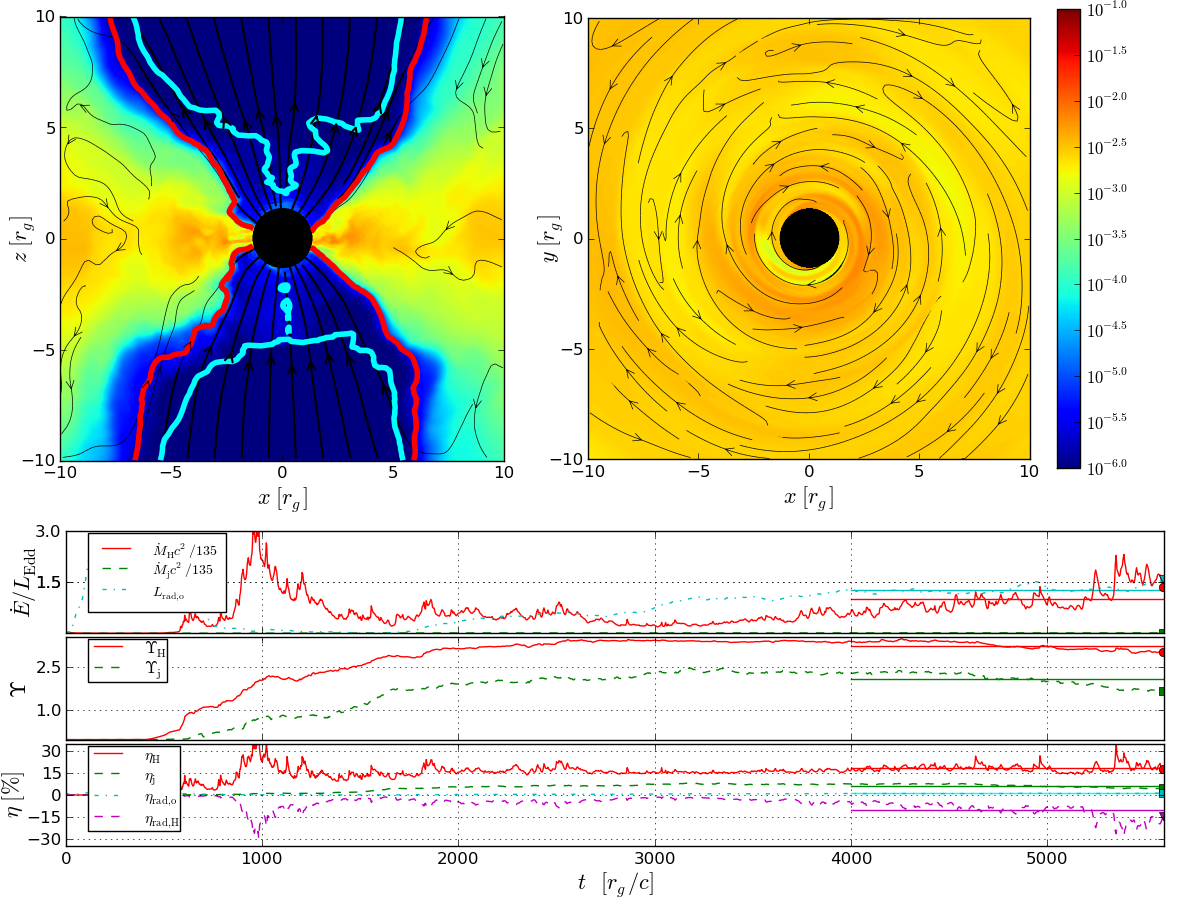}
\caption{Evolved snapshot (see Supporting Information and \href{http://www.youtube.com/watch?v=q2DeKxUHae4}{this link} for the movie)
  of the fiducial model at $t\approx 5600r_g/c$ showing log of
  rest-mass density in color (see legend on right) in both the $z-x$
  plane at $y=0$ (top-left panel) and $y-x$ plane at $z=0$ (top-right
  panel).  Black lines trace field lines, where thicker black lines
  show where field is lightly mass-loaded.  In the top-left panel, the
  thick red line corresponds to where $b^2/\rhorest=1$, while the
  thick cyan line corresponds to where the optical depth (away from
  the polar axis) reaches unity (i.e. $\tau_a=1$ as given by
  Eq.~(\ref{taua})).  The bottom panel has 3 subpanels.  The top
  subpanel shows $\Mdot$ through the BH ($\dot{M}_{\rm H}$), out in
  the jet ($\dot{M}_{\rm j}$, at $r=50r_g$), and radiative luminosity
  ($L_{\rm rad,o}$, from optically thin region at $r=50r_g$) with
  legend.  All quantities have been normalized by the Eddington
  luminosity ($L_{\rm Edd}$), where in addition all mass fluxes are
  normalized by the time-averaged value of $\dot{M}_{\rm H}/L_{\rm
    Edd}\approx 135$ so that all quantities can be shown on a single
  panel.  The middle subpanel shows $\Upsilon$ for similar conditions.
  The bottom subpanel shows the efficiency ($\eta$) for similar
  conditions, where $\eta_{\rm rad, o}$ corresponds to the radiative
  efficiency from the radiation escaping out of the optically thin
  regions at large radii, while $\eta_{\rm rad, H}$ is for the
  radiation that ends up trapped and absorbed by the black hole (so it
  is negative).  Horizontal solid lines of the same colors show the
  averages over the averaging period, while square/triangle/circle
  tickers are placed at the given time and values.  In summary, for
  super-Eddington accretion at $\dot{M}c^2/L_{\rm Edd}\sim 100$, the
  total BH efficiency is moderate at $\eta\sim 20\%$, while the
  radiative efficiency is quite low at $\eta_{\rm rad, o}\sim 1\%$.}
\label{evolvedmovie}
\end{figure*}

Fig.~\ref{evolvedmovie} shows a typical snapshot for the rest-mass
density, field lines, and fluxes ($\Mdot$, $\Upsilon$, and $\eta$)
on the BH, through $r=50r_g$ in the jet, and at $r=50r_g$ in the
magnetized wind.  The BH's magnetic flux dominates the mass influx
with $\Upsilon_{\rm H}\approx 3$ during the quasi-steady-state period.
Because $\Upsilon\gtrsim 1$, one expects the Blandford-Znajek (BZ)
effect to be activated, and the energy extraction efficiency is
moderate at $\eta\sim 20\%$.  Much of the energy extracted from the BH
reaches the jet at large radii (i.e. $\eta_{\rm j}\sim 0.3\eta_{\rm
  H}$).

\begin{figure}
\centering
\includegraphics[width=3.1in,clip]{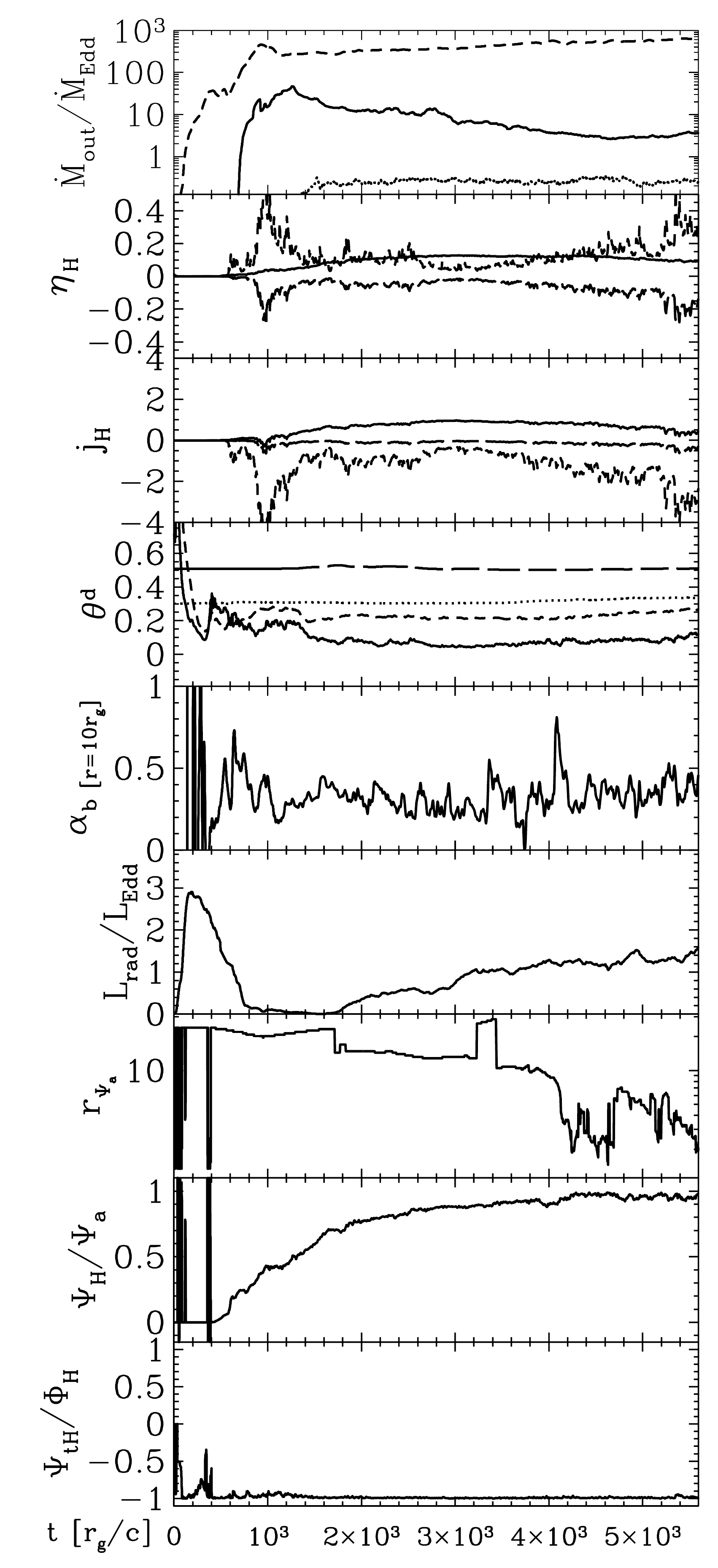}
\caption{Quantities vs. time.  Top Panel: $\dot{M}_{\rm out}$ for
  magnetized wind ($\dot{M}_{\rm mw,o}$, solid line, and mostly middle
  line), entire wind ($\dot{M}_{\rm w,o}$, short-dashed line, and
  upper-most line), and jet ($\dot{M}_{\rm j}$, dotted line, and
  lowest line).  Next Panel: BH efficiency ($\eta_{\rm H}$) for EM
  (solid line, mostly middle line), MAKE (short-dashed line, mostly
  upper line) terms, RAD (long-dashed line, lower line) terms. Next
  Panel: BH specific angular momentum ($\jmath_{\rm H}$) for EM (solid
  line, upper line), MAKE (short-dashed line, lower line) terms, and
  RAD (long-dashed line, middle line) terms. Next Panel: $\theta^d$ at
  $r=\{r_{\rm H}/r_g,5,20,100\}r_g$ with, respectively, lines:
  \{solid, short-dashed, dotted, long dashed\} corresponding to the
  lowest to upper-most lines. Next panel: $\alpha_b$ at $r=10r_g$.
  Next Panel: Radiative luminosity at large distances ($L_{\rm
    rad,o}$, at $r=50r_g$ from the optically thin region).  Next
  Panel: $r_{\Psi_a}$ for the radius out to where there is the same
  magnetic polarity as on the hole (solid line). Next panel: Magnetic
  flux on the BH per unit flux available in the flow with the same
  polarity: $\Psi_{\rm H}(t)/\Psi_a(t)$.  Bottom panel: $\Psi_{\rm
    tH}(t)/\Phi_{\rm H}(t)\sim 1/l$, for $l$ mode of vector spherical
  harmonic multipole expansion of $\Avpotvec_\phi$.  In summary, the
  flow has reached a quasi-steady-state at late times.  While the
  black hole efficiency is order $20\%$, the radiation emitted at
  large radius from the optically thin region only has an efficiency
  of order $1\%$ and is of order the Eddington luminosity.}
\label{fluxvst}
\end{figure}

Fig.~\ref{fluxvst} shows various quantities vs. time.  All
quantities are in a quasi-steady-state for $t\gtrsim 3000r_g/c$.  The
mass ejected in the circulating wind ($\dot{M}_{\rm w,o}$ dominates
the magnetized wind ($\dot{M}_{\rm mw,o}$) and jet ($\dot{M}_{\rm j}$)
at large radii ($r_{\rm o}=50r_g$ here); see \S\ref{integrations} for
definitions of various outflow components.  The MAKE term most often
dominates the EM term in $\eta_{\rm H}$ and ${\jmath}_{\rm H}$.  The
RAD term is always negative, indicating absorption of positive energy
radiation into the BH.  The flow has a moderate average total
efficiency of $\eta_{\rm H}\sim 20\%$.  Note that the MAKE term is
composed of a particle term (i.e. $\eta^{\rm PAKE}=1+\uvec_t$) and an
enthalpy term (i.e. $\eta^{\rm EN}=\uvec_t(\ug+\pg)/\rho_0$).

The $\alpha$-viscosity parameter holds steady at about $\alpha_b\sim
0.5$.  $\Upsilon\sim 3$ in the pure inflow ($u_r<0$ only) available at
large radii.  The value is similar to that on the black hole.  This
can be understood by looking at the value of $r_{\Psi_a}\sim r_{\rm
  H}$ by $t\sim 3000$, which shows that most of the magnetic flux that
is available is already on the horizon.  This is also evident by
looking at $\Psi_{\rm H}(t)/\Psi_a(t)$ (i.e. ratio of time-dependent
fluxes) corresponding to [the flux on the hole] per unit [flux on the
  hole plus available of the same polarity just beyond the hole].
$\Psi_{\rm H}(t)/\Psi_a(t)\sim 1$ is reached by $t\sim 3000$, after
which there is no more magnetic flux available to feed the black hole
or disk.  Finally, $|\Psi_{\rm tH}(t)/\Phi_{\rm H}(t)|\sim 1$, which
shows that the horizon's field is dipolar ($l\approx 1$).

\subsection{Time-Averaged Radial ($r$) Dependence}\label{sec:velvisc}

\begin{figure}
\centering
\includegraphics[width=3.2in,clip]{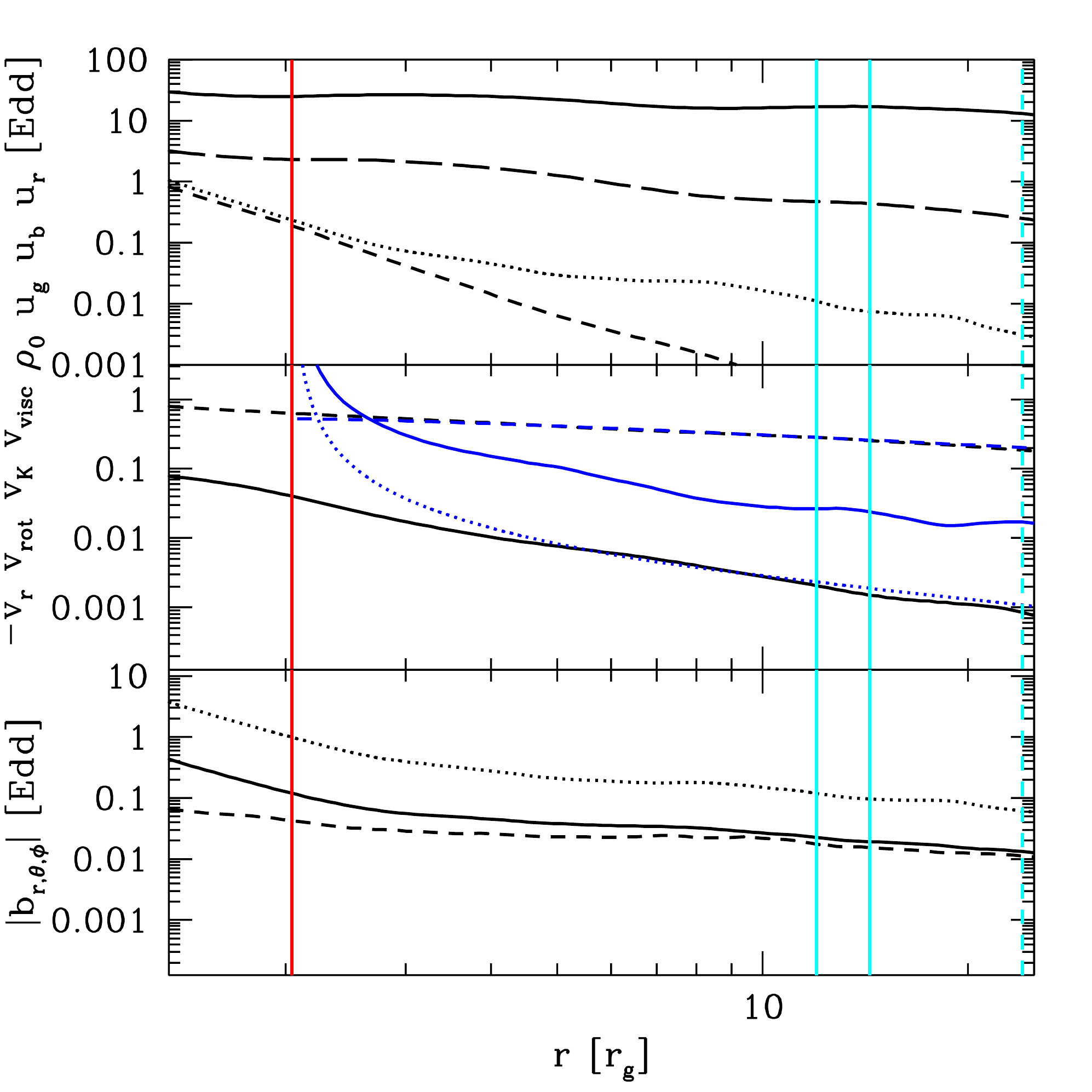}
\caption{The time-angle-averaged densities, 3-velocities, and 4-field
  strengths using a density-weighted average.  Top panel shows
  rest-mass density ($\rho_0$) as black solid line, internal energy
  density ($\ug$) as black short-dashed line, magnetic energy density
  ($u_b$) as black dotted line, and radiation energy density in the
  radiation frame ($\bar{E}$) as long-dashed line.  Middle panel shows
  negative radial velocity ($-v_r$) as black solid line, rotational
  velocity ($v_{\rm rot}$) as black short-dashed line, Keplerian
  rotational velocity ($v_{\rm K}$) as blue short-dashed line, and
  $\alpha$-viscosity theory radial velocity ($v_{\rm visc}$) when
  using $p_{\rm b}$ in denominator for $\alpha=\alpha_b$ (blue solid
  line) and when choosing a fixed $\alpha(\theta^d)^2=0.003$ (blue
  dotted line).  Bottom panel shows comoving 4-field spatial
  components with $r$, $\theta$, and $\phi$ components shows as solid,
  short-dashed, and dotted black lines, respectively.  The vertical
  red line marks the ISCO.  Vertical solid cyan lines show range from
  $r=12r_g$ to $3$ inflow times.  The short-dashed vertical cyan line
  marks a single inflow time.  In summary, the flow near the black
  hole has a flat rest-mass density profile, and the magnetic energy
  density and radiation energy density dominate the internal energy
  density.  Also, the rotational velocity is very close to Keplerian.}
\label{rhovelvsr}
\end{figure}

Fig.~\ref{rhovelvsr} shows the time-averaged densities,
3-velocities, and comoving 4-fields vs. radius using a
density-weighted average to focus on heavy disk material.  The
solution is in inflow equilibrium ($3$ inflow times; see
section~\ref{sec_infloweq}) only out to $r\sim 14r_g$.  Beyond the BH,
the rest-mass density is quite flat as expected for a flow supported
by radiation pressure at super-Eddington accretion rates.  The
rotational velocity is very close to Keplerian.

The GR viscosity estimate for $v_r$ denoted $v_{\rm visc}$ (see above
Eq.~(\ref{alphaeq})) overestimates the simulation $v_r$ when using the
$\alpha$-viscosity with total pressure, where a better match is
obtained using only magnetic pressure.  If we set
$\alpha(\theta^d)^2\to 0.003$ at all radii, then $|v_{\rm visc}|\approx
|v_r|$ outside the ISCO and inside the inflow equilibrium region.

\begin{figure}
\centering
\includegraphics[width=2.9in,clip]{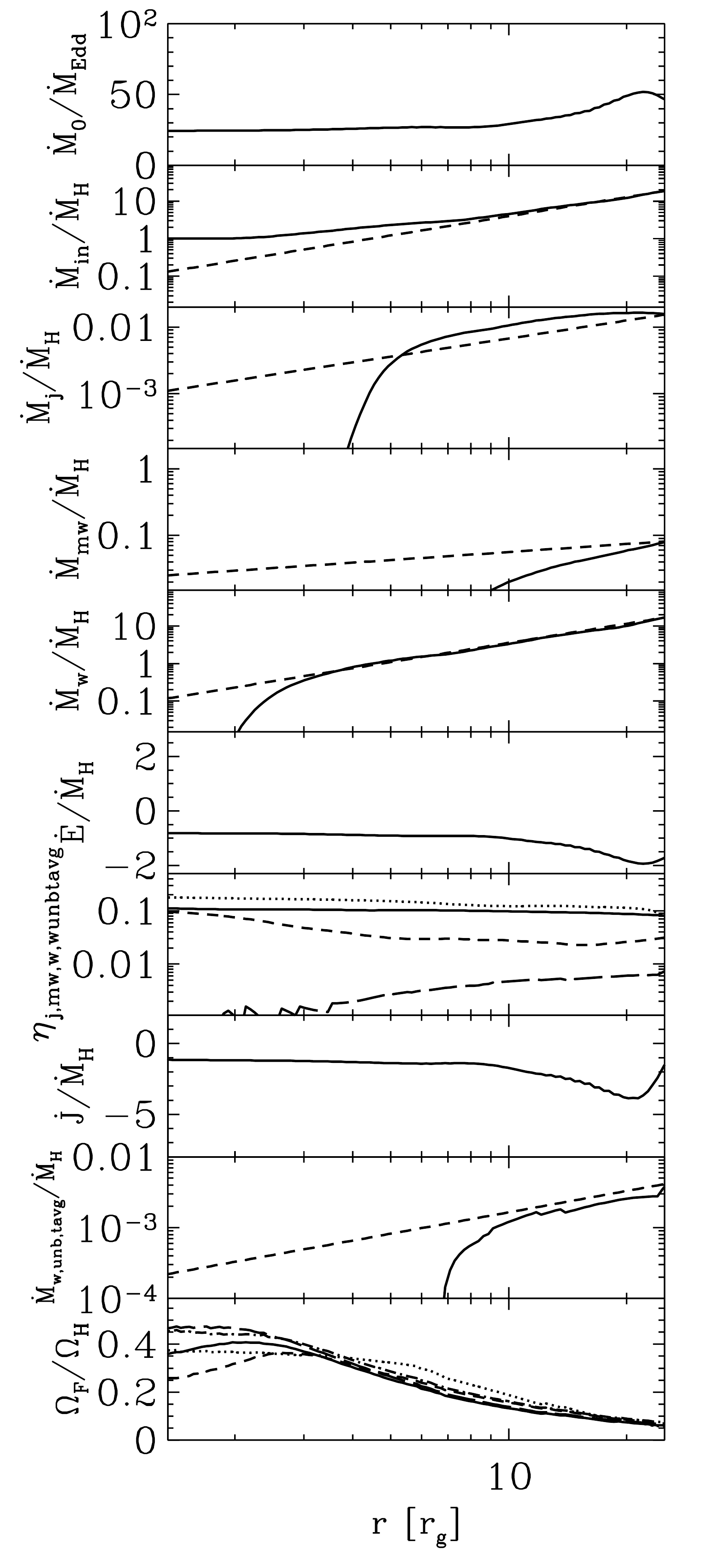}
\caption{The time-averaged angle-integrated fluxes.  From top to
  bottom, panels are: Total mass accretion rate ($\Mdot$), inflow rate
  ($\dot{M}_{\rm in}$), jet outflow rate ($\dot{M}_{\rm j}$),
  magnetized wind outflow rate ($\dot{M}_{\rm mw}$), entire wind
  outflow rate ($\dot{M}_{\rm w}$), total specific energy accretion
  rate ($\dot{E}/\dot{M}_{\rm H}$), efficiency for the jet (solid
  line) magnetized wind (short-dashed line) and wind (dotted line) and
  true unbound wind (long-dashed line, lowest line), total specific
  angular momentum accretion rate ($\jmath=\dot{J}/\dot{M}_{\rm H}$),
  true unbound wind mass outflow rate $\dot{M}_{\rm{}w,unb,tavg}$
  (solid line) and its power-law fit with power-law index $1.0$
  (short-dashed line), and field line angular rotation frequency per
  unit BH angular frequency ($\Omega_{\rm F}/\Omega_{\rm H}$) for
  time-averaged versions of $\Omega^d_{\rm F}$ (solid line),
  $\Omega^c_{\rm F}$ (short-dashed line), $\Omega^e_{\rm F}$ (dotted
  line), $|\Omega^d_{\rm F}|$ (long-dashed line), and $|\Omega^c_{\rm
    F}|$ (dot-short-dashed line).  These $\Omega_{\rm F}$ are averaged
  within the disk+corona part of the flow.  Power-law fits for mass
  inflow and outflow rates are shown as short-dashed lines.  In
  summary, inflow equilibrium is achieved out to $r\sim 14r_g$, and
  there is a small true net mass outflow of unbound material.}
\label{fluxvsr}
\end{figure}

Fig.~\ref{fluxvsr} shows the fluxes (see section~\ref{fluxes})
vs. radius as well as the field line angular rotation frequency
$\Omega_{\rm F}$ (using various definitions defined in
section~\ref{magneticfluxdiag}).  These quantities are associated with
conserved quantities such that ratios of total fluxes would be
constant along flow-field lines in stationary ideal MHD.  The total
fluxes are constant out to $r\sim 14r_g$, the inflow equilibrium
radius for this short duration simulations.  Also shown are the
components (inflow, jet, magnetized wind, and entire wind) of the mass
and energy flow.  The mass inflow and outflow at large radii somewhat
follow power-laws after sufficient averaging over turbulent eddies.
The jet efficiency is order $10\%$ and is constant at large radii.

Power-law fits over the outer-radial domain (including the region not
actually in inflow equilibrium) for the mass flow rates are
$\Mdot\propto r^{1.7}$ for the inflow and entire wind, $\Mdot\propto
r^{0.9}$ for the jet.  A fit of $\Mdot\propto r^{0.4}$ is shown for
the magnetized wind, but the radial range in equilibrium is not
sufficient to check this fit.

The quantity $\dot{M}_{\rm{}w,unb,tavg}$ is the true unbound wind
computed from time-$\phi$-averaged versions of fluxes and the
time-$\phi$ averaged value of $u_t(\rhorest+\ug+\pg)/\rhorest<-1$,
such that any circulation is eliminated from the calculation
\citep{2012MNRAS.426.3241N,2013MNRAS.436.3856S}.  This includes both
ingoing and outgoing flow (i.e. we don't choose the flow component
based upon $u^r$), so this gives a conservative estimate of how much
(if any) net unbound mass is flowing out.  This calculation works
because most of the disk starts (and remains) thermally bound.  A fit
of $\dot{M}_{\rm{}w,unb,tavg}\propto r^{1}$ is shown for the true
unbound wind, but the radial range in equilibrium is not sufficient to
check this fit.

The field line angular frequency $\Omega_{\rm F}\sim \Omega_{\rm H}/4$
(as in BZ77's paraboloidal model) in the disk+corona+wind
(i.e. ``fdc'' averaging, for full flow except the highly-magnetized
jet).

\begin{figure}
\centering
\includegraphics[width=2.9in,clip]{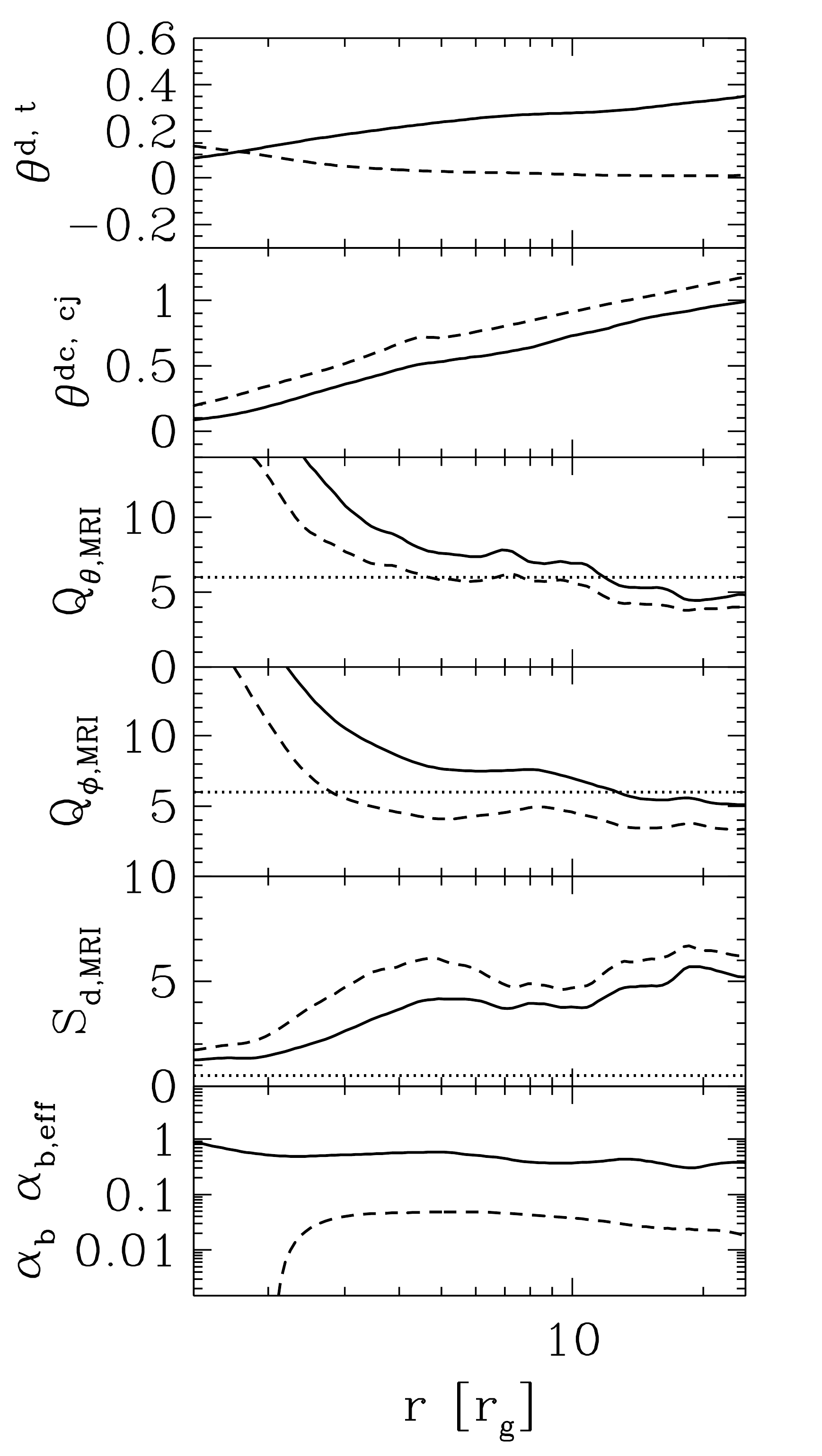}
\caption{Other time-angle-averaged quantities.  From top to bottom,
  panels are: density half-angular thickness ($\theta^d$, solid line)
  and thermal half-angular thickness ($\theta^t$, short-dashed line),
  disk-corona interface angle ($\theta^{dc}$, solid line) and
  corona-jet interface angle ($\theta^{cj}$, short-dashed line),
  number of cells per fastest growing MRI wavelength ($\Qone$, solid
  line ; $\Qoneweak$, short-dashed line) with $\Qone=6$ shown as
  dotted line above where the vertical ($\theta$) MRI is resolved,
  number of cells per fastest growing MRI wavelength ($\Qthree$, solid
  line ; $\Qthreeweak$, short-dashed line) with $\Qthree=6$ shown as
  dotted line above where the azimuthal ($\phi$) MRI is resolved,
  number of fastest growing MRI wavelengths across the full disk
  thickness ($\Qtwo$, solid line ; $\Qtwoweak$, short-dashed line)
  with $\Qtwo=1/2$ shown as dotted line below where the MRI is
  suppressed, and viscosity parameter ($\alpha_b$, solid line ;
  $\alpha_{b,\rm eff}$, short-dashed line).  In summary, the
  linear MRI is active and numerically marginally resolved.}
\label{othersvsr}
\end{figure}

Fig.~\ref{othersvsr} shows the time-averages for the disk's
geometric half-angular thickness ($\theta^d$), the thermal
half-thickness ($\theta^t$, using the density-weighted average), flow
interface angular locations, resolution of the MRI wavelength, and
approximate $\alpha$ viscosity parameter.  The disk-corona and
corona-jet interfaces trace the path of the well-collimated jet out to
large radii.  The $\Qone\gtrsim 6$ and $\Qthree\gtrsim 10$ as required
to marginally resolve the MRI \citep{sano04}.  The value of $\Qtwo\sim
4$, indicating that the magnetic field has not reached the saturated
state where a magnetically arrested disk (MAD) or magnetically choked
accretion flow (MCAF, that occurs at high spin) would form.

\subsection{Time-Averaged Angular ($\theta$) Dependence}\label{sec:thetadep}

\begin{figure}
\centering
\includegraphics[width=2.9in,clip]{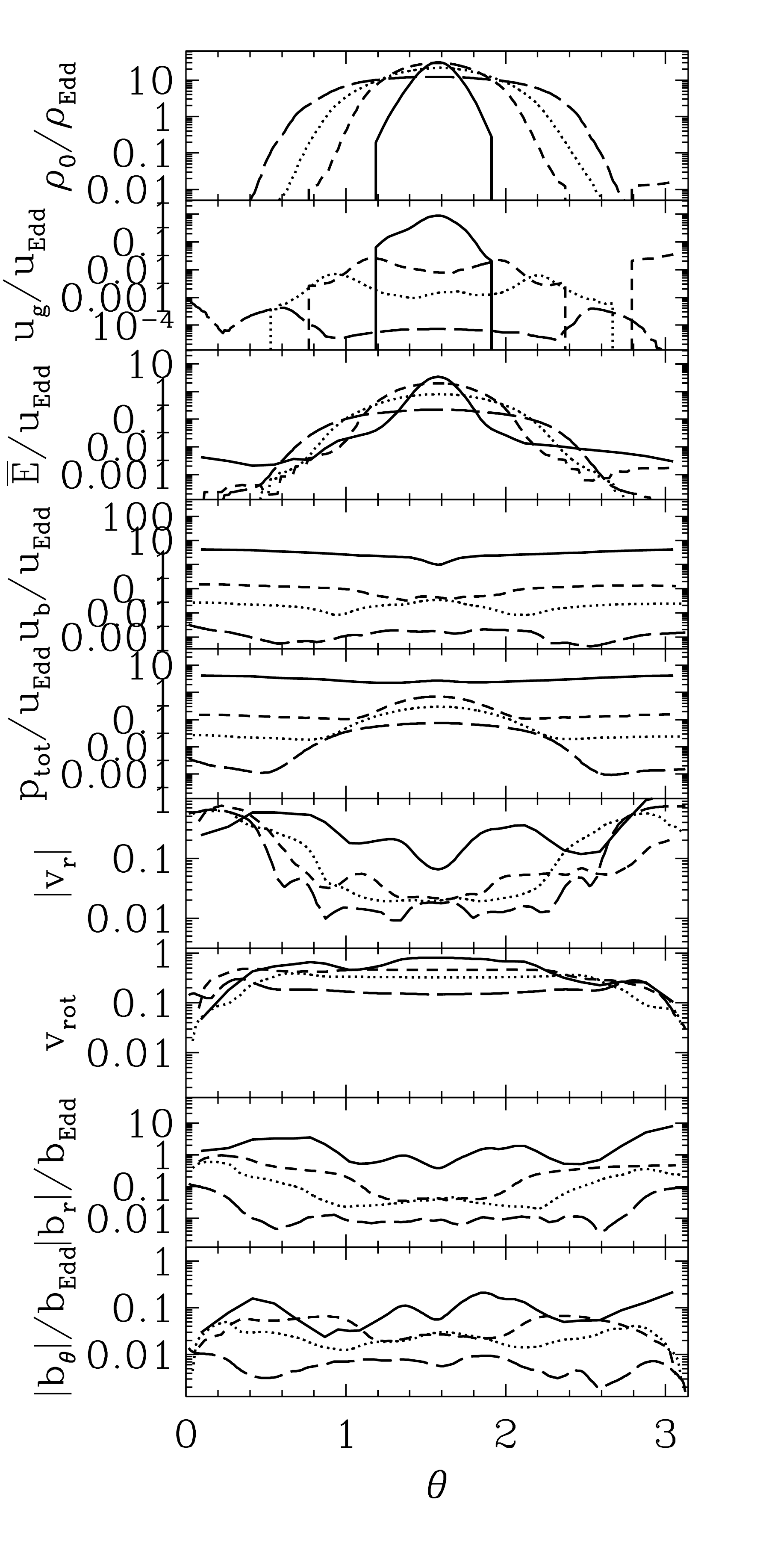}
\caption{Similar quantities as in Fig.~\ref{rhovelvsr}, except
  plotted vs. $\theta$ at $r=\{r_{\rm H}/r_g,4,8,30\}r_g$
  (respectively: solid, short-dashed, dotted, and long dashed lines).
  If numerical density floors were activated at some space-time point,
  then $\rho_0=\ug=0$ was set there.  In summary, the disk is broader
  than Gaussian and is supported by radiation pressure.}
\label{rhovelvsh}
\end{figure}

Fig.~\ref{rhovelvsh} is similar to Fig.~\ref{rhovelvsr} but for
quantities vs. $\theta$ at four different radii.  The time-averaged
density is flatter than a Gaussian distribution.  The figure also
shows that the total pressure is roughly constant with angle as
dominated by the radiation pressure.

\begin{figure}
\centering
\includegraphics[width=2.9in,clip]{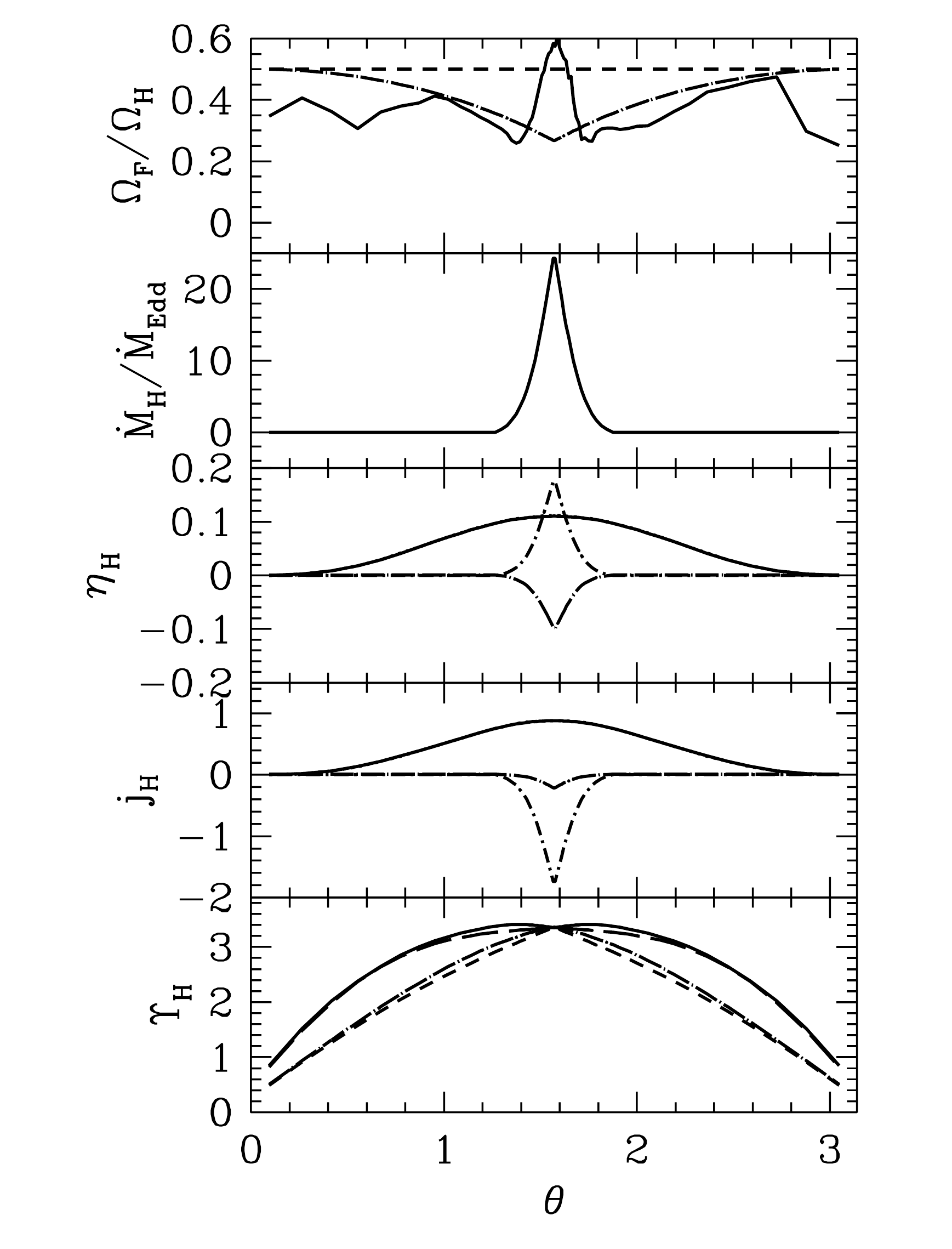}
\caption{Time-$\phi$-averaged quantities and flux integrals on the
  horizon as a function of $\theta$.  From top to bottom: 1) Field
  line rotational angular frequency ($\Omega_{\rm F}/\Omega_{\rm H}$)
  for simulation (solid line), 1st-order-in-spin accurate value for
  monopolar (short-dashed line) and paraboloidal (dot-long-dashed
  line) BZ solutions ; 2) Rest-mass flux ($\dot{M}_{\rm H}$) ; 3)
  Electromagnetic (EM, solid line) and matter (MA, dot-short-dashed
  line, which rises at equator) efficiency and radiation (RAD,
  dot-long-dashed line, which drops at equator) efficiency ($\eta_{\rm
    H}$), along with the full BZ-type EM formula without any
  renormalization (dotted line, which overlaps very well with solid
  line) ; 4) Electromagnetic (EM, solid line) and matter (MA,
  dot-short-dashed line, which drops the most at equator) and
  radiation (RAD, dot-long-dashed line, which drops the least at
  equator) specific angular momentum flux (${\jmath}_{\rm
    H}=\dot{J}_{\rm H}/\dot{M}_{\rm H}$), along with the full BZ-type
  EM formula without any renormalization (dotted line, which overlaps
  very well with solid line); 5) Gammie parameter ($\Upsilon_{\rm H}$)
  for the simulation (solid line), and the BZ model for the cases:
  0th-order-in-spin accurate monopolar field (short-dashed line),
  0th-order-in-spin accurate paraboloidal field (dot-long-dashed
  line), and 2nd-order-in-spin accurate monopolar field (long-dashed
  line).  These BZ versions are normalized so total magnetic flux is
  the same as in the simulation.  Notice how the 2nd-order-in-spin
  accurate monopolar BZ model fits the simulation result quite well.
  For the last 3 panels, the divisor is (implicitly) $\dot{M}_{\rm H}$
  that has been fully angle-integrated to a single value.  So,
  $\eta_{\rm H}$, $j_{\rm H}$, and $\Upsilon_{\rm H}$ show the angular
  dependence of $\dot{E}_{\rm H}$, $\dot{J}_{\rm H}$, and $\Psi_{\rm
    H}$, respectively.  In summary, the agreement between the
  simulation and the BZ picture is excellent.}
\label{horizonflux}
\end{figure}

Fig.~\ref{horizonflux} shows the horizon's values of quantities
related to the BZ effect \citep{bz77}.  The simulation's fluxes are
computed via Eq.~(\ref{fluxtheta}).  The ``full BZ-type EM formula''
referred to in the figure uses the EM energy flux computed from
equation~33 in \citet{2004ApJ...611..977M}, which only assumes stationarity and
axisymmetry (rather than also small spin in BZ77) and uses the
simulation's $\Omega_{\rm F}(\theta)$ and $\Bvec^r(\theta)$ on the
horizon.  This figure shows that most of the horizon is highly
magnetized due to accretion occurring through a magnetically
compressed inflow.

The agreement between the simulations and the BZ picture is excellent
for the highly magnetized regions, where roughly $\Omega_{\rm F}\sim
\Omega_{\rm H}/2$ near the disk-jet interface (here, $\Omega_{\rm F}$
is the time-average of Eq.~(\ref{omegaeq})).  While the simulation is
roughly consistent with BZ's paraboloidal solution, the equatorial
$\Omega_{\rm F}$ is somewhat suppressed due to the disk inflow, as
expected \citep{1999ApJ...522L..57G}.  Also, near the polar axes,
$\Omega_{\rm F}$ is affected by ideal MHD effects and numerical floor
mass injection.

\subsection{Space-Time Averaged Fluxes, Viscosities, Numerical Quality Factors}\label{sec:tables}

Lastly, we show a summary of diagnostics computed as described in
section~\ref{sec:diagnostics}, and a similar set of diagnostics were
computed in \citet{2012MNRAS.423.3083M} and can be compared.  The diagnostics were
taken from time-averages computed between $t=4000M$ and $5596M$ for
this model we identify as A0.94BpN100L20, which identifies it as
having spin (A) approximately $a/M\approx 0.94$, a magnetic field (B)
that is poloidal (p) and the field is normalized (N) to have
$\beta_{\rm min}\approx 100$ at $t=0$, and the mass accretion rate per
unit Eddington (L) is of order $\Mdot/\Mdotedd\sim 20$ (the
$\Mdot/L_{\rm Edd}\sim 100$).  Quantities labelled $i$ for
``inner'' are measured at $r\sim 10r_g$, while quantities labelled $o$
are measured at $r\sim 50r_g$.  The mass inflow ($\dot{M}_{\rm in,o}$)
and total wind mass flow ($\dot{M}_{\rm mw,o}$) measurements at the
outer radius are not in equilibrium due to the short duration of the
simulation, but we keep the measurements for comparison with tables in
\citet{2012MNRAS.423.3083M}.

For the rest-mass fluxes and ejection rates, the rest-mass fluxes are
normalized by $\Mdotedd$, while the luminosity in the last
column is normalized by $\dot{L}_{\rm Edd}$.  These values show that
the mass flow is super-Eddington, while the radiative output at large
distances (here measured at $r=50r_g$ in the optically thin region) is
only near the Eddington rate.

The wind quantities like $\dot{M}_{\rm{}mw,i}$ (magnetized unbound
wind at $r=10r_g$), $\dot{M}_{\rm{}mw,o}$ (magnetized unbound wind at
$r=50r_g$), $\dot{M}_{\rm{}w,i}$ (total wind at $r=10r_g$), and
$\dot{M}_{\rm{}w,o}$ (total wind at $r=50r_g$) were computed based
upon measurements of fluxes at each instant having $u^r>0$.  However,
the flow circulates and much of that motion cancels-out.

So we also compute the quantities like $\dot{M}_{\rm{}w,unb,tavg,i}$
(unbound wind at $r=10r_g$) and $\dot{M}_{\rm{}w,unb,tavg,o}$ (unbound
wind at $r=50r_g$) based upon first time-$\phi$-averaging the fluxes,
time-$\phi$-averaging $u_t(\rhorest+\ug+\pg)/\rhorest$ that is $<-1$
for an unbound flow, and only {\it then} computing the spatial
integral.  This avoids including any short-period circulations and
measures the residual outflow from the accretion flow
\citep{2012MNRAS.426.3241N,2013MNRAS.436.3856S}.  This includes both
ingoing and outgoing flow (i.e. we don't choose the flow component
based upon $u^r$), so this gives a fairly conservative estimate of how
much (if any) net unbound mass is flowing out.  Not using $u^r>0$ as a
restriction works because most of the disk starts (and remains)
thermally bound and so the disk does not contribute to this
measurement.

The true unbound wind at $r\sim 50r_g$ is
$\dot{M}_{\rm{}w,unb,tavg,o}\sim 0.1\Mdotedd$, which is only about
$1\%$ of the mass that reaches the black hole.  This is probably an
upper limit (see \citealt{2012MNRAS.426.3241N,2013MNRAS.436.3856S}).

The efficiency values show that the black hole has an efficiency of
about $\eta_{\rm H}\approx 19\%$, which is quite similar to expected
for standard thin disk theory given by the Novikov-Thorne value of
$\eta_{\rm NT}\approx 18\%$.  Unlike the MCAF models in
\citet{2012MNRAS.423.3083M} and MAD models in \citet{tnm11,tmn12,2012JPhCS.372a2040T}, these models involve weak magnetic fields
similar to many other models published in the literature \citep{mb09}.
Interestingly, in this simulation, the MA term dominates the EM term
for energy extraction from the BH, with the radiation absorbed
contributing to a significant decrement in efficiency.  About half of
the EM energy extracted goes into the EM jet driven by the BZ
mechanism.  The true unbound wind has an efficiency of
$\eta_{\rm{}w,unb,tavg,o}\sim 1\%$ by $r\sim 50r_g$, which is as
efficient as the radiation (but this is probably an upper limit, see
\citealt{2012MNRAS.426.3241N,2013MNRAS.436.3856S}).

The spin-up rates show that the black hole is spinning down, but only
mildly so compared to the extreme spin down occurring in MAD/MCAF
simulations \citep{2012MNRAS.423.3083M}.  The spin-down is dominated by the EM, MA,
PA terms with negligible contribution by the radiation absorbed by the
BH.  The spin-down for the spin chosen is comparable to those for
weakly magnetized disks
\citep{2004ApJ...611..977M,2004ApJ...602..312G}, which have a
spin-equilibrium value of $a/M\sim 0.9$.  So we would expect a similar
spin-equilibrium for our models.

The $\alpha$ viscosities are order $\alpha\sim 0.03$, which is lower
than those in MADs/MCAFs where $\alpha\sim 1$.  The quality factors
show that our simulation marginally resolves the turbulent structures
in the disk in all directions and for both the mass and field
components.  The simulation marginally resolves the MRI with $Q_{\rm
  MRI}\gtrsim 10$.  The field has not reached the MAD/MCAF state as
indicated by $\Qtwo\gtrsim 1$ such that there are still about $4$ full
wavelengths that can fit vertically inside the disk.  This is also
consistent with the relatively low $\Upsilon\sim 3$.  Even though the
field is ordered and dipolar ($|\Phi_{\rm H}/\Psi_{\rm fH}|\sim 1$),
it is relatively weak due to the limited available magnetic flux in
the initial conditions (as often used by many researchers).

\begin{table*}
\begin{center}
\caption{Physical Diagnostics and Numerical Quality Measurements for model A0.94BpN100L20}
\begin{tabular}[h]{|r|r|r|r|r|r|r|r|r|r|r|}
\multicolumn{11}{c}{Rest-Mass Accretion/Ejection Rates and Radiative Luminosity per unit Eddington}\\
\hline
              $\dot{M}_{\rm{}H}$  &   $\dot{M}_{\rm{}in,i}-\dot{M}_{\rm{}H}$  &   $\dot{M}_{\rm{}in,o}-\dot{M}_{\rm{}H}$  &    $\dot{M}_{\rm{}j}$  &    $\dot{M}_{\rm{}mw,i}$  &     $\dot{M}_{\rm{}mw,o}$  &    $\dot{M}_{\rm{}w,i}$  &   $\dot{M}_{\rm{}w,o}$ &    $\dot{M}_{\rm{}w,unb,tavg,i}$  &   $\dot{M}_{\rm{}w,unb,tavg,o}$  &    $L_{\rm{}rad,o}$  \\
\hline
                   24  &                                       85  &                                       440  &                   0.26  &                    0.48  &                      3.1  &                     80  &                     560  &    0.03 & 0.11 &                   1.3  \\  %  rad1
\hline
\hline
\end{tabular}
\begin{tabular}[h]{|r|r|r|r|r|r|r|r|r|r|r|r|r|r|}
\multicolumn{14}{c}{Percent Energy Efficiency: BH, Jet, Winds, Radiation, and NT}\\
\hline
              $\eta_{\rm{}H}$  &     $\eta^{\rm{}EM}_{\rm{}H}$  &   $\eta^{\rm{}MAKE}_{\rm{}H}$  &   $\eta^{\rm{}PAKE}_{\rm{}H}$  &     $\eta^{\rm{}EN}_{\rm{}H}$  &      $\eta^{\rm{}RAD}_{\rm{}H}$  &      $\eta_{\rm{}j}$  &    $\eta^{\rm{}EM}_j$  &     $\eta^{\rm{}MAKE}_{\rm{}j}$  &        $\eta_{\rm{}mw,o}$  &     $\eta_{\rm{}w,o}$  &  $\eta_{\rm{}w,unb,tavg,o}$  &   $\eta^{\rm{}RAD}_{\rm{}o}$  &     $\eta_{\rm{}NT}$  \\
\hline
                18.5  &                          11  &                            18  &                            22.4  &                          -4.42  &                           -10.5  &                6.3  &                   5.38  &                            0.92  &                                5.14  &                  1.97  &   0.82 &        0.94  &                    17.9  \\  %  rad1
\hline
\hline
\end{tabular}
\begin{tabular}[h]{|r|r|r|r|r|r|r|r|r|r|r|r|r|r|r|}
\multicolumn{14}{c}{Spin-Up Parameter: BH, Jet, Winds, Radiation, and NT}\\
\hline
              $s_{\rm{}H}$  &       $s^{\rm{}EM}_{\rm{}H}$  &       $s^{\rm{}MA}_{\rm{}H}$  &      $s^{\rm{}PA}_{\rm{}H}$  &      $s^{\rm{}EN}_{\rm{}H}$  &     $s^{\rm{}RAD}_{\rm{}H}$  &      $s_{\rm{}j}$  &      $s^{\rm{}EM}_j$  &       $s^{\rm{}MA}_{\rm{}j}$  &         $s_{\rm{}mw,o}$  &      $s_{\rm{}w,o}$&      $s_{\rm{}w,unb,tavg,o}$  &     $s^{\rm{}RAD}_{\rm{}o}$  &    $s_{\rm{}NT}$  \\
\hline
             -0.162  &                       -0.448  &                       0.255  &                       0.245  &                       0.01  &                        0.031  &             -2.17  &                -0.269  &                       -1.91  &                            -2.35  &               -26.8  &  -0.2  &        0.016  &                0.411  \\  %  rad1
\hline
\hline
\end{tabular}
\begin{tabular}[h]{|r|r|r|r|r|r|r|r|r|r|r|r|}
\multicolumn{12}{c}{Viscosities, Grid Cells per Correlation lengths and MRI Wavelengths, MRI Wavelengths per full Disk Height, and Radii for MRI Suppression}\\
\hline
              $\alpha_{b,\rm{}eff}$  &      $\alpha_{b,\rm{}eff2}$  &      $\alpha_b$  &     $\alpha_{b,\rm{}M2}$  &     $\alpha_{b,\rm{}mag}$  &     $\bfrac{Q_{n,\rm{}cor,}}{{}_{\{\rho_0,b^2\}}}$  &    $\bfrac{Q_{l,\rm{}cor,}}{{}_{\{\rho_0,b^2\}}}$  &  $\bfrac{Q_{m,\rm{}cor,}}{{}_{\{\rho_0,b^2\}}}$  &  $Q_{\theta,\rm{}MRI,\{i,  o\}}$  &  $Q_{\phi,\rm{}MRI,\{i,  o\}}$  &  $S_{\rm{}d,\rm{}MRI,\{i,  o\}}$  &  $\bfrac{r_{\{S_{\rm{}d},S_{\rm{}d,\rm{}weak}\}}}{{\  }_{\rm{}MRI=1/2}}$  \\
\hline
                      0.029  &                       0.029  &           0.42  &                     0.28  &                      0.28  &                                               21,  16                                              &  14,                                              10  &                         8,     6  &                       6,     6  &                         12,     12  &                                                    4,                  4   &  -,  -  \\  %  rad1
\hline
\hline
\end{tabular}
\begin{tabular}[h]{|r|r|r|r|r|r|r|r|r|r|r|r|r|r|}
\multicolumn{14}{c}{Absolute Magnetic Flux per unit: Rest-Mass Fluxes, Initial Magnetic Fluxes, Available Magnetic Fluxes, and BH Magnetic Flux}\\
\hline
              $\Upsilon_{\rm{}H}$  &    $\Upsilon_{\rm{}in,i}$  &     $\Upsilon_{\rm{}in,o}$  &     $\Upsilon_{\rm{}j}$  &    $\Upsilon_{\rm{}mw,i}$  &    $\Upsilon_{\rm{}mw,o}$  &    $\Upsilon_{\rm{}w,i}$  &    $\Upsilon_{\rm{}w,o}$  &    $\left|\frac{\Psi_{\rm{}H}}{\Psi_1(t=0)}\right|$  &     $\left|\frac{\Psi_{\rm{}H}}{\Psi_2(t=0)}\right|$  &  $\left|\frac{\Psi_{\rm{}H}}{\Psi_3(t=0)}\right|$  &  $\left|\frac{\Psi_{\rm{}H}}{\Psi_a}\right|$  &     $\left|\frac{\Psi_{\rm{}H}}{\Psi_s}\right|$  &     $\left|\frac{\Phi_{\rm{}H}}{\Psi_{\rm{}fH}}\right|$  \\
\hline
                    3.2  &                       0.47  &                       0.91  &                    2.1  &                       4.3  &                       3.4  &                      1.1  &                      1.7  &                                                 0.72  &                                                 0  &                                                 0  &                                            0.96  &                                            0.96  &                                                    1   \\  %  rad1
\hline
\hline
\end{tabular}
\end{center}
\label{tbls}
\end{table*}

%%% Local Variables:
%%% mode: latex
%%% TeX-master: "ms"
%%% End:

\section{Summary}
\label{sec:summary}

We have incorporated the M1 closure for radiation into
the code HARMRAD.  Several radiative tests demonstrate the accuracy,
robustness, and speed of the method in both the optically thin and
thick regimes in both unmagnetized and magnetized regimes.

We also performed a relatively low-resolution short-duration fully 3D
simulation of super-Eddington accretion onto a rotating black hole
with spin $a=0.9375$ with $\Mdot\sim 100L_{\rm Edd}/c^2\sim
20\Mdotedd$.  The magnetic field was chosen to be sub-MAD levels to
focus on the effects of radiative physics.  The disk is essentially
Keplerian, with a substantial negative matter energy flux through the
horizon.  There is no sign of the global
Papaloizou-Pringle instability \citep{1984MNRAS.208..721P}, as likely
due to turbulence \citep{1987MNRAS.227..975B}, but the
magneto-rotational instability \citep{1998RvMP...70....1B} drives
turbulent accretion and winds. There is an electromagnetically-driven
jet in the polar regions, which we show is powered by the
Blandford-Znajek (BZ) mechanism.  The radiative luminosity is
essentially near Eddington, but both the electromagnetic jet and
radiative emission are geometrically beamed in the polar regions, so
the isotropic equivalent luminosity for the polar regions is
super-Eddington.

As applied to the cosmological growth of black holes, as discussed in
the introduction and given by Eq.~(\ref{BHgrow}), the fiducial 3D
model with $a/M=0.9375$ has a radiative efficiency of $\eta_{\rm
  rad}\approx 1\%$ while the accretion efficiency remains high at
$\eta_{\rm acc}\approx 20\%$ near Novikov-Thorne values expected for
an optically thick geometrically thin disk.  Over cosmological times,
this allows growth from $\MBH=10M_{\odot}$ to arbitrarily high $\MBH$,
which suggests that, even for modestly super-Eddington rates of
$\Mdot\sim 100L_{\rm Edd}/c^2\sim 20\Mdotedd$, the black
hole spin does not limit black hole mass growth over cosmological
times.  In addition, while the spin-up parameter of $s \approx -0.16$
shows the BH of spin $a/M=0.9375$ will spin down, this rate is similar
to obtained in other comparable non-MAD GRMHD simulations \citep{2004ApJ...611..977M} for
which the spin equilibrium is $a/M\sim 0.9$.  So we expect the black
hole to remain rapidly rotating during this super-Eddington accretion.

These simulations show that the GRRMHD code HARMRAD is capable of
accurately simulating optically thick or thin problems, including
accretion flows around rotating black holes.  However, the adopted M1
closure assumes the radiation is isotropic in a single frame, which
limits the ability of the closure to handle general anisotropic
radiation distributions.  Also, our current M1 scheme is based upon
frequency integrated quantities.  However, the M1 closure improves
upon the flux-limited diffusion approximation.
We also include no Componization \citep{2013MNRAS.433.1054Y}.

In addition, while the code is stable, accurate, and fast even with
our use of the high-resolution PPM scheme, in some cases PPM exhibits
non-uniform (due to random grid-scale oscillations) convergence
behavior as well as mild noise when the LAXF Riemann solver is used at
sharp discontinuities.  We plan to continue to improve PPM and the
Riemann solve used to keep the high resolution of PPM while avoiding
the oscillations and non-uniform convergent behavior.  We also plan to
explore the full higher-order IMEX methods provided in
section~\ref{sec:imex}, which are presented for easy implementation
into HARM/HARMRAD.

\section*{Acknowledgments}

We thank James~M.~Stone and Omer~Blaes for useful discussions.  RN and
AS were supported in part by NSF grant AST1312651. AT was supported
by NASA through the Einstein Fellowship Program, grant
PF3-140115.  We acknowledge NSF support via XSEDE
resources, NICS Kraken and Nautilus under grant numbers TG-PHY120005
(JCM), TG-AST100040 (AT), TG-AST080026N (RN and AS), and NASA support via High-End
Computing (HEC) Program through the NASA Advanced Supercomputing (NAS)
Division at Ames Research Center (JCM, AS, and RN) that provided
access to the Pleiades supercomputer.

\section*{SUPPORTING INFORMATION}

Additional Supporting Information may be found in the online version
of this article: Movie file. Movie of the fiducial model A0.94BpN100L20
showing the animated version of Fig.~\ref{evolvedmovie} (see caption
alongside movie file for more detail) available at \href{http://www.youtube.com/watch?v=q2DeKxUHae4}{http://www.youtube.com/watch?v=q2DeKxUHae4}.

\appendix

\bibliographystyle{mnras}
{\small
%\bibliography{mybibnew}

}

\label{lastpage}
\end{document}